\documentclass[twocolumn]{aastex62}

\received{4 April 2019}
\revised{29 August 2019}
\accepted{29 August 2019}
\submitjournal{ApJ}

\shorttitle{Aluminium-26 from massive binary stars}
\shortauthors{Brinkman et al.}

\begin{document}

\title{ALUMINIUM-26 FROM MASSIVE BINARY STARS I: NON-ROTATING MODELS\footnote{This paper is dedicated to the celebration of the 100th birthday of Margaret Burbidge, in recognition of the outstanding contributions she has made to nuclear astrophysics (Burbidge et al. 1957). }}

\correspondingauthor{Hannah Brinkman}
\email{hannah.brinkman@csfk.mta.hu}

\author{H. E. Brinkman}
\affil{Konkoly Observatory,\\Research Centre for Astronomy and Earth Sciences, Hungarian Academy of Sciences\\ Konkoly Thege Miklos ut 15-17, H-1121 Budapest, Hungary}
\affil{Graduate School of Physics, University of Szeged,  Dom ter 9, Szeged, 6720 Hungary}

\author{C. L. Doherty}
\affiliation{Konkoly Observatory,\\Research Centre for Astronomy and Earth Sciences, Hungarian Academy of Sciences\\ Konkoly Thege Miklos ut 15-17, H-1121 Budapest, Hungary}
\affiliation{Monash Centre for Astrophysics (MoCA), School of Physics and Astronomy, Monash University, Victoria 3800, Australia}

\author{O. R. Pols}
\affiliation{Department of Astrophysics/IMAPP, Radboud University, PO Box 9010, 6500 GL Nijmegen, The Netherlands}

\author{E. T. Li}
\affiliation{College of Physics \& Energy, Shenzhen University, China}

\author{B. C\^{o}t\'{e}}
\affiliation{Konkoly Observatory,\\Research Centre for Astronomy and Earth Sciences, Hungarian Academy of Sciences\\ Konkoly Thege Miklos ut 15-17, H-1121 Budapest, Hungary}
\affiliation{Joint Institute for Nuclear Astrophysics - Center for the Evolution of the Elements, USA}

\author{M. Lugaro}
\affiliation{Konkoly Observatory,\\Research Centre for Astronomy and Earth Sciences, Hungarian Academy of Sciences\\ Konkoly Thege Miklos ut 15-17, H-1121 Budapest, Hungary}
\affiliation{Monash Centre for Astrophysics (MoCA), School of Physics and Astronomy, Monash University, Victoria 3800, Australia}

\begin{abstract}
\noindent Aluminium-26 is a short-lived radionuclide with a half-life of 0.72\,Myr, which is observed today in the Galaxy via $\gamma$-ray spectroscopy and is inferred to have been present in the early Solar System via analysis of meteorites. Massive stars are considered the main contributors of $^{26}$Al. Although most massive stars are found in binary systems, the effect, however, of binary interactions on the $^{26}$Al yields have not been investigated since Braun\,\&\,Langer\,(1995). Here we aim to fill this gap. We have used the MESA stellar evolution code to compute massive (10\,M$ _{\odot}$\,$\leq$\,M\,$\leq$80\,M$ _{\odot} $), non-rotating, single and binary stars of solar metallicity (Z=0.014). We computed the wind yields for the single stars and for the binary systems where mass transfer plays a major role. Depending on the initial mass of the primary star and orbital period, the $^{26}$Al yield can either increase or decrease in a binary system. For binary systems with primary masses up to $ \sim $35-40\,M$ _{\odot} $, the yield can increase significantly, especially at the lower mass-end, while above $\sim $45\,M$ _{\odot}$ the yield becomes similar to the single star yield or even decreases. Our preliminary results show that compared to supernova explosions, the contribution of mass-loss in binary systems to the total $^{26}$Al abundance produced by a stellar population is minor. On the other hand, if massive star mass-loss is the origin of $^{26}$Al in the early Solar System, our results will have significant implications for the identification of the potential stellar, or stellar population, source.
\end{abstract}


\section{Introduction} \label{intro}
Aluminium-26 ($ ^{26} $Al), a short-lived radionuclide with a half-life of 0.72\,Myr, decays to an excited state of $ ^{26} $Mg, and the quick, subsequent decay to the ground state releases a $ \gamma $-photon at an energy of 1.81\,MeV. These photons, and thus the decay of $ ^{26} $Al, have been detected in the Galaxy in $ \gamma  $-ray spectroscopic observations by the COMPTEL and INTEGRAL satellites. From these observations it has been determined that the Galaxy contains about 2-3\,M$ _{\odot} $ of $ ^{26} $Al \citep{Diehl2013}. Considering the short half-life of this isotope, the production of $ ^{26} $Al is an ongoing process in the Galaxy. By mapping the distribution of the observed $ \gamma $-rays, it has been shown that most of the $ ^{26} $Al is confined to the plane of the Galaxy, and there are clumps that coincide with known OB-associations, i.e. groups of stars with masses $ \geq $10\,M$ _{\odot} $ (see Figure\,16 of \citealt{Diehl2013}), such as the Cygnus region \citep{MartinCygnus}, the Scorpius-Centaurus region \citep{DiehlScoCen}, and the Carina region \citep{Voss2012}. The observations indicate that massive stars are the main source of $ ^{26} $Al in the Galaxy. These stars expel the $ ^{26} $Al isotope through winds and supernova explosions.\\
\indent $ ^{26} $Al was also present in the early Solar System as inferred from $ ^{26} $Mg excess in meteorites \citep{Lee1977}. From this excess the $ ^{26} $Al/$ ^{27} $Al ratio at the time of the formation of the solar system is determined. This ratio is currently reported to be (5.23$ \pm$0.13)$ \times $10$ ^{-5} $ \citep{Jacobsen2008}. $ ^{26} $Al played an important role in the early stages of our Solar System because its decay has been linked to the heating of planetesimals \citep{Lichtenberg2016}, the first rocks with sizes between 10-100 km, from which the rocky planets such as our Earth are believed to have formed. Due to the large amount of $ ^{26} $Al present in the early Solar System, its radioactive heating was dominant over the contribution of other radionuclides, however its origin is still unclear. The abundance of $ ^{26} $Al in the early Solar System was higher than the $ \gamma $-ray observed abundance of $ ^{26} $Al currently in the interstellar medium, and an extra source of $ ^{26} $Al is needed (\citealt{Lugaro2018, Benoit2019}).\\ 
\indent Furthermore, there appears to be a discrepancy between the $ ^{60} $Fe/$ ^{26} $Al ratio from supernova models on the one hand and $ \gamma $-ray observations on the other hand, and the early Solar System \citep[Section\,3.5]{Lugaro2018}. The $ ^{60} $Fe/$ ^{26} $Al abundance ratio from the $\gamma$-ray observations is approximately 0.55, while in the early Solar System it is about 3-300 times lower. In current supernova models, this ratio is about 3-10 times higher \citep{AustinSNe, Sukhbold2016}. This discrepancy suggests either an underproduction of $ ^{60} $Fe in the early Solar System compared to the Galactic average, or an extra source of $ ^{26} $Al.\\
\indent $ ^{26} $Al is significantly produced in all stars with an initial mass above $ \sim $2\,M$_{\odot}  $ by proton captures on $ ^{25} $Mg during core and shell hydrogen-burning. For massive stars in particular there are two additional production phases: carbon/neon convective shell burning, and explosive neon burning during the supernova \citep{LandC2006}. The $ ^{26} $Al produced during the core hydrogen-burning stage is mainly ejected through stellar winds driven by radiative pressure. Only the $ ^{26} $Al remaining in the envelope, which has not decayed or been destroyed by then, is expelled by the supernova explosion. Some of the $ ^{26} $Al produced during shell hydrogen-burning is expelled over time by the winds as well. The $ ^{26} $Al produced in the other phases is expelled during the supernova explosion. The stellar winds, depending on the initial mass and metallicity, can be rather strong, leading to mass-loss rates ($ \dot{M} $) ranging from 10$ ^{-7} $\,M$ _{\odot} $/yr to 10$ ^{-4} $\,M$ _{\odot} $/yr. In some cases these winds can drive off the entire envelope of the star. For the most massive stars the winds can be so strong that when the star leaves the main sequence, the hydrogen burning shell is stripped away as well. What is left of the star is an exposed helium core, also known as a Wolf-Rayet star.  Through these two mechanisms, winds and supernova explosions, massive stars are considered the main contributors of $ ^{26} $Al in the Galaxy. For overviews of massive star evolution and their supernovae, see e.g. \cite{LangerReview} and \cite{WoosleyHeger2002}.\\
\indent So far most research has been focussed on calculating the $ ^{26} $Al yields of massive single stars, both rotating and non-rotating, including their winds and supernova explosions, see for example \cite{LandC2006, LandC2018, WandH2007} and \cite{Ekstrom2012}. However, most massive stars are found in binary systems (see \citealt{DucheneandKraus} for a review), and are close enough to interact with each other. \cite{Sana2012} find that more than 70\% of all O-type stars ($\gtrsim$ 15\,M$ _{\odot} $) interact with their companion during their lifetimes. The binary systems show a strong preference for close orbits and have a uniform distribution of mass ratios. More than 25\% of O-stars will interact with a companion before the end of hydrogen burning in their cores \citep{Sana2012}. The recent detections of gravitational waves from merging binary black holes and binary neutron stars \citep{BBHMAbbott2016,BNSMAbbott2017} have highlighted the astrophysical relevance of massive binaries. Binary stars are therefore in important subject to study.\\
\indent Interestingly, the binary interactions can also influence the $ ^{26} $Al yields from massive stars, as originally proposed by \cite{BraunandLanger}. The binary interaction process in close binaries that has the strongest impact on the $ ^{26} $Al wind yields is mass transfer, which can radically alter the way stars in binaries lose mass as compared to single stars. Mass transfer between the stars can change the time at which mass loss starts as compared to a single star, as well as the amount of mass that is lost from the star. Because of this difference, investigating massive binary systems is important for understanding both the Galactic distribution of $ ^{26} $Al, as well as the possible stellar sources that produced the $ ^{26} $Al in the early Solar System.\\
\indent Even though binaries are crucial to address these issues, no yields are available except for two systems in a brief conference proceedings paper by \cite{BraunandLanger}. The results found by these authors 24 years ago are in urgent need of re-examination, update, and expansion, especially given the predominance of interacting binaries among massive stars discussed above. The ubiquity of massive binaries can have a potentially huge impact on element production and galactic chemical evolution. This paper aims to demonstrate the potential that is offered by massive binary systems. With this objective, we present here the $ ^{26} $Al wind yields from single, massive non-rotating stars, and combined yields of wind and mass transfer for binary systems. In Section\,\ref{method} we describe the method and the physical input. In Section\,\ref{OtherMasses} we present the results of the simulations for stars with masses from 10 to 80\,M$ _{\odot} $. In Section\,\ref{sec:Discussion} we discuss the impact of different aspects of binary evolution, the influence of reaction rates on the yields, and implications of this study on Galactic $ \gamma $-rays and the early Solar System. In Section\,\ref{sec:Conclusion} we end with our conclusions.
\section{Method and input physics} \label{method}
We have used version 10398 of the MESA stellar evolution code \citep{MESA1,MESA2,MESA3,MESA4} to calculate non-rotating massive star models. Several other stellar evolution codes allow for modelling binary evolution, see for an overview \citet[Table\,2]{DeMarcoIzzard2017}. Here, we have decided to use MESA because it is a widely supported tool, and because it includes the option of future work on the supernova explosions. The input physics we used for the single stars and the binary systems is described in the next section.\\
\indent The MESA input files used for the simulations presented here are available as online material.\\
\indent Our focus is on the $^{26}$Al yield from winds and mass transfer. To calculate these yields, we need to integrate over time because unlike a supernova explosion, winds take place over a longer timespan. Because of this time-dependence, a part of the $ ^{26} $Al decays in the interstellar medium after it is expelled from the star. We have not taken this decay into account in these time-integrated yields. For the calculation of the yield we have evolved the stars up to the onset of carbon burning. At this point the further evolution will only take a few thousand years or less, and this is not enough time for either the winds to expel much more mass from the star, or for $ ^{26} $Al to decay further. Because $ ^{26} $Al is destroyed during helium burning by neutron-capture reactions, there is almost no $ ^{26} $Al left in the stellar core at the onset of carbon burning. If the stars do not reach the onset of carbon burning, the simulations were stopped after 10$ ^{4} $ time steps.
The initial masses for our models are 10, 15, 20, 25, 30, 35, 40, 45, 50, 60, 70, and 80\,M$ _{\odot} $. All except for 70 and 80\,M$ _{\odot} $ are also the primary masses for the binary systems, where primary always refers to the initially more massive star. The masses of the secondary stars are set by the mass ratio, $q\,=\, \frac{M_{2}}{M_{1}} $, in which $M _{1} $ is the primary mass and $M _{2} $ is the secondary mass. The mass ratio used here is $q\,=\,0.9$, unless otherwise indicated.
\subsection{Input physics}\label{Input}
Within MESA, a large number of choices can be made for the input physics. Here we briefly describe and motivate the main choices we made. The MESA input files used for the simulations are available as online material.\\
\indent The initial composition used in this study is solar with $Z$=0.014, following \cite{Asplund2009}. For the initial helium content we have used $Y$=0.28. Our nuclear network contains all the relevant isotopes for hydrogen, helium burning, and early carbon burning, as well as all relevant isotopes connected to the production and destruction of $ ^{26} $Al, including the ground and isomeric states of $ ^{26} $Al. For a visual representation of the isotopes connected to $ ^{26} $Al production and destruction, see Figure\,7a of \cite{Lugaro2018}.\\
\indent The network contains the following 63 isotopes: n, $ ^{1,2} $H, $ ^{3,4} $He, $ ^{6,7} $Li, $^{7,8}$Be, $ ^{8} $B, $ ^{12-14} $C, $ ^{13-15} $N, $ ^{14-19} $O, $ ^{17-20} $F, $ ^{19-23} $Ne, $ ^{21-24} $Na,$ ^{23-26} $ Mg, $ ^{25} $Al, $ ^{26} $Al$ _{g} $,$ ^{26} $Al$ _{m} $,$ ^{27,28} $Al, $ ^{27-30} $Si, $ ^{28-31} $P, $ ^{30-34} $S, $ ^{32-35} $Cl, and $ ^{56,57} $Fe.
The reaction rates used are based on the thermonuclear rates by \cite{NACRE} (NACRE-rates), and when not available by \cite{Caughlan1988}, see Section 4.4 of \cite{MESA1}. As of version 10398 of the MESA code, the isomeric states are implemented. The reaction rates for $ ^{26} $Al$ _{g}\,\leftrightarrow\,^{26} $Al$ _{m} $ are taken from \cite{gupta01}. The other reactions involving the $ ^{26} $Al isomers are taken from the JINA reaclib \citep{JINA2010}, using \cite{straniero13} and \cite{iliadis10}, which are the most up-to-date rates for $ ^{26} $Al. In Section\,\ref{RATES} we discuss the uncertainties in the reaction rates and their influence on the $ ^{26} $Al production. There we will also discuss briefly how using the JINA rates instead of the NACRE rates affects the yields.\\
\indent To establish the location of convective boundaries, we have used the Ledoux criterion. Several free parameters need to be set in MESA to model convection and mixing. The mixing-length parameter, $ \alpha_{mlt} $, was set to 1.5. Semi-convection, which occurs in a region that is stable according to the Ledoux criterion but unstable according to the Schwarzschild criterion, is modelled by a free parameter $ \alpha_{sc} $. We use $ \alpha_{sc} $=0.1, which we found to give results that best resemble the non-rotating tracks in the Hertzsprung-Russell diagram by \citet[see their Figure 6]{MandM2000}. We made use of convective overshooting via the "step-overshoot" scheme, where the overshoot region extends above the convective border by a length $l$, given by $l=H _{p} \cdot \alpha_{ov}$. In the equation, $H _{p} $ is the pressure scale height at the convective border and $ \alpha_{ov} $ is a free parameter. The exact value of this parameter is uncertain (see e.g. \citealt[Section\,2.4]{Brott2011}). We have chosen to use $ \alpha_{ov} $=0.2, which gives moderate overshooting and is within the uncertainties given by \cite{Brott2011}. This value is commonly used by other groups to compute massive stars (\citealt{OvershootClaret} and references therein).\\
\indent We have used the wind mass-loss scheme as described in \cite{Brott2011} and as it is implemented by \cite{Schootemeijer2018} for MESA, who use a combination of the prescription by \cite{NieuwenhuijzendeJager1990,Hamann1995,Vink2000} and \cite{Vink2001}. This scheme is commonly used to parametrise mass loss from massive stars.\\
\indent We did not include stellar rotation in our models, because we want to avoid the associated additional complications and uncertainties, for more details see e.g. \cite{BookMaeder2009}. Instead our focus in this first, exploratory study is on the potential impact of binary interactions on the yields. In a future paper we plan to include rotational mixing in our models, see also Section\,\ref{sec:Conclusion}.
\subsection{Exploration of the orbital period}\label{SNB}
\begin{figure*}[t]
	\includegraphics[width=1.0\textwidth]{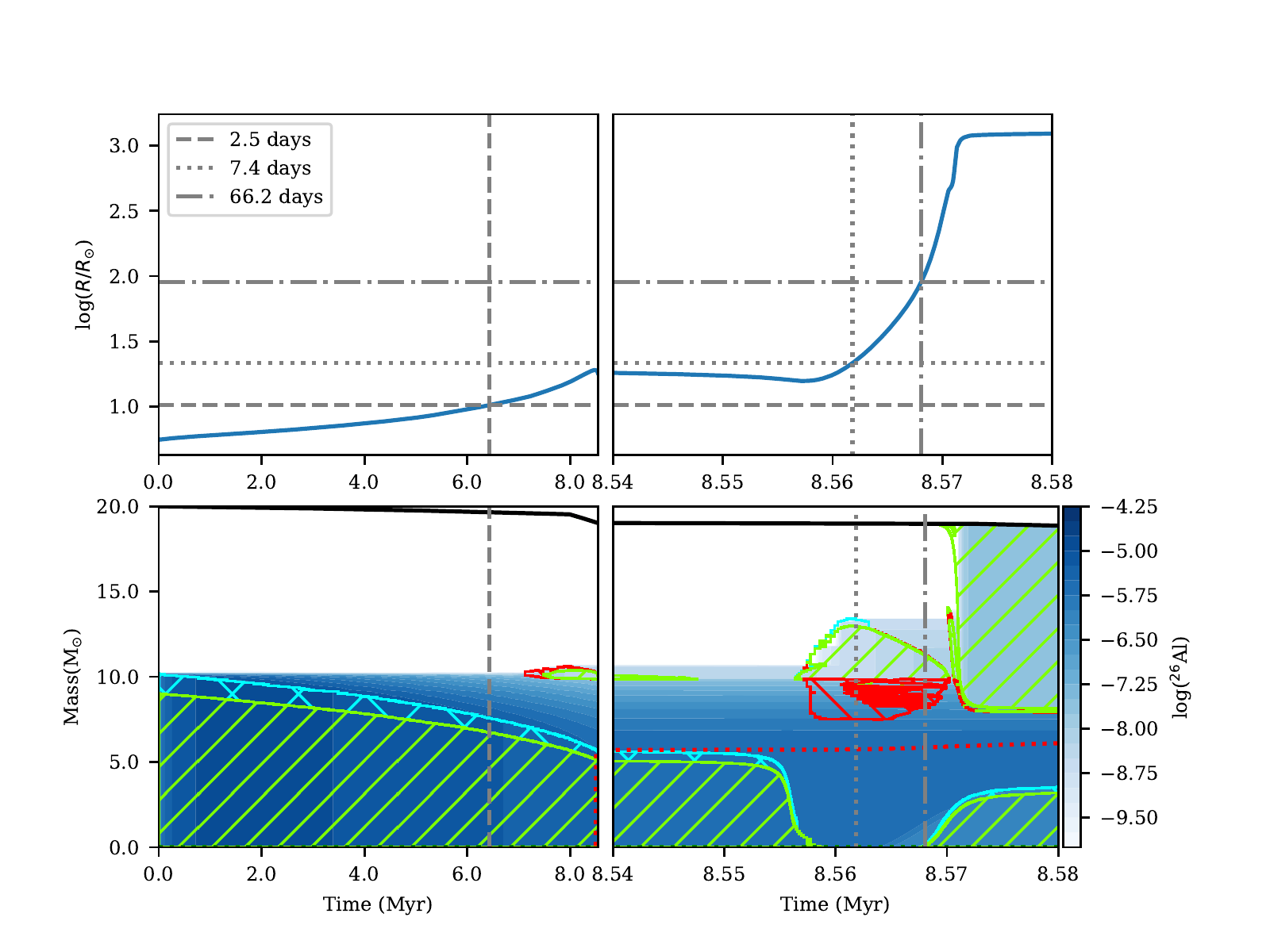}
	\caption{Principle of the semi-numerical binary scheme. The upper panels show the radius evolution of a 20\,M$ _{\odot} $ star over time. The left panel shows the slow expansion of the radius during the main sequence. The right panel shows the rapid expansion at the end of hydrogen burning and the onset of helium burning. The horizontal lines represent the size of the Roche lobe for three periods, 2.5, 7.4, and 66.2 days. The time when the size of the Roche lobe crosses the radius (grey, vertical lines), represents the time when the mass transfer would start if the system was a binary and we strip the envelope here. The bottom panels present two parts of a Kippenhahn-diagram\,(KHD) for a 20\,M$ _{\odot} $ star. The green shaded areas correspond to areas of convection. The cyan shaded areas correspond to overshooting. The red shaded areas correspond to semi-convection. The red dotted line in the lower right panel indicates the hydrogen depleted core, or helium core, where the hydrogen content is below 0.01 and the helium content is above 0.1. The colour scale shows the $ ^{26} $Al mass fraction as a function of the mass coordinate and time. The vertical lines in the lower panels correspond to the vertical lines in the top panels, showing where in the evolution the envelope is stripped.}
	\label{Semi-Numerical}
\end{figure*}
To start to explore the influence of the initial period of the binary systems on the $ ^{26} $Al yields for a given initial mass of the primary star, we first apply the simulations of the single stars in an analytical binary scheme, which we call the semi-numerical binary (SNB) scheme. In this scheme, we vary the two parameters we expect to have the greatest impact on the yields: the primary mass, $M _{1} $, and the orbital period, $P$, while we keep the mass ratio constant, $q=0.9$. With Kepler's Third law, \begin{equation}
	\frac{a^{3}}{P^{2}} = \frac{G(M_{1}+M_{2})}{4\pi^{2}}, \label{Eq1}
\end{equation}
and Eggleton's approximation of the size of the Roche lobe \citep{Eggleton},
\begin{equation}
	\frac{R_{L1}}{a} = \frac{0.49q^{-\frac{2}{3}}}{0.6q^{-\frac{2}{3}}+ln(1+q^{-\frac{1}{3}})}, \label{Eq2}
\end{equation}
the point in time when the primary overflows its Roche lobe can be determined.\\
Combined with the fully evolved single star, this allows an initial estimation of the amount of $ ^{26} $Al that can potentially be the ejected by a binary system, which we then tested with the full numerical models described in Section 2.3.\\
\indent For periods ranging from a few days to $ \sim $100 days, we  calculate the size of the Roche lobe. At the time when the stellar radius equals the size of the Roche lobe, we assume that the full envelope of the star is stripped away. Depending on the evolutionary stage, this is either down to the upper border of the overshoot region for mass transfer during hydrogen burning, or down to the top of the hydrogen-depleted or helium core for mass transfer after hydrogen burning. The helium core is defined as the part of the star where the hydrogen abundance is below 0.01 and the helium abundance is above 0.1. For these stripped regions, we calculate the $ ^{26} $Al yield by summing the amount of $ ^{26} $Al in all the cells stripped away. We illustrate this procedure in Figure\,\ref{Semi-Numerical}. The vertical dashed line shows where the envelope is stripped for an initial period of 2.5 days. The envelope is stripped down to the cyan dashed area, representing the overshoot region. The dotted line and the dashed-dotted line show the same for periods of 7.4 and 66.2 days, respectively.\\
\indent The limits for the period were chosen based on the stellar radius of the primary stars and in such a way that mass transfer occurs either during hydrogen burning, commonly referred to as Case A mass transfer, or after hydrogen burning, but before the central ignition of helium, commonly referred to as Case B mass transfer \citep{KippenhahnWeigert1967}. The smallest possible orbit was calculated by assuming that the star fills its Roche lobe directly at the zero-age main sequence. This value was multiplied by 2 to give the shortest period we used, which is around 2-3 days. The longest period was chosen to be that of de widest orbit in which the system undergoes Case B mass transfer without having a convective envelope, which is $ \sim$100 days. The convective envelope develops during helium burning, and these mass transfer cases are between Case B and Case C. Case C mass transfer refers to all mass transfer that occurs after core helium burning has finished. The reason for choosing to consider only Case A and Case B mass transfer is that the $ ^{26} $Al in the envelope decays after hydrogen burning. This means that if the mass transfer occurs after He burning, the yields will be almost identical to the single star yields. Since we want to study the impact of binary interactions on the yields, and massive star binaries occur preferentially in close orbits \citep{Sana2012}, we focus on relatively close orbital periods.\\
\indent The change in the size of the Roche lobe due to mass loss through winds of either star is not taken into account in this approach. Therefore, the size of the Roche lobe is determined only by the initial configuration of the system. The duration of the mass transfer phase is not taken into account either, as the envelope is stripped instantaneously.
\subsection{Numerical binary input physics}\label{binary}
After exploring the parameter space with SNBs, we made a selection of systems to use for a fully numerical more computationally demanding binary simulation with MESA. We expect the results to differ because the assumption that we made in the SNB models that the full envelope is stripped is not physical, even with fully non-conservative mass-transfer, and because the SNB scheme does not take the change in the binary parameters into account when the stars are losing mass. Also, more importantly, in the SNB scheme the envelope is stripped instantaneously, while in reality this happens gradually over time. This time dependence of the mass loss affects the $ ^{26} $Al yields because of further decay within the star. We selected the systems such that both cases of mass transfer are covered, and we selected specifically the latest Case A system and the earliest Case B system. The periods are more sparsely sampled than for the SNBs, selecting 3-6 periods for each primary mass, ranging from a few to $ \sim $100 days. Because the closest orbit in which Roche lobe overflow occurs is determined by the initial radius of the primary star, the shortest period is different for each primary mass since the radius scales with mass. Therefore, the set of periods used depends on the primary mass (the periods used can be found in Appendix\,\ref{BigTable}). In these simulations the focus is only on the yields from the primary star, i.e., the secondary star is evolved as well but is not taken into account for the $ ^{26} $Al yields.\\
\indent One of the most important, but also a very uncertain parameters in binary evolution is the efficiency of the mass transfer. The mass-transfer efficiency is the fraction of the mass lost by the primary star that is accreted by the secondary star. In MESA the parameter $ \beta $ is defined as fraction of mass lost from the system such that the mass-transfer efficiency is equal to 1-$ \beta $. In this paper we will follow the definition as used in the MESA code, but note that it is also quite common in the literature to use $ \beta $ for the mass-transfer efficiency itself. With the definition used in MESA, $ \beta $=1 means that no mass is accreted by the secondary, also known as fully non-conservative mass transfer, while $ \beta $=0 means that all mass is accreted by the secondary, which is referred to as conservative mass transfer. Observational evidence suggests that different values for $ \beta $ occur in nature, from rather conservative to non-conservative mass transfer (e.g. \cite{Pols1991,DeMink2007,Schootemeijer2018b}). The main evidence that for massive binaries ($ \gtrsim $20\,M$ _{\odot} $) the mass-transfer efficiency is low, comes from from attempts to reproduce the Galactic population of Wolf-Rayet+O-star binaries. These systems require a $ \beta \gtrsim$0.8 \citep{Petrovic2005,Shao2016}. In this paper, we chose to use fully non-conservative mass transfer, $ \beta $=1. Using this gives us an upper limit to the amount of $ ^{26} $Al ejected from the binary systems. For a few selected systems we varied mass-transfer efficiency (Section\,\ref{Betas}) as well as the mass ratio (Section\,\ref{Qs}). We did not vary the eccentricity of the systems because we assume that all systems are circularized by tidal friction \citep{Zahn1977} by the time the interaction takes place.
\section{Results}\label{OtherMasses}
\indent In this section we first compare the two models by \cite{BraunandLanger} to our results for the same primary masses. We also compare these models to the results by \cite{LandC2006, LandC2018, WandH2007} and \cite{Ekstrom2012}. Subsequently we present the results for the other primary masses and compare these to the results from the literature.
\subsection{20\,M$ _{\odot} $ and 50\,M$ _{\odot} $}
\begin{figure*}\begin{center}
	\includegraphics[trim = 2mm 1mm 9mm 13mm, clip, width=0.45\textwidth]{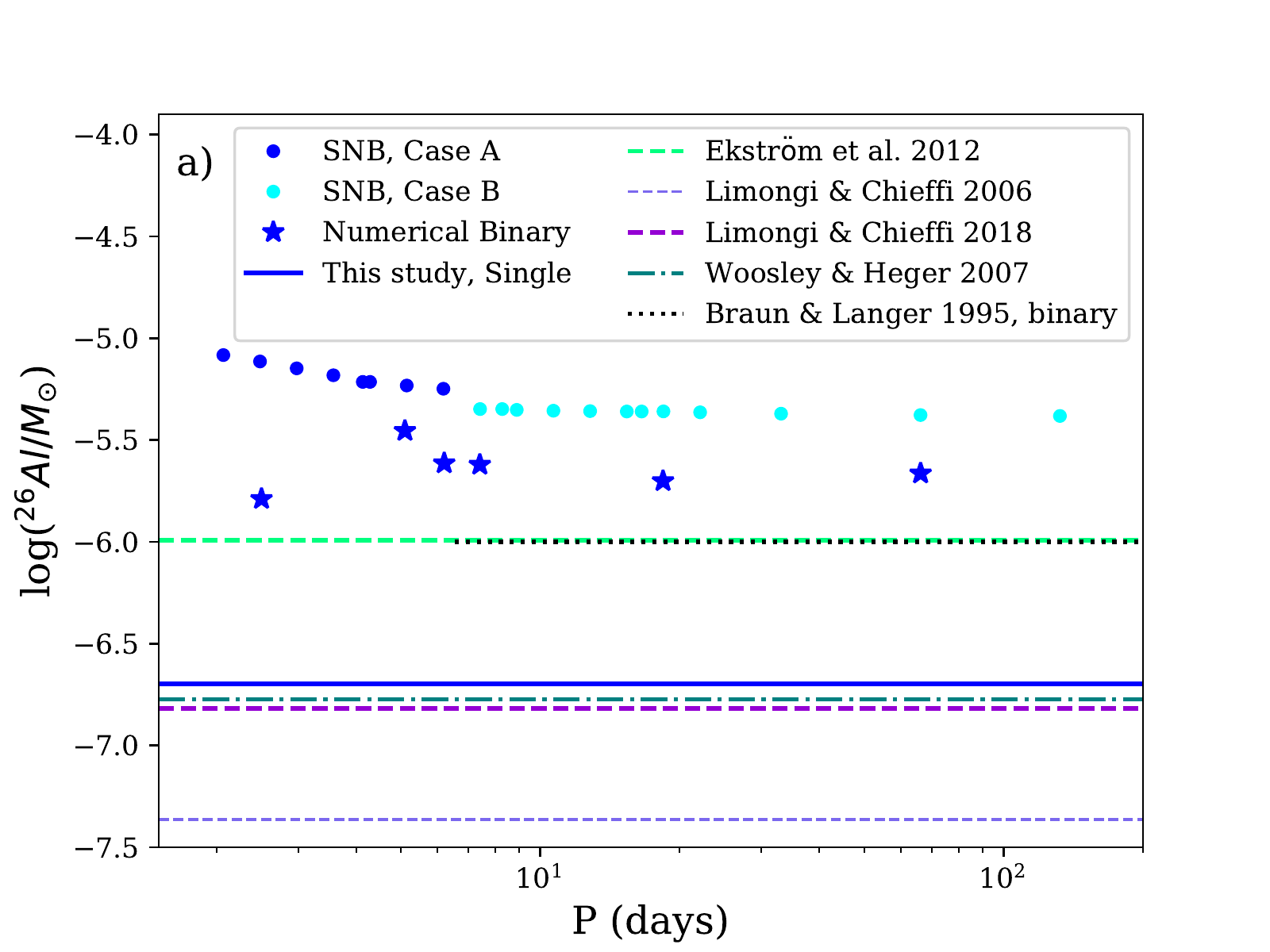}
	\includegraphics[trim = 2mm 1mm 9mm 13mm, clip, width=0.45\textwidth]{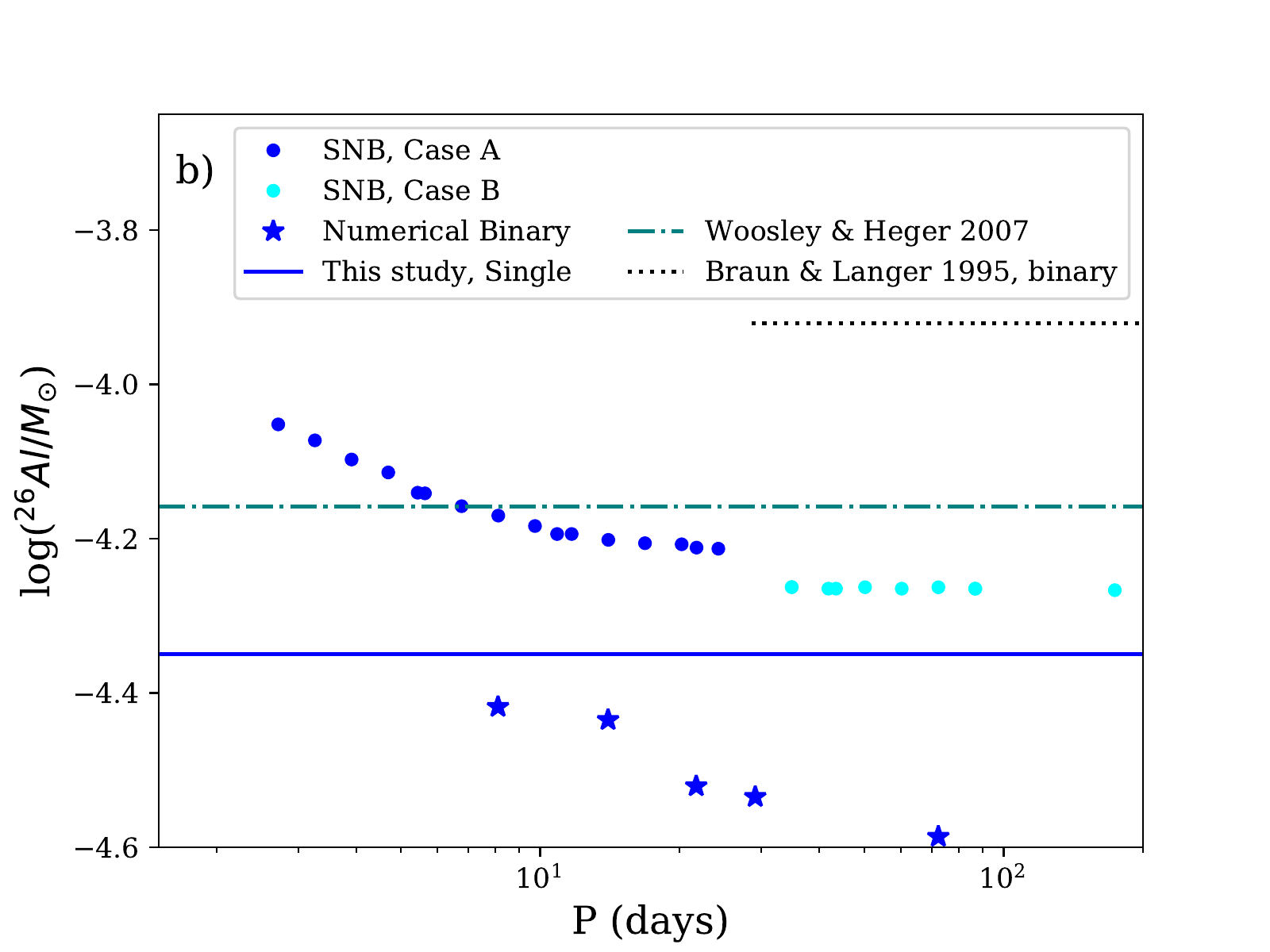}\\
 \end{center}
 \caption{(a) $ ^{26} $Al yields for a 20\,M$ _{\odot} $ single, non-rotating star from this study (blue solid line) and various other studies. Note the difference in te scale, where the left panel covers more than 3 orders of magnitude and the right panel less than one order of magnitude. Please note that the yield from \cite{Ekstrom2012} is almost identical to the yield by \cite{BraunandLanger}. The black dotted line indicates the yield from the binary by \cite{BraunandLanger} given as a line, because the period of this Case B system is unknown. The dots are the yields of our SNBs, blue for Case A and cyan for Case B. The stars indicate the yields for the numerical binaries. (b) Same as Figure\,\ref{20and50}a for the 50\,M$ _{\odot} $ primary.}
\label{20and50}
\end{figure*}
\indent \cite{BraunandLanger} present $ ^{26} $Al yields for two binaries in their paper with initial primary masses of 20\,M$ _{\odot} $ and 50\,M$ _{\odot} $. Here we discuss our results for the same masses. In Appendix\,\ref{20MSUN} and Appendix\,\ref{50MSUN} a detailed description of the 20\,M$ _{\odot} $ and 50\,M$ _{\odot} $ stars respectively and a selection of binary systems with these primary masses can be found. In Appendix\,\ref{BigTable} the $ ^{26} $Al yields for all systems are provided, as well as information on the evolutionary stages of all stars.
\subsubsection{20\,M$ _{\odot} $}\label{20results}
Figure\,\ref{20and50}a shows the $ ^{26} $Al yields for a 20\,M$ _{\odot} $ single star from our models as well as from the literature, the binaries and the SNBs with a companion of 18\,M$ _{\odot} $. The yields from the SNBs (small dots in Figure\,\ref{20and50}) are between a factor of 20-40 higher than the yield of our single star model. The yields decrease until an orbital period of $ \sim $6.2 days and then plateaus for the SNBs. This is caused by a transition in the type of mass transfer, the systems with periods shorter than $ \sim $6.2 days are Case A mass transfer (blue dots in Figure\,\ref{20and50}), those with longer orbital periods are Case B (cyan dots in Figure\,\ref{20and50}). The yields from SNB Case A systems are sensitive to the period, decreasing with increasing orbital period. Since the half-life of $ ^{26} $Al is 0.72\,Myr, part of the $ ^{26} $Al that was present in the core and expelled by the shorter period Case A systems has decayed by the time mass transfer takes place in the widest Case A system. The Case B systems have a lower yield than the Case A systems by a factor up to 2 because for the Case B systems more $ ^{26} $Al has already decayed relative to the Case A systems. However, because mass transfer occurs at a very similar time, between 8.56-8.57\,Myr, the orbital period has almost no influence on the yield from these systems.\\
\indent Compared to their SNB counterparts, the Case B systems follow a similar trend. There is only a small variation in the yields between the shortest and widest orbital period systems. The $ ^{26} $Al yields are lower for the numerical binary systems in all cases. This is because the orbital adjustment during the mass-transfer phase changes the size of the Roche lobe (see Equations \ref{Eq1} and \ref{Eq2}). This causes the star to detach ($ R\,<\,R_{L} $) from its Roche lobe when a smaller amount of mass is lost compared to the SNBs. The Case A systems follow a different trend than their SNB counterparts. For systems with short orbital periods the change in the orbital period plays a larger role. Unlike the SNBs, where the mass is lost in one instantaneous event, the numerical Case A binary systems go through two phases of mass transfer prior to the end of helium burning (Case A and Case AB, see Figure\,\ref{20ALL}d in Appendix\,\ref{20MSUN}). Close to the end of helium burning of the primary star in the closest period systems, the secondary star starts to evolve off the main-sequence and fills its Roche lobe, starting a phase of reverse mass-transfer (see Section\,\ref{RevMT}). This will end the evolution of the system in our set-up. The shorter the period, the earlier the reverse Roche lobe overflow will take place, leading to a lower yield. For the wider Case A systems, which do not go through reverse mass-transfer prior to the end of helium burning of primary, the $ ^{26} $Al yield goes down again due to the internal decay of the $ ^{26} $Al, just as for the SNBs.\\
\indent To put our single and binary yields in perspective, Figure\,\ref{20and50}a also shows the results the results by \cite{LandC2006, LandC2018, WandH2007}; and \cite{Ekstrom2012}. Our wind yield is in agreement with the yield found by \cite{WandH2007} and by \cite{LandC2018}. The yield given by \cite{Ekstrom2012} is much higher than ours, by almost an order of magnitude. This is because, as they mention in their Section\,2.6.2, they have artificially increased their mass-loss rate. They remark that this leads to an order of magnitude larger mass-loss rate, when averaged over time, than the rate by \cite{deJager1988}. This explains the large offset between the results of \cite{Ekstrom2012} and ours. The yields for our binary systems, both numerical and semi-numerical, are still higher than this single star result.\\
\indent We compare our results to the yield given by \cite{BraunandLanger}. The wind yield they find is comparable to the single star yield found by \cite{Ekstrom2012}. Compared to the Case B binary yields we present, this is a factor of $ \sim $2 smaller. While the mass-transfer efficiency for the systems is the same and we both used the Ledoux criterion, the metallicity is slightly different (Z=0.02) and the mass-loss rates are different as well. Another difference is that the secondary is not fully evolved but treated as a point mass. However, it is difficult to say exactly which of these differences leads to the difference in the yield.
\subsubsection{50\,M$ _{\odot} $}\label{50results}
Figure\,\ref{20and50}b shows the results for the 50\,M$ _{\odot} $ single star model, as well as for the binaries and the SNBs with a companion of 45\,M$ _{\odot} $. Note the difference in the vertical scale compared to Figure\,\ref{20and50}a. The $ ^{26} $Al yields from the SNBs are in general up to a factor of 2 higher higher than the $ ^{26} $Al yield of our single star model. This increase is significantly less than for the 20\,M$ _{\odot} $ binaries, where the increase is between a factor of 20 and 40 (Figure\,\ref{20and50}a).\\
\indent At the other hand, as can be seen in Figure\,\ref{20and50}b, the yields from the numerical binary systems are lower than for our single star model up to a factor of 2. However, we note that neither the single star, nor any of the binary systems reached the end of helium burning. During helium burning, more mass will be lost from all 50\,M$ _{\odot} $ stars. This can lead to a yield higher than shown in Figure\,\ref{20and50}. The increase will be the largest for the systems that stopped at an earlier stage of helium burning. This could push a few of the binary systems closer to the single star yield, but some might remain below it. Considering this, we confirm the conclusion of \cite{BraunandLanger} that, above a given mass, the effect of binary interaction no longer significantly increases the $ ^{26} $Al yield, but keeps it at the same level or decreases it compared to the single star yield. This is because the mass-loss rate by wind is comparable to the mass-loss rates due to binary interaction at this stellar mass. This leads to a similar amount of mass lost from the stars independent of whether the star has a companion or not. The general trend is that for 50\,M$ _{\odot} $ the binary interaction does not increase the $ ^{26} $Al yield.\\
\indent We compare our results to those given by \cite{BraunandLanger}. Their wind yield for the binary system is higher than for all the SNBs. Compared to our single star, the binary system has a yield  a factor of $ \sim $2.5 larger, and it is a factor of $ \sim $3-4.5 larger than our binary systems. As stated in Section\,\ref{20results}, it is difficult to say exactly which of the differences between the model set-ups is the source of the difference in the yield.
\subsection{Other primary masses}
\begin{figure}
	\includegraphics[width=0.5\textwidth]{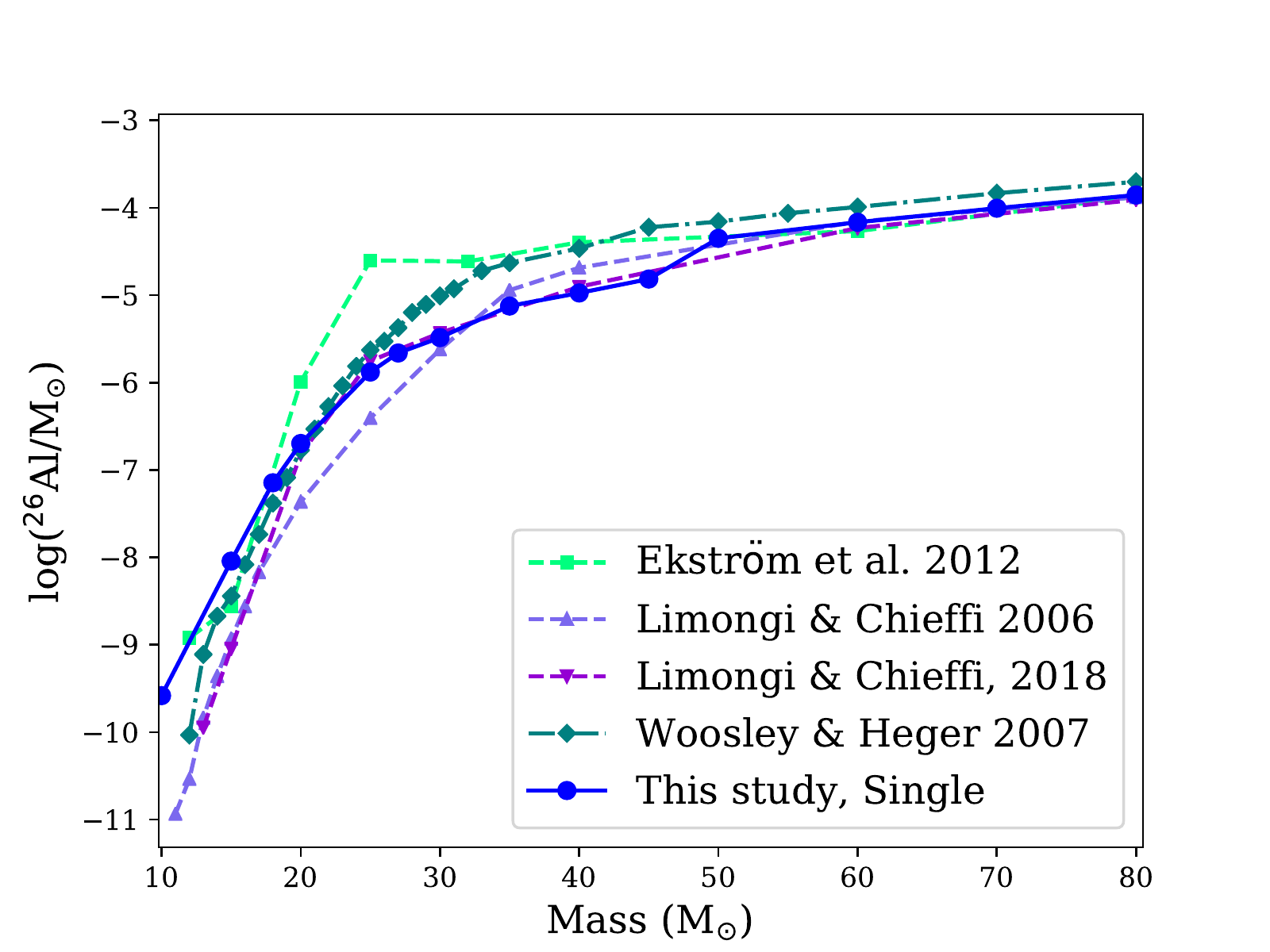}
	\caption{Non-rotating single star wind yields from this study and four other studies.}
	\label{AllTogether}
\end{figure}
\begin{figure*}\begin{center}
	\includegraphics[trim = 2mm 5mm 9mm 13mm, clip, width=0.45\textwidth]{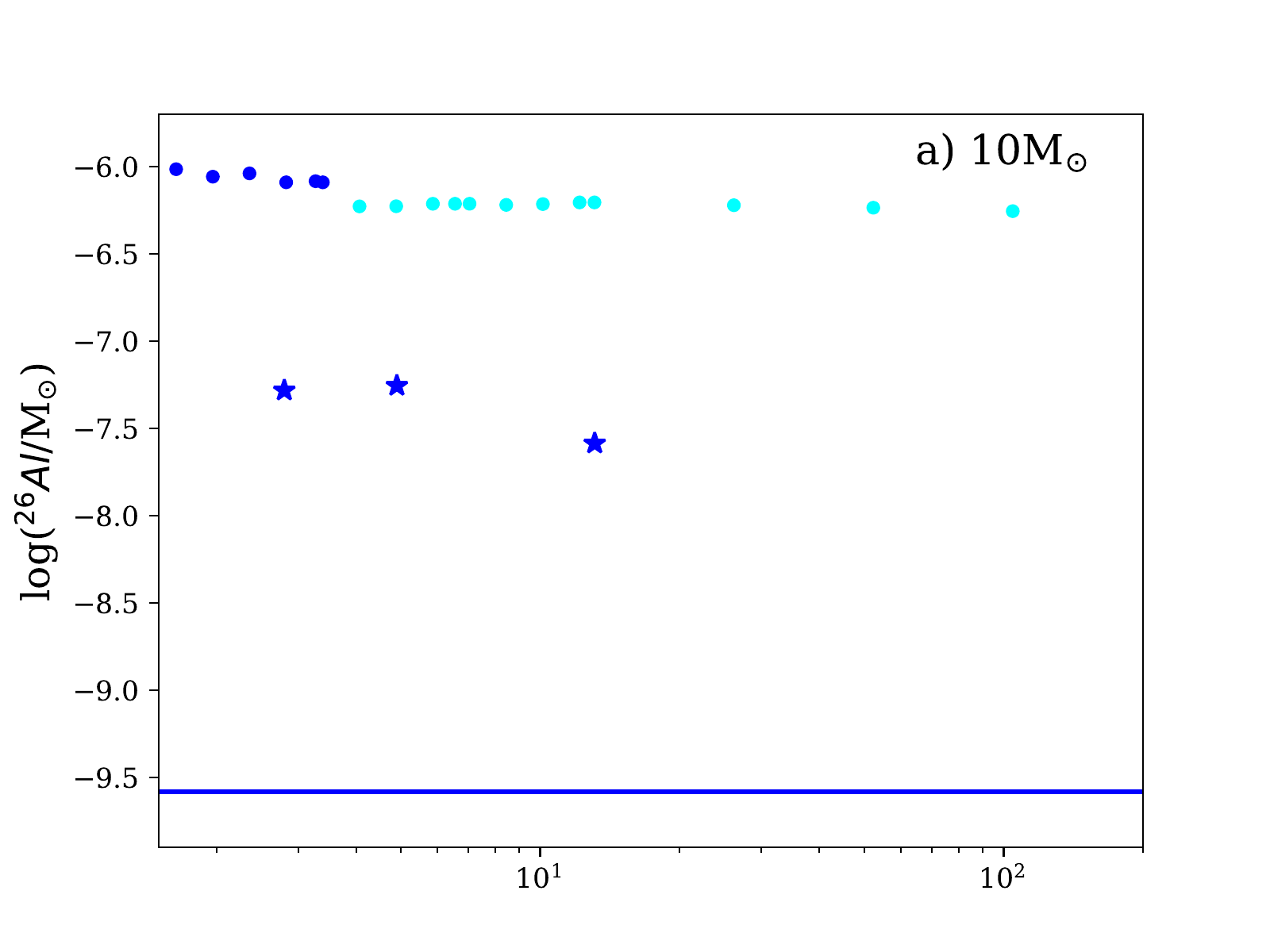}
	\includegraphics[trim = 2mm 5mm 9mm 13mm, clip, width=0.45\textwidth]{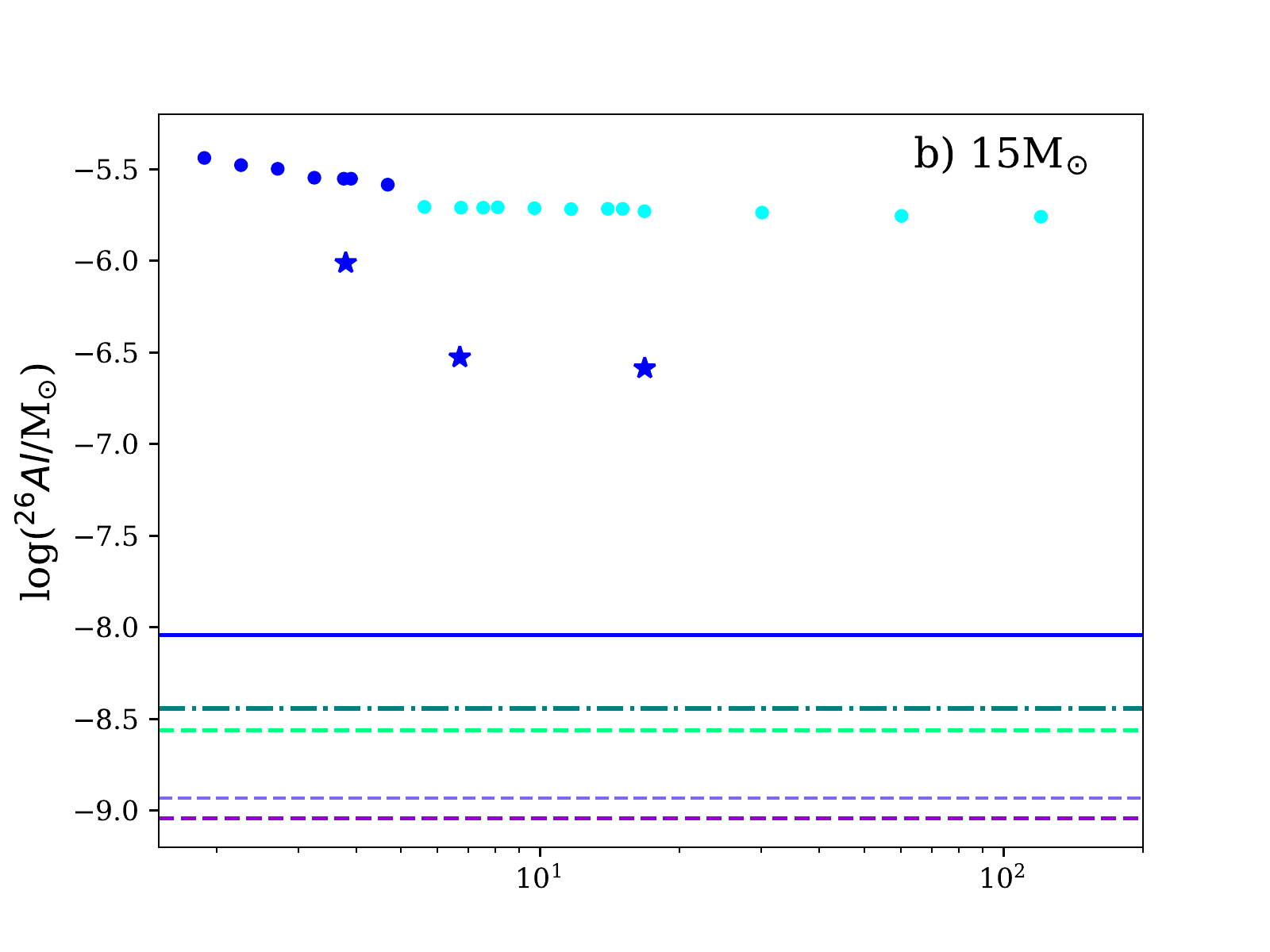}\\
    \includegraphics[trim = 2mm 5mm 9mm 13mm, clip,width=0.45\textwidth]{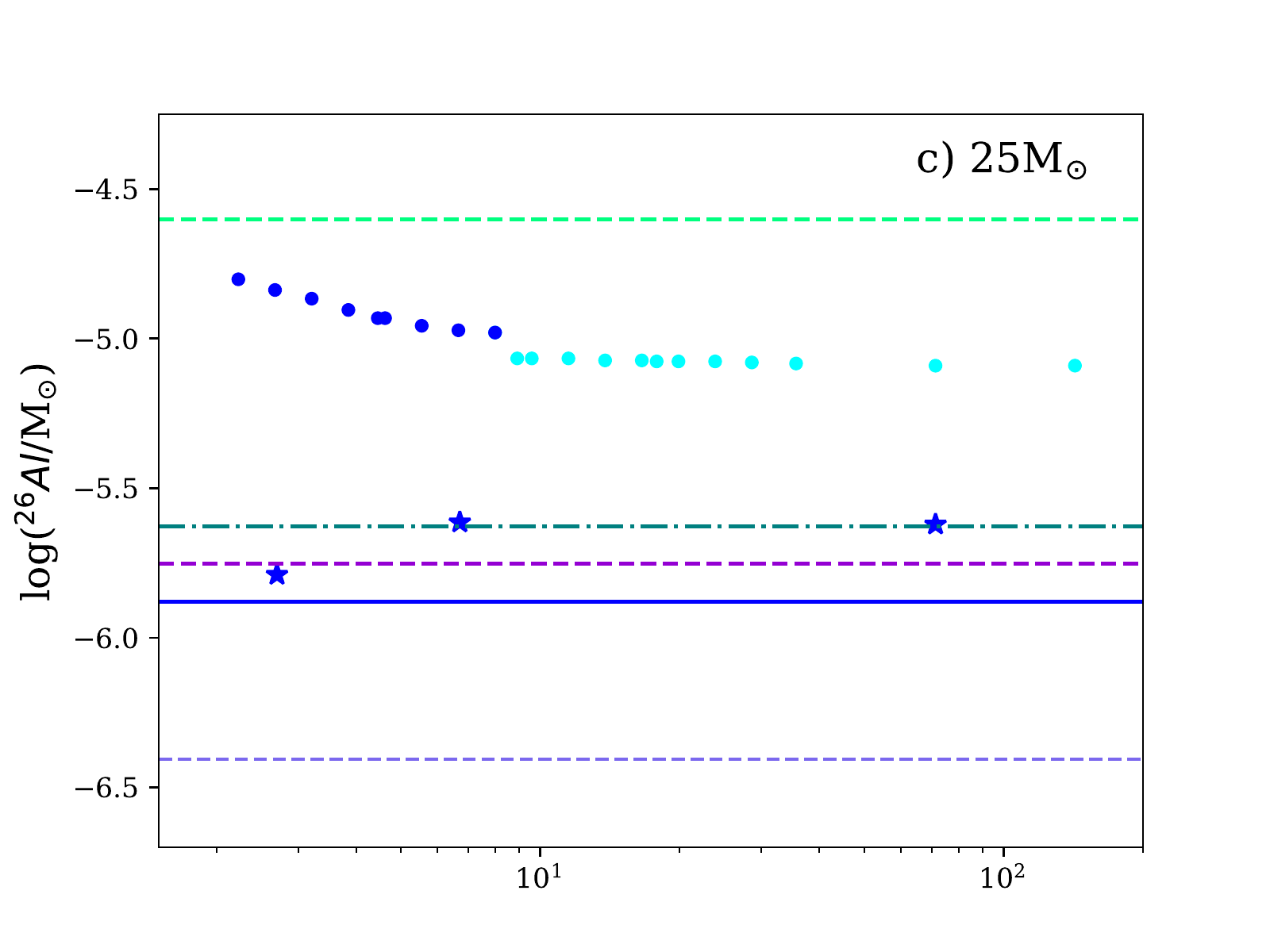}
    \includegraphics[trim = 2mm 5mm 9mm 13mm, clip,width=0.45\textwidth]{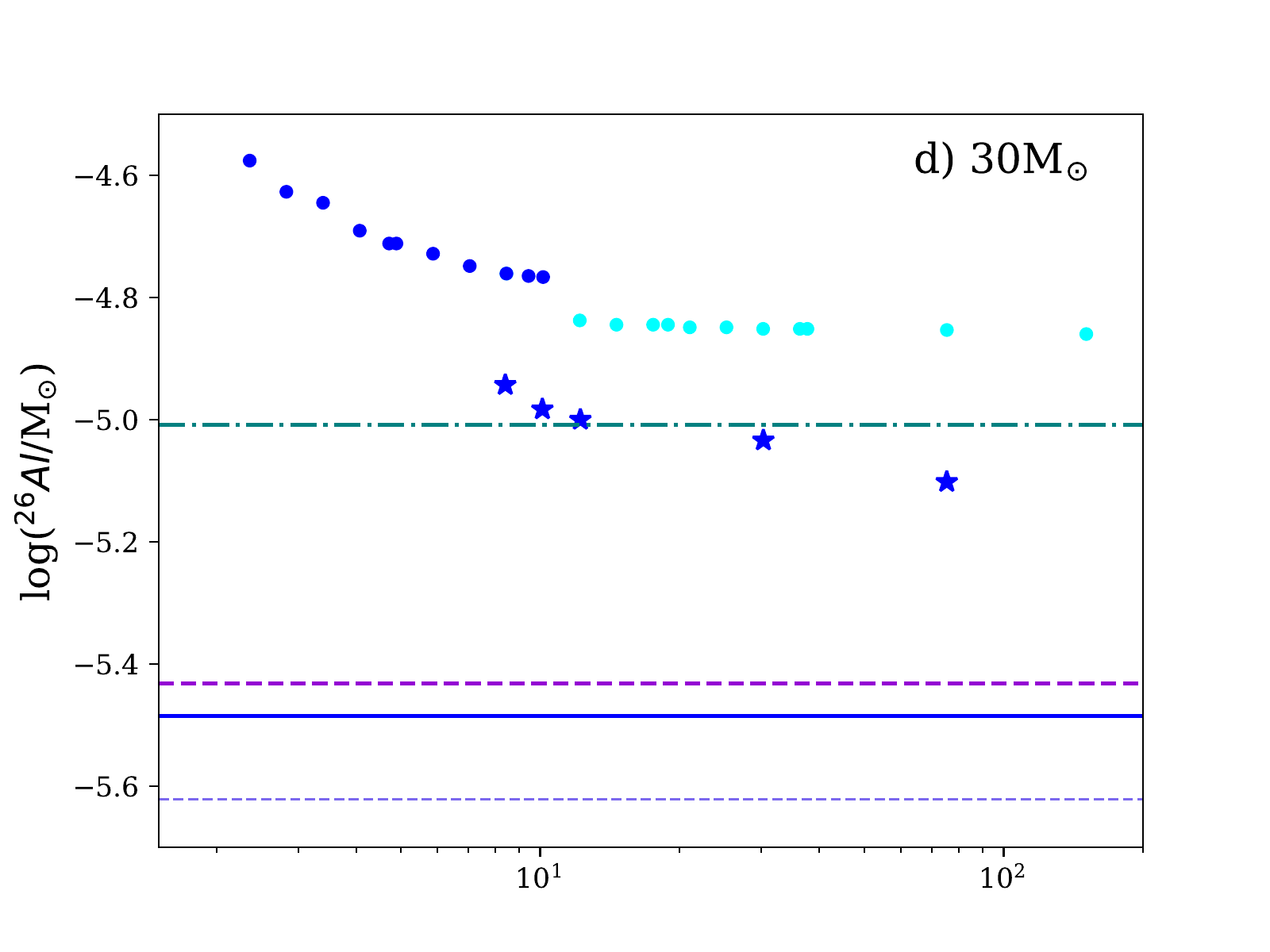}\\
	\includegraphics[trim = 2mm 5mm 9mm 13mm, clip,width=0.45\textwidth]{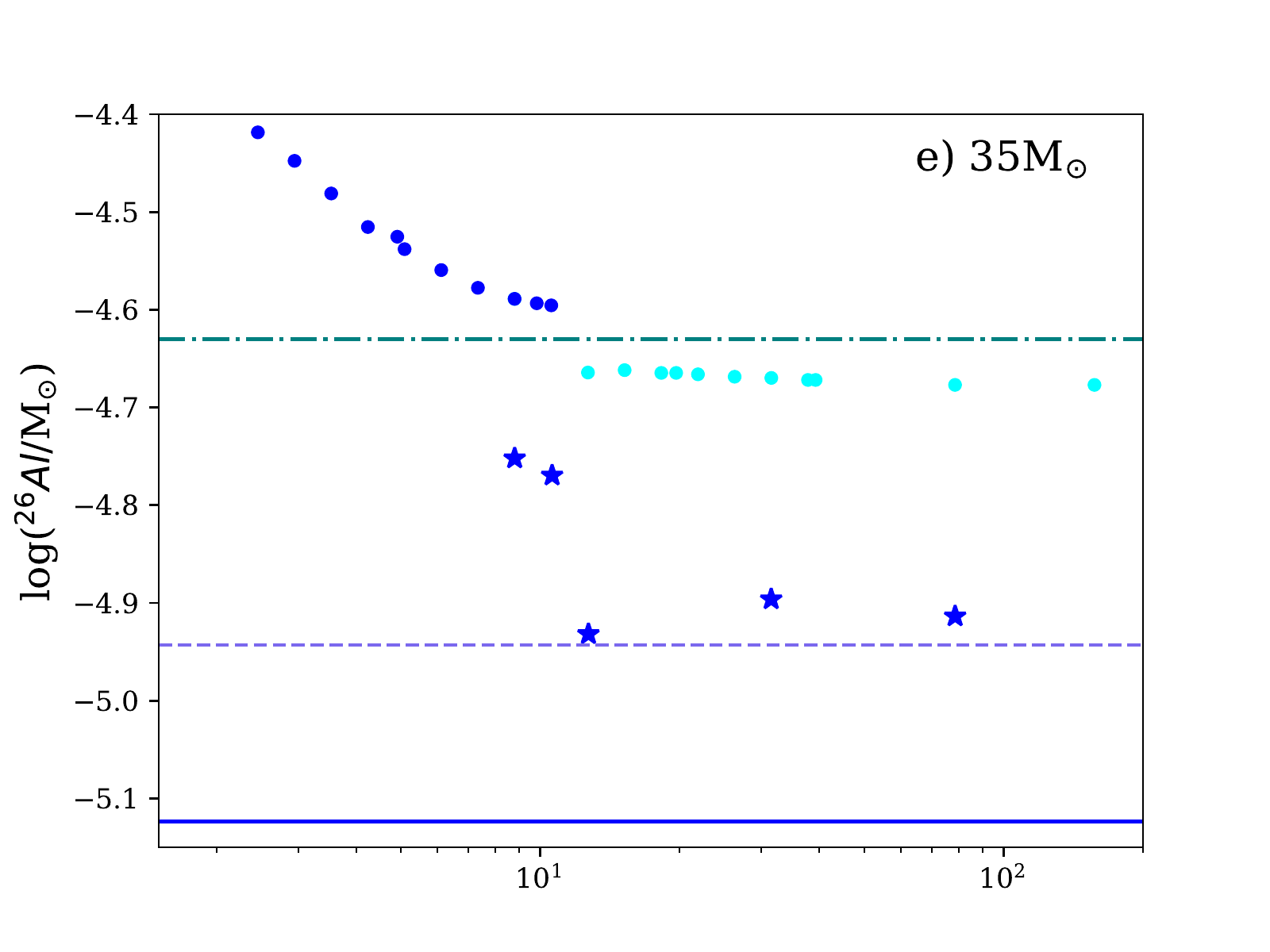}
	\includegraphics[trim = 2mm 5mm 9mm 13mm, clip,width=0.45\textwidth]{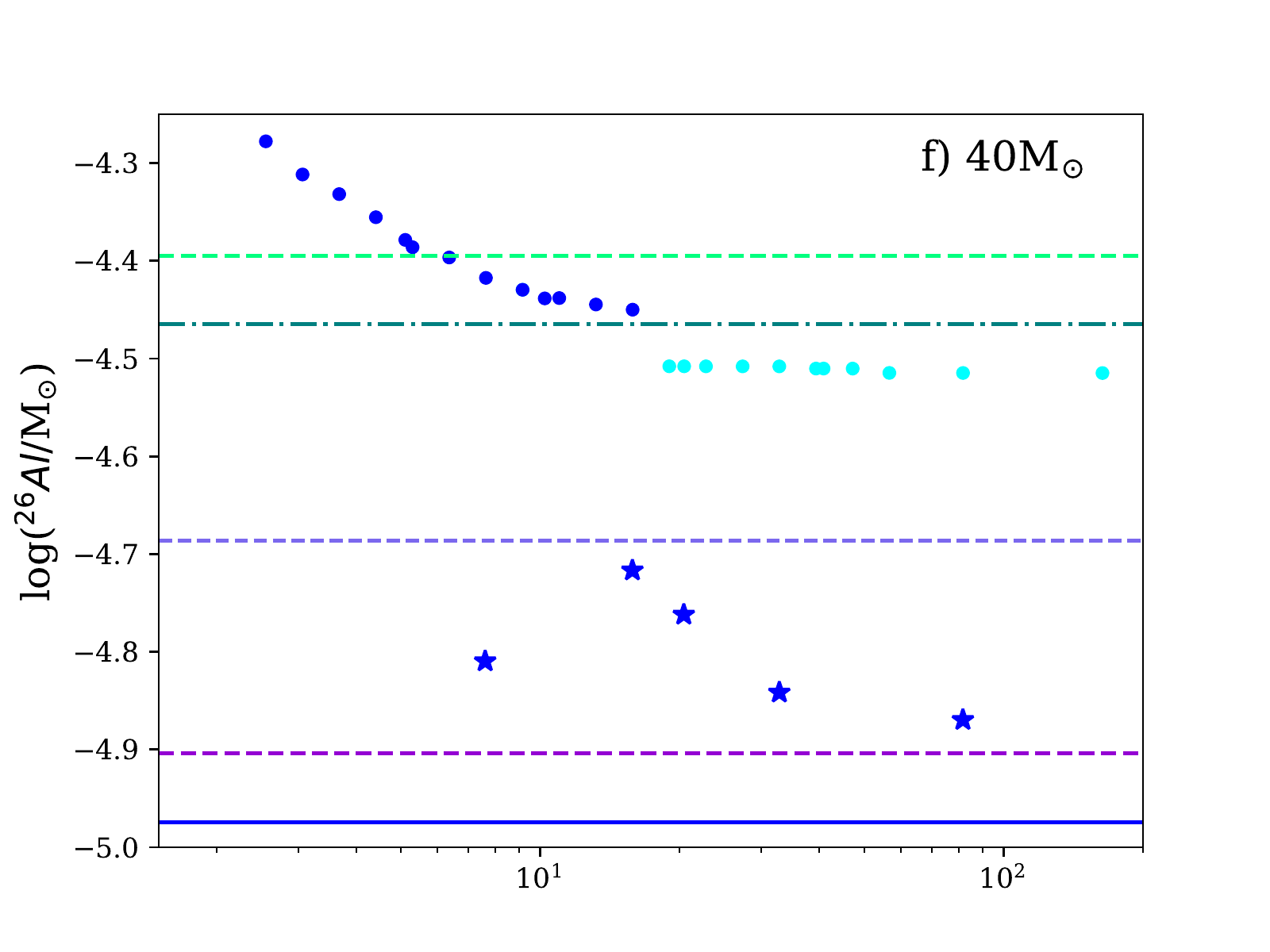}\\
	\includegraphics[trim = 2mm 1mm 9mm 13mm, clip,width=0.45\textwidth]{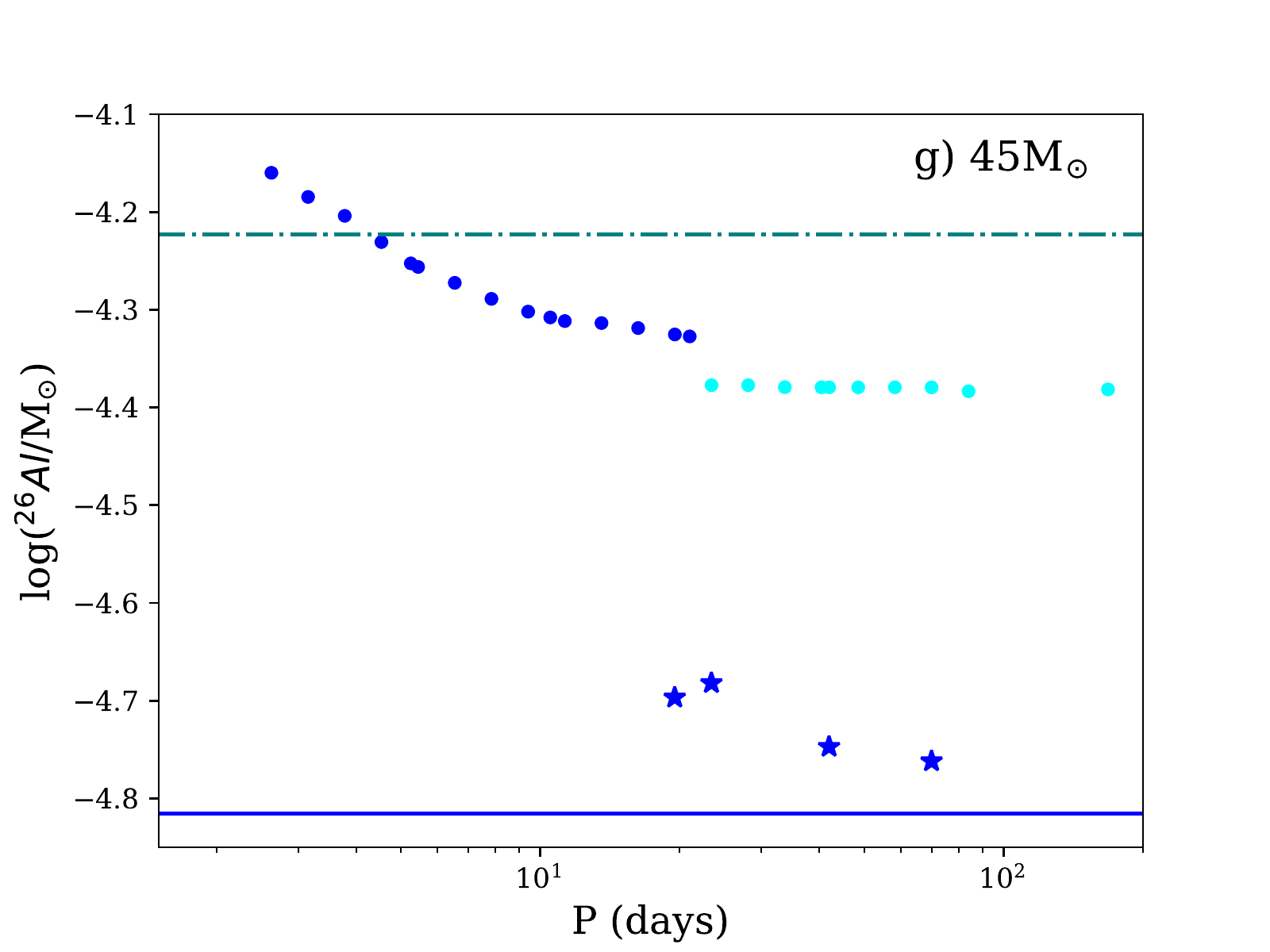}
	\includegraphics[trim = 2mm 1mm 9mm 13mm, clip,width=0.45\textwidth]{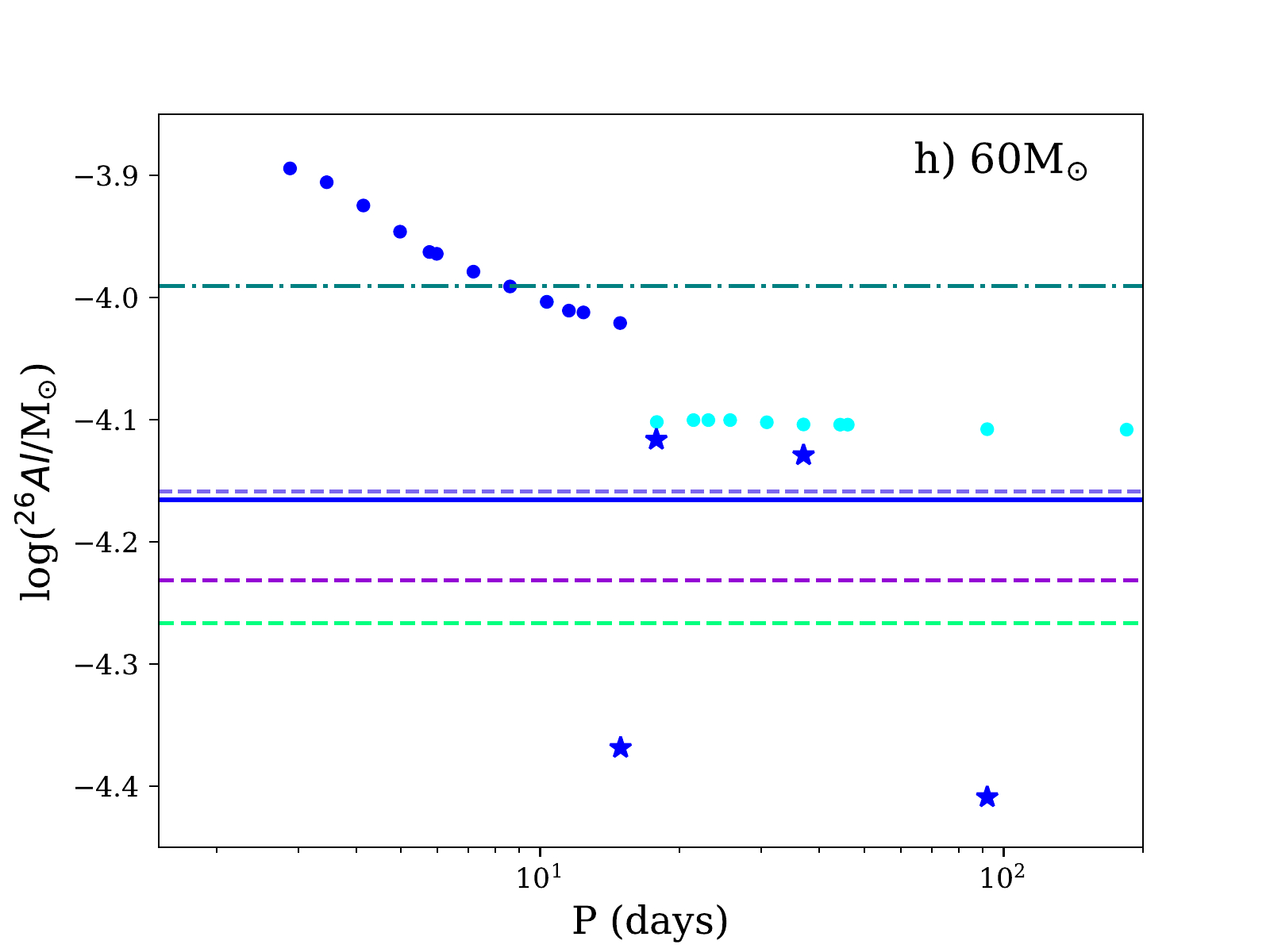}\vspace{-2mm}
	\end{center}
	\caption{Same as Figure\,\ref{20and50}, and the colours and the symbols indicate the same as in these two panels. Note that while the horizontal axis is the same for all of them, the vertical axis changes from 2 orders of magnitude to 0.5 orders of magnitude.}
	\label{TogetherALL}
\end{figure*}
Figure \ref{AllTogether} shows the $ ^{26} $Al wind yields for the single stars from our study, as well as from the literature. All studies show the same trend of increasing $ ^{26} $Al yield with increasing stellar mass. There is a spread in the yields that decreases towards the highest masses. This illustrates the variation in the yields coming from single star evolution. For the lower mass end of our mass range, below 20\,M$ _{\odot} $, our result is higher than for the other studies. From 20\,M$ _{\odot} $ to 30\,M$ _{\odot} $ our single star yields are comparable to most of the other yields. Above 30\,M$ _{\odot} $ our yields are slightly lower than \cite{WandH2007}, but comparable to \cite{LandC2018}. An explanation for this difference is that not all of our stars completed helium burning and a small amount of extra envelope lost can increase the $ ^{26} $Al yields. From this figure it is also clear that the 20\,M$ _{\odot} $ and 25\,M$ _{\odot} $ models of \cite{Ekstrom2012} give much higher yields compared to all other studies, for the reasons already discussed in Section\,\ref{20results} for the 20\,M$ _{\odot} $ case.\\
\indent Figure\,\ref{TogetherALL} shows our results for different primary masses. It is important to notice the difference in the vertical scale for the different primary masses. For the lowest masses in our range (10 and 15\,M$ _{\odot}$), the variations are the largest and the vertical scale covers three orders of magnitude. This is due to the fact that for these masses the mass-loss rate for the single star is very small. For example, for the 10\,M$ _{\odot}$ single star, it is between $3\times 10^{-9}$- $3\times 10^{-8}$\,M$ _{\odot} $/yr on the main sequence, and has a maximum value of $2 \times 10 ^{-6}$ \,M$ _{\odot} $/yr at the end of the simulation. This is roughly one and two orders of magnitude lower than for the 20\,M$_{\odot}$ and 50\,M$_{\odot}$ single stars, respectively. This difference makes the effect of binarity on the mass lost much more noticeable, relatively to the single star, for the lowest masses in our range. As a consequence, when the stellar mass increases the variations in the $^{26}$Al yield decrease. For the lowest masses (10 and 15 \,M$ _{\odot}$) all the binary yields are larger than the single star yields from all the studies reported here. For stellar masses of 25 and 30\,M$ _{\odot}$, the vertical scale covers two and one order of magnitudes, respectively, and the numerical binary yields are in some cases very similar to the single star yields, or even lower, when considering the 25\,M$ _{\odot}$ yield from \cite{Ekstrom2012}. For masses from 35\,M$ _{\odot}$ and above the vertical scale covers a range of yield variation of a factor 3 to 5 only, and in several instances the binary yields, both numerical and semi-numerical, are similar to or even lower than the single star yields, as predicted by \cite{BraunandLanger}. For these cases, binarity effectively produces variations in the yields that cannot be distinguished from variations due to uncertainties in the stellar models of single stars.\\
\indent For all the masses considered, the yields from the SNBs are always, by construction, higher than the numerical yields. The SNB yields show a clear pattern, where Case A mass-transfer produces yields that decrease smoothly with orbital period and are always higher than those of the Case B mass-transfer models, where the yields are not affected by the period. For the numerical binaries, the Case B systems follow the same trend as their SNB counterparts while for the Case A systems, the yields of very short orbital period models are affected by reverse mass-transfer, as explained in Section\,\ref{20results} (see also Section\,\ref{RevMT}).\\
\indent In summary, the increase in the $^{26}$Al yield in the binary system, relative to the single star, decreases with the mass - with typical multiplication factors of roughly 150, 50, 10-20, 5, 3, and 2 for stars of mass 10, 15, 20, 25, 30, 35\,M$ _{\odot}$, and no significant changes for higher masses. The yields per system are tabulated in Appendix\,\ref{BigTable}.
\section{Discussion}\label{sec:Discussion}
In this section we discuss a few aspects of binary evolution that could influence the $ ^{26} $Al yields of the systems: the effect of the secondary and reverse mass-transfer, mass-transfer efficiency, and mass ratio. We also discuss the impact of the reaction rates responsible for the production of $ ^{26} $Al, and finally, we present some implications of the results for Galactic and Solar System evolution.
\subsection{The effect of the secondary and reverse mass-transfer}\label{RevMT}
Many of the binary systems described in this paper experience reverse mass transfer before the end of the simulation (see Appendix\,\ref{20MsunCaseA} and Appendix\,\ref{BigTable} for details). During this phase mass is transferred from the secondary star to the primary star. The question that arises is how the further evolution of the system will affect the $ ^{26} $Al yield, both of the primary star and the secondary star. Reverse mass-transfer is likely to result in a common-envelope phase, where the envelope of the star filling its Roche lobe engulfs both stars and the orbit shrinks substantially \citep{Ivanova2013}. If the system survives this phase as a close binary, the envelope of the secondary is expelled from the system. This can significantly increase the yield of the binary system as a whole. Furthermore, the close binary system that is left after this phase could eject more mass, and thus $ ^{26} $Al, by either winds or further mass transfer in a close orbit. In case the common-envelope phase results in the merger of the binary into a single star, part of the envelope may still be ejected, and the merged object could eject more $ ^{26} $Al through stellar-wind mass loss.\\
\indent The above considerations are likely to be relevant for a large fraction of the systems we simulated, and not only those in which we found reverse mass transfer during the evolution of the primary. Many of our systems will experience reverse mass-transfer and a common-envelope phase at a later stage, after the primary star has finished its evolution and has become a compact object. This requires that the binary system is still bound after the supernova explosion, which depends on the dynamics of the supernova explosion and the resulting kick the compact object will receive. Altogether, the further evolution of the systems including that of the secondary star is quite complicated and subject to many uncertainties, and the resulting $ ^{26} $Al yields are hard to predict but potentially very important. The complete problem of the effect of binary evolution on the $ ^{26} $Al yields can only be explored by a combination of binary population synthesis, which incorporates all these effects and which allows for exploration of their uncertainties (e.g \citealt{Izzard2006, Izzard2018}), and further detailed binary calculations of selected interesting cases.
\subsection{Mass-transfer efficiency}\label{Betas}
\begin{figure}
	\includegraphics[width=0.5\textwidth]{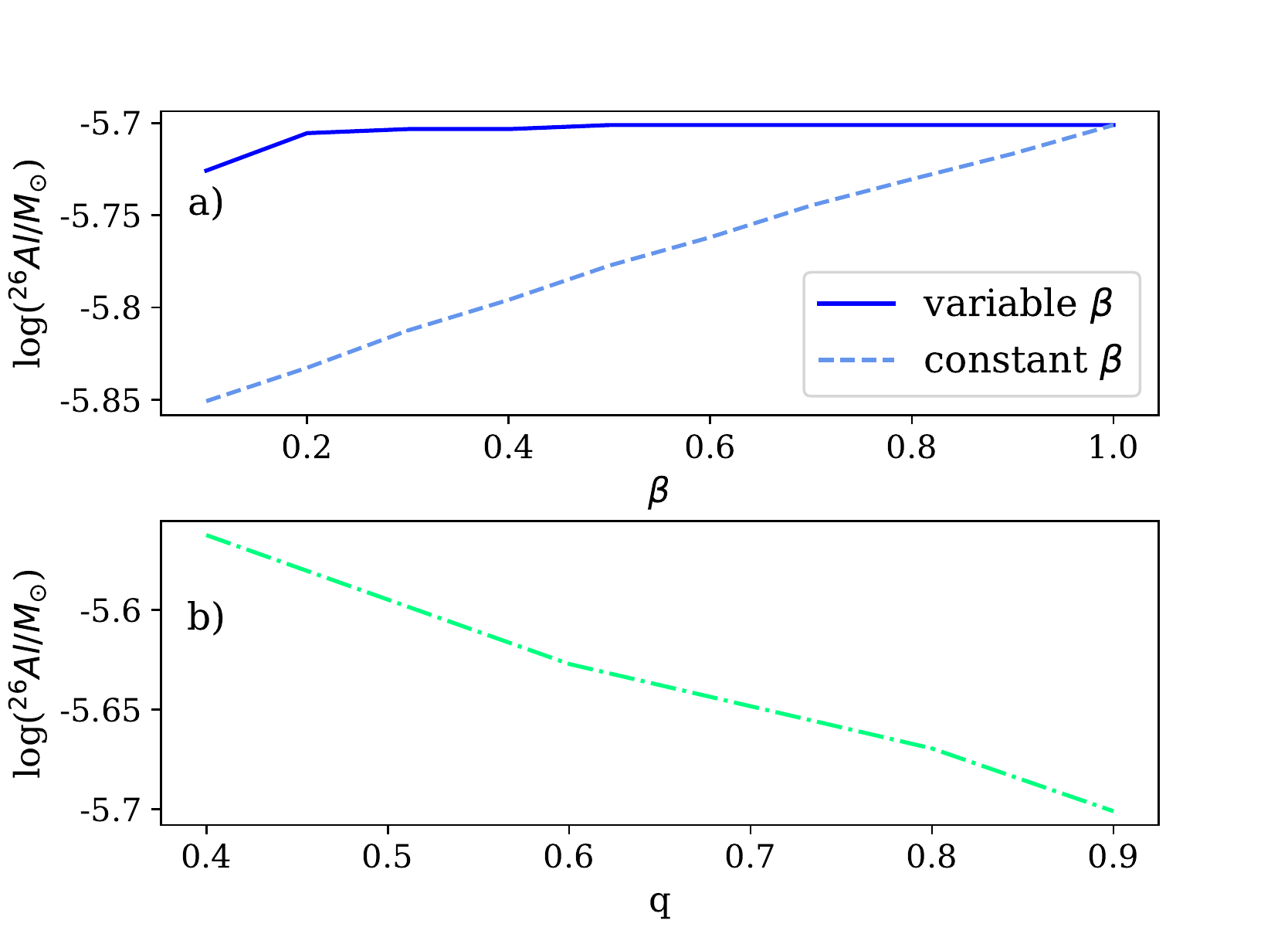}
	\caption{(a) effect of varying the mass-transfer efficiency, $ 1-\beta $. The solid blue line is for a variable $ \beta $ and the dashed blue line is for a constant $ \beta $. (b) effect of varying the mass ratio, $q$, on the $ ^{26} $Al yield.}
	\label{BetasandQ}
\end{figure}

Up to now we have assumed that the mass transfer between the stars of the binary is fully non-conservative, meaning that all the mass transferred from the primary to the secondary is subsequently lost from the system. In reality it is unclear how much mass is accreted and how much is lost, see Section\,\ref{binary}.\\
\indent In order to estimate the influence of the mass-transfer efficiency on our results, we take the following simplified approach. We take the binary system with a primary mass of 20\,M$ _{\odot} $ and a secondary mass of 18\,M$ _{\odot} $ at a period of 18.4 days. We use a semi-numerical scheme to calculate the $ ^{26} $Al yield for different values of $ \beta $, using the mass stripped from the primary star in the numerical binary simulation for this system. We assume that the $ ^{26} $Al yield due to wind is not affected by $ \beta $. We use two approaches; i) we assume that initially the mass transfer is fully conservative until a fraction of 1-$ \beta $ of the total transferred mass is accreted, and the remaining part of the transferred mass is lost from the system, ii) we assume that $ \beta $ is constant in time during the mass-transfer phase. The mass accreted by the secondary is not added to the yield. We have not taken into account the changes of the orbit as a result of the change in mass of the secondary.\\
\indent In Figure\,\ref{BetasandQ}a the results of the two approaches are shown as a function of $ \beta $, where the yield for $ \beta $=1 is the same as in Appendix\,\ref{BigTable}. Using the first approach, the yield is almost independent of $ \beta $, because most of the $ ^{26} $Al is located in the deeper layers of the star, and mostly $ ^{26} $Al-poor material is transferred. Using the second approach the effect is quite modest, mostly because most of the $ ^{26} $Al yield (1.34$ \times $10$ ^{-6} $\,M$ _{\odot} $) comes from the wind during helium burning after the mass-transfer phase. However, further investigation of the mass-transfer efficiency with detailed simulations is needed, and will be done in future work.
\subsection{The influence of the mass ratio}\label{Qs}
Apart from the initial primary mass and orbital period, the outcome of the binary evolution also depends on the initial mass-ratio. All systems presented so far have an initial mass-ratio of $ q$=0.9, but this parameter can take on a wide range of values between close to 0 to 1. Here we briefly show the influence of the initial mass ratio on the final $ ^{26} $Al yield for a few systems with a primary mass of 20\,M$ _{\odot} $ and a period of 18.4 days. For a fixed primary mass, a smaller mass ratio results in a smaller separation at the same orbital period (Equation\,\ref{Eq1}), and in a larger ratio of the Roche-lobe radius to the separation (Equation\,\ref{Eq2}). These effects nearly cancel each other, leading to only a slightly larger Roche-lobe radius for the same orbital period. This has only a very small effect on the moment when mass transfer occurs in the system and  on the $ ^{26} $Al yield. A more substantial change in the yield comes from the different adjustment of the orbit to mass loss. For the system with a mass ratio of 0.4, the orbit shrinks during the Roche lobe overflow, and then expands again. At the end of the simulation, the period is $ \sim $15 days. For the system with a mass ratio of 0.6, the orbit shrinks only a little, and at the end of the simulation, the orbit has expanded to a period of $ \sim $50 days. For the other two mass ratios, 0.8 and 0.9, the orbit only expands, ending with periods of $ \sim $85 and $ \sim $101 days, respectively. This different response to the Roche-lobe overflow leads to slightly more mass loss (by up to 0.5\,M$ _{\odot} $) for systems with a lower mass ratio during the mass-transfer phase. However, what causes the main difference to the yields is that systems with lower mass ratios lose more mass at the end of the helium-burning phase, where the $ ^{26} $Al-rich region is stripped. Combined, this leads to the $ ^{26} $Al yield increasing with decreasing initial mass ratio, as shown in Figure\,\ref{BetasandQ}b. Because the mass ratio affects the orbital evolution of the system, this should be considered in future work, especially when considering the possibility of reverse mass-transfer.
\subsection{Effects of the reaction rates}\label{RATES}
\begin{figure}
	\includegraphics[width=0.5\textwidth]{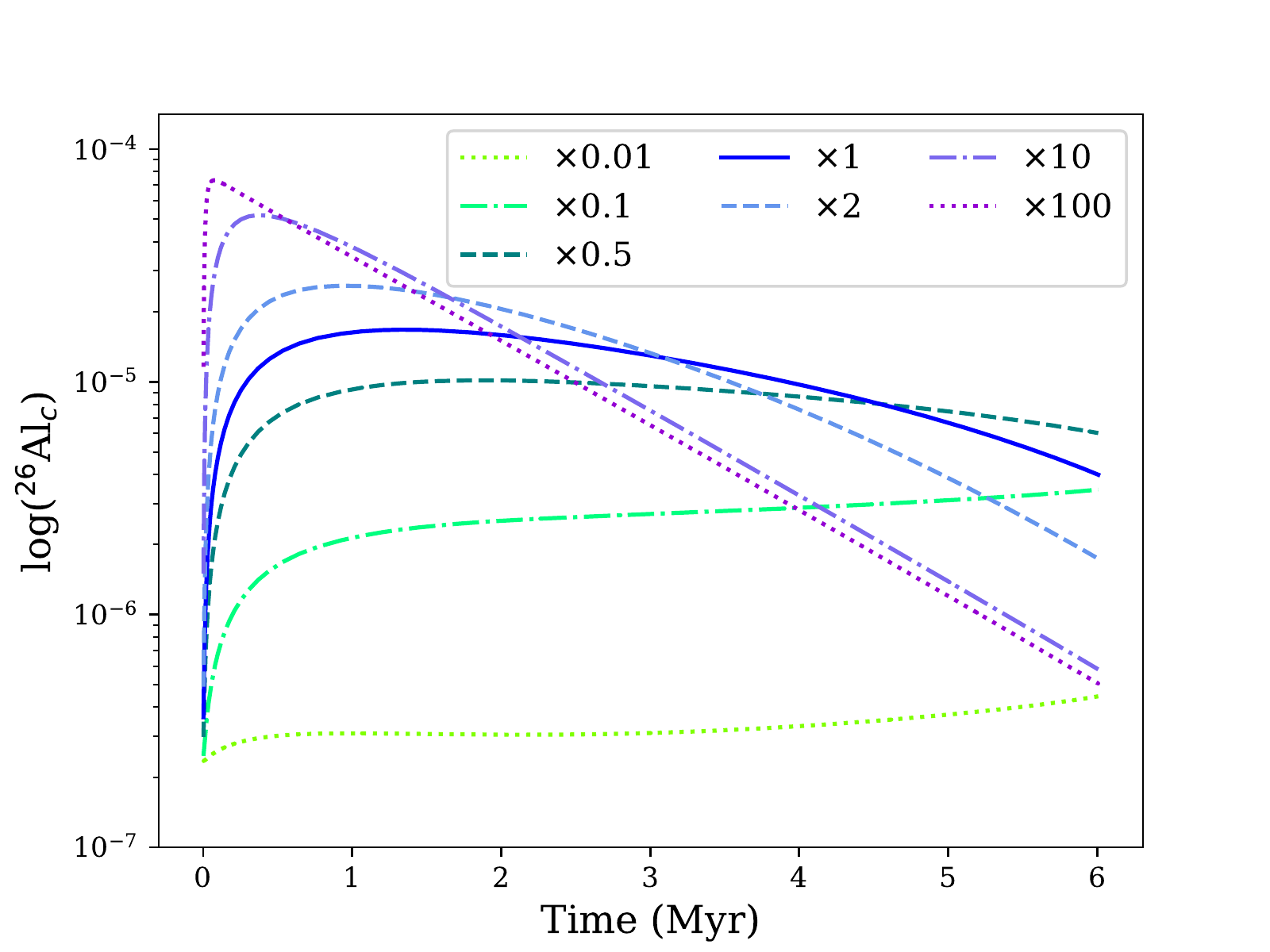}
	\caption{Time evolution of the mass fraction of $^{26}$Al$_g$ in the core of a 30\,M$ _{\odot} $ stellar model during the core hydrogen burning-phase. The $^{25}$Mg(p,$\gamma$)$^{26}$Al$_g$ rate is varied.}
	\label{fig:rates_time}
\end{figure}
\begin{figure*}
	\includegraphics[width=1\textwidth]{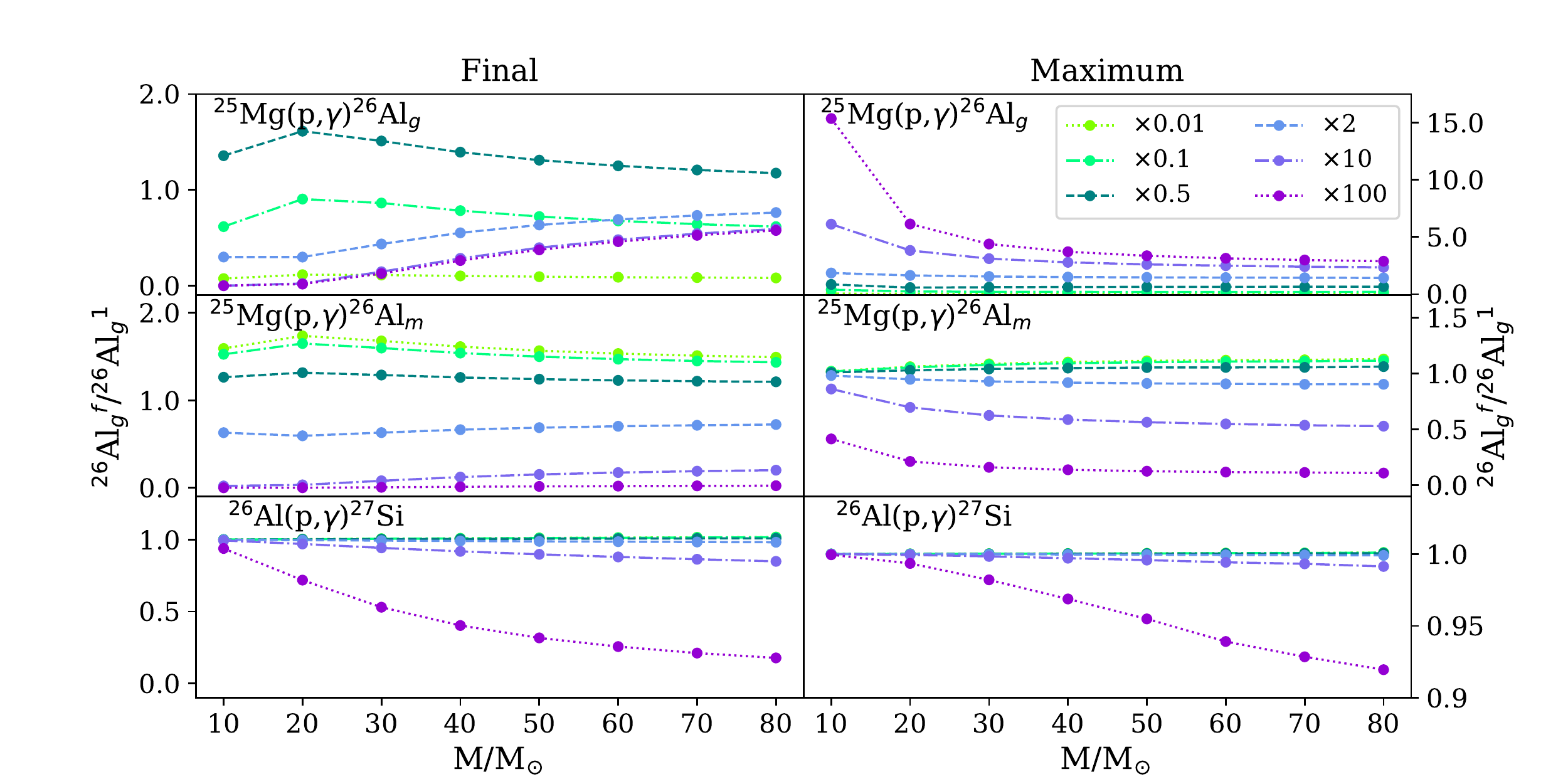}
	\caption{Effect of varying the three reaction rates discussed in Section\,\ref{RATES} for different initial stellar masses. The vertical axis shows the ratio of $^{26}$Al$_g$ from the modified models (superscript f) to the standard models (superscript 1) within the core. In the left panels we show the ratios of the final abundances (end of hydrogen burning), in the right panels the ratios of the maximum abundances.}
	\label{fig:rates_masses}
\end{figure*}
\begin{figure}
	\includegraphics[width=0.5\textwidth]{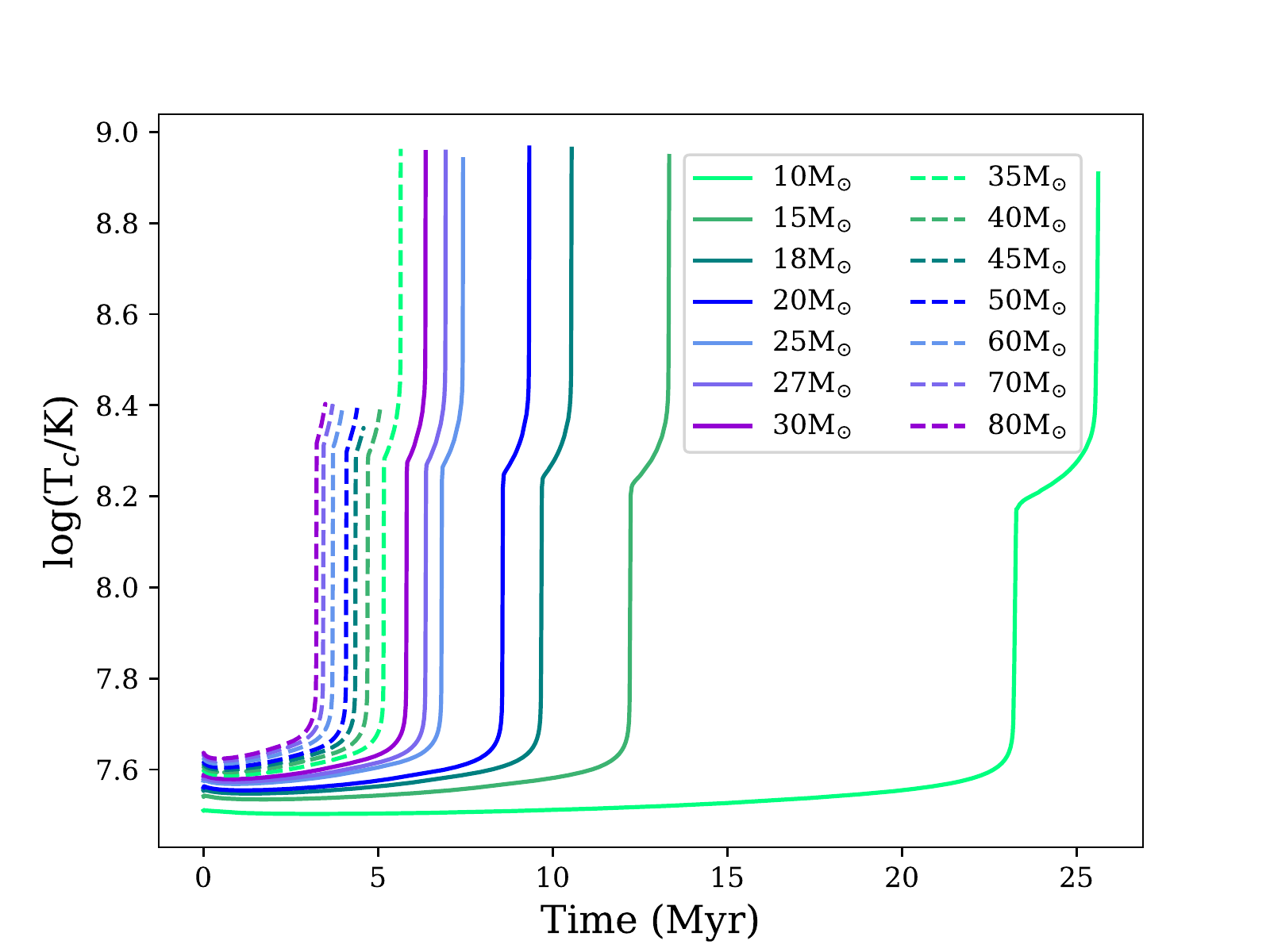}
	\caption{Evolution of the central temperature in the core of our single stars as function of time and for different masses. The sharp increase in the temperature towards the end of the evolution corresponds to the end of H burning and the contraction of the core. The temperature range relevant to the production and destruction of $^{26}$Al is that before this sharp increase.}
	\label{fig:T_masses}
\end{figure}
We investigated the effect of varying the rates of the three reactions that are crucial for the production of $^{26}$Al in the H-burning core of massive stars. These are the production channel of $^{26}$Al$_g$, $^{25}$Mg(p,$\gamma$)$^{26}$Al$_g$, the competing channel producing the isomer 
$^{26}$Al$_m$, that quickly decays into $^{26}$Mg, $^{25}$Mg(p,$\gamma$)$^{26}$Al$_m$, and the main destruction channel of $^{26}$Al$_g$ in H-burning conditions, $^{26}$Al$_g$(p,$\gamma$)$^{27}$Si. The 
rates of the first two reactions are from the experiment performed underground by the LUNA collaboration \citep{straniero13}, provided with an uncertainty of roughly 40\% and 30\%, respectively, at a typical temperature of 50\,MK. The third reaction is from the compilation of \citet{iliadis10}, and the uncertainty at typical activation temperatures of 50, 60, and 70\,MK is of a factor of 34, 20, and 8, respectively.\\
\indent We tested variations of such rates by running the full evolutionary MESA models for the single stars and each time we multiplied one of the rates above by a multiplication factor ranging from 0.01 to 100, kept constant in the whole range of temperatures. This tested range is much larger than the uncertainties quoted above. However, for the $^{25}$Mg(p,$\gamma$) reaction, of the two isolated narrow resonances at 57\,keV and 92\,keV that dominate the rate from roughly 30-80\,MK, only the 92\,keV resonance was directly measured by LUNA, while the 58\,keV resonance is still inaccessible to direct experiment. Only indirect reaction data \citep{iliadis96} and theoretical values \citep{li15} are available to calculate the 58\,keV resonance contribution. Furthermore, the relative importance of the two $^{25}$Mg(p,$\gamma$) reaction channels is strongly affected by the value of the feeding factor to the ground state of $^{26}$Al, which describes the probability of the $^{25}$Mg(p,$\gamma$)$^{26}$Al resonances to decay through complex $\gamma$ cascades to the ground state. The LUNA rates include the feeding factor of 0.6 for the 92\,keV resonance provided by \citet{strieder12} with an uncertainty 
of roughly 30\%. However, there are large discrepancies between this and the 
previous values, and there is no recent information on the feeding factor for 
the 58\,keV resonance. The rate of the $^{26}$Al$_g$(p,$\gamma$)$^{27}$Si reaction at 
low temperatures is strongly influenced by unobserved, low-energy 
resonances, whose contributions may modify the rate beyond the currently given 
lower and upper limits. Finally, the given reaction rates do not include the 
possible contribution of electron screening, except for the contribution of the 92\,keV resonance in the $ ^{25} $Mg(p,$ \gamma $)-reaction.\\
\indent In Figure~\ref{fig:rates_time} we show how variations of the 
$^{25}$Mg(p,$\gamma$)$^{26}$Al$_g$ reaction rates affect the time evolution of 
$^{26}$Al in the core of a 30\,M$ _{\odot} $ star. There is a striking difference between 
how the rate variations affect the maximum and the final values achieved. The 
maximum value varies by orders of magnitude as the rate varies, and it 
is reached earlier in time as the rate increases: for the standard case 
(multiplication factor of 1) it is reached within 2\,Myr, for the highest 
multiplication factor of 100 it is reached almost immediately, while for the 
lowest factor of 0.01, it is reached only at the end of the H-burning phase. On the 
other hand, the final value is also controlled by the decay of $^{26}$Al$_g$: 
if $^{26}$Mg is converted into a large $^{26}$Al$_g$ abundance very early 
in time (multiplication factor of 100), there is more time and a larger 
initial abundance to decay and the final $^{26}$Al$_g$ is similar to the case when the rate is multiplied by 0.01. The standard case gives the same final $ ^{26} $Al abundance as the case with a multiplication factor of 0.1, while the case 
that results in the highest final abundance (roughly a factor of 1.5 higher than 
the standard) corresponds to a multiplication factor of 0.5.\\
\indent Keeping these trends in mind we show in Figure~\ref{fig:rates_masses} the 
variations of the maximum and final abundances in the core during hydrogen burning for models of different masses. To help interpreting these results we also show in Figure~\ref{fig:T_masses} the 
evolution of the central temperature during the computed evolution of all our 
single star models. In the case of the $^{25}$Mg(p,$\gamma$)$^{26}$Al$_g$ the 
variations in the final abundance do not show a significant trend with mass 
and the highest value is achieved for every mass by the 0.5 multiplication 
factor. For the maximum abundance, instead, variations are more pronounced as the 
mass decreases and the rate increases. When the 
$^{26}$Al$_g$(p,$\gamma$)$^{27}$Si is varied, significant differences appear only 
when the rate is multiplied by a factor of 100 and these differences are 
strongly dependent on the initial mass. Also, the maximum values are not 
strongly affected in this case as the reaction becomes more activated later in 
the evolution as the temperature increases with time (Figure\,\ref{fig:T_masses}).\\
\indent When comparing to the work of \citet{iliadis11}, in particular their Sec\,3.4 
reporting a sensitive study for a 80\,M$ _{\odot} $ star during core H burning, we find 
qualitatively similar results, although quantitatively there are differences. In the 
case of the $^{26}$Al$_g$(p,$\gamma$)$^{27}$Si reaction multiplied by 100 we find a 
ratio with respect to the standard case of 0.18, while \citet{iliadis11} report 0.017. This is 
probably related to the fact that our temperature is somewhat lower than that 
reported in Figure\,10 of \citet{iliadis11}: at a mass fraction of H of $10^{-4}$ we find 
a temperature of $\simeq 75$\,MK, instead of $\simeq$ 80\,MK. Some differences may also 
be related to the fact that \citet{iliadis11} used a post-processing method, while we 
calculate a full evolutionary model for each rate variation. In the case of the 
$^{25}$Mg+p reactions, it is not possible to make an exact comparison since we use
the new rates by \citet{straniero13}. However, we qualitatively agree on the result 
that variations larger than a factor of two appear only when the rates are changed by 
two orders of magnitude.\\
\indent Finally, the temperature dependence of the decay rate of $^{26}$Al is included in our 
calculations. This dependence arises as the ground and isomeric states of $^{26}$Al, 
which are prohibited from communicating with each other due to the large spin difference,
may communicate in hot stellar plasma via $\gamma$ transitions involving higher-lying 
energy levels. While the resulting effect is still debated 
\citep{gupta01,reifarth18,banerjee18}, at the temperature range of interest here,
between 30\,MK and 80\,MK, the communication between the two states is very weak, much 
lower than the $\beta$-decay rates. We have verified nevertheless that using two 
very different rates \citep{gupta01,reifarth18} we see no change in the $^{26}$ Al 
abundance from our calculations.\\
\indent For the other reactions we have used the NACRE rates, as mentioned in Section\,\ref{Input}. We have tested the influence of changing from the NACRE rates to the most recent JINA rates on the $ ^{26} $Al wind yields for single star models. Changing the rates leads to a decrease in the $ ^{26} $Al yield as compared to the NACRE rates, by most a factor of 2, mostly due to the new $ ^{14} $N(p,$ \gamma $) rate. The change in the $ ^{14} $N(p,$ \gamma $)-rate has more influence on the $ ^{26} $Al yield than the rates described before because this rate influences the structure and evolution of the star on the main-sequence, where the $ ^{26} $Al is produced.
\subsection{Implications for Galactic and Solar System evolution}
\begin{figure}
	\begin{tabular}{l}
	\hspace*{0.01cm}
	\vspace*{-0.1cm}
	\includegraphics[width=0.45\textwidth]{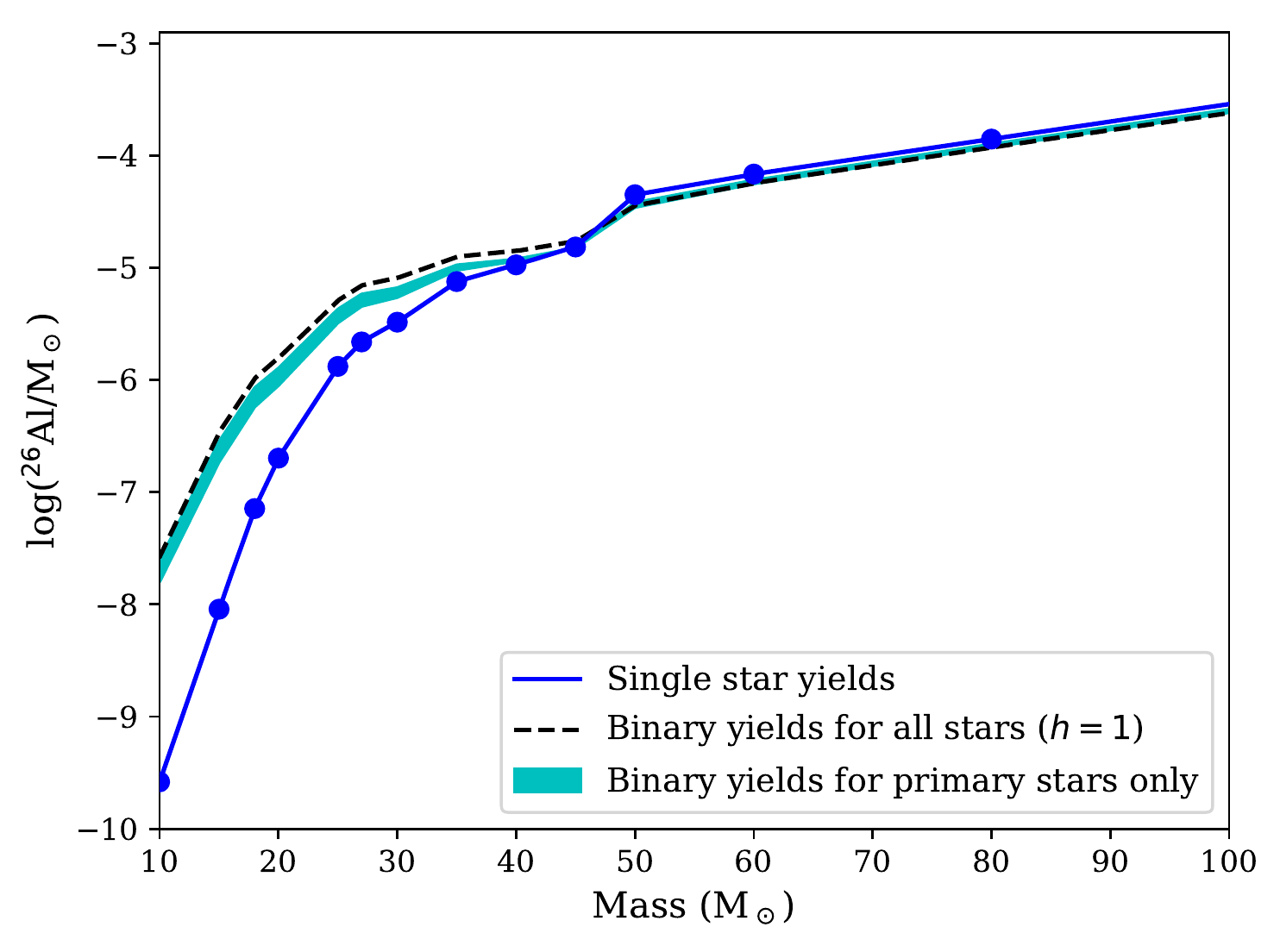} \\
	\includegraphics[width=0.453\textwidth]{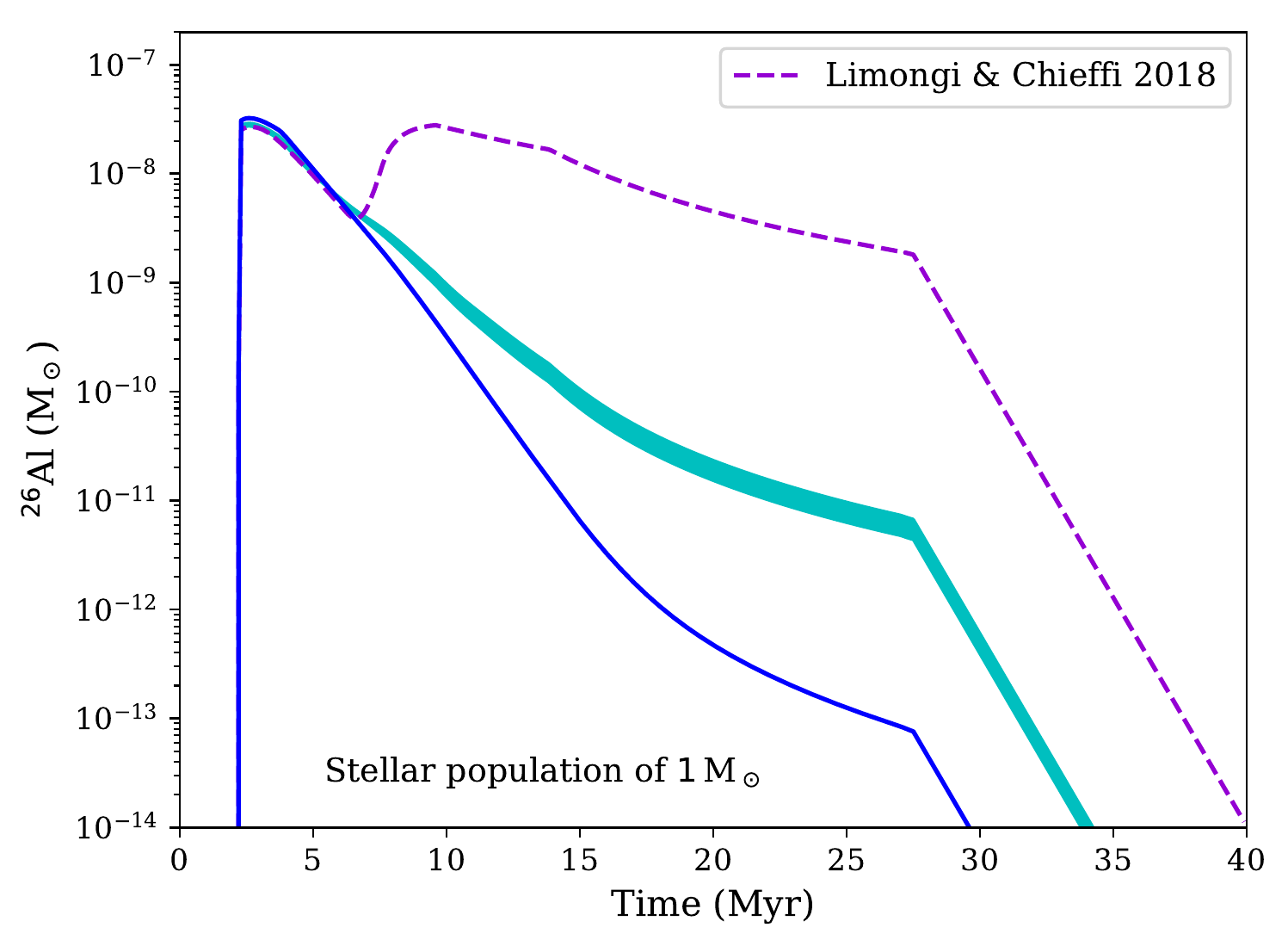}
	\end{tabular}
	\caption{Upper panel: Effective $^{26}$Al wind yields as a function of stellar initial mass for 
	single stars (blue solid line) and for stars in binary systems (cyan band). The blue dots
	represent our single star models. The thickness of the cyan band
	represents the range of solutions when assuming a binary fraction $h$ between 0.5 and 0.9. The
	dashed black line represents the extreme case with $h=1$ where the binary yields are applied to every star.\\ Bottom panel: Evolution of the mass of $^{26}$Al ejected into the interstellar medium by a simple stellar population
	of 1\,M$_\odot$ as a function of time since the formation of the stellar population. The mass of $ ^{26} $Al shown in this panel includes radioactive decay once the it has been ejected by stars, which is why the mass is decreasing over time. For comparison, the dashed violet line shows the result using the yields of \cite{LandC2018} for the single stars, which include both the wind and the supernova contribution. The treatment of radioactive decay is included and is explained in \cite{Benoit2019}.}
	\label{fig_SSP}
\end{figure}
To bring our study of $ ^{26} $Al from massive binary stars into a Galactic chemical evolution context, we must integrate our yields with the stellar initial mass function. To properly include binary yields in such a context, all possible mass ratios and orbital periods should be considered. In this section we use a simpler and preliminary approach, because we did not explore the complete parameter space for binary systems. Therefore, the results presented in this section are first-order approximations. First we determine the average increase in the $ ^{26} $Al yield from binary systems as compared to the single star yields (see Table\,\ref{OneBigTable} in Appendix\,\ref{BigTable}), which we call the `binary enhancement factor'. The values we have used for this enhancement factor are given in Table\,\ref{BinaryEnhancement}. Second, we calculate an `effective binary $^{26}$Al yield' as a function of initial stellar mass by assuming that a fraction $h$ of all massive stars are primary stars in a binary system. This fraction is connected to $f_\mathrm{binary}$, the fraction of all massive stars that are part of a binary system either as a primary or a secondary star,
\begin{equation}
f_\mathrm{binary}=\frac{2h}{(1+h)}.
\end{equation}

\begin{table}
\caption{The binary enhancement factor for the different primary masses.}\label{BinaryEnhancement}
\begin{center}
\begin{tabular}{|c|c|c|c|}
\hline 
M$ _{ini} $ (M$ _{\odot} $) & factor &M$ _{ini} $ (M$ _{\odot} $) & factor \\ 
\hline 
10 & 150 &35 & 2\\
15 & 50 &40 & 1.5\\
20 & 10 &45 & 1.25 \\
25 & 5 &50 & 1\\
30 & 3 &60 & 1\\
\hline 
\end{tabular} 
\end{center}
\end{table}
For any given stellar mass, the effective yield is defined as:
\begin{equation}
Y_\mathrm{eff} = \frac{(1-h)Y_\mathrm{single} + h(Y_\mathrm{prim} + \langle Y_\mathrm{sec}\rangle)}{1+h\langle q\rangle},
\label{Benoitsequation}
\end{equation}
\noindent where $Y_\mathrm{single}$, $Y_\mathrm{prim}$, and $Y_\mathrm{sec}$ are the yields of a single star and of the primary and secondary stars of a binary system, respectively, depending on the mass of the star.
The denominator factor is explained in the next paragraph.
For all primary stars, we use the enhanced binary yield obtained by multiplying the yields of the single star by the binary enhancement factor. To calculate $Y_\mathrm{sec}$, we assume an average mass ratio of $\langle q\rangle=0.5$ with a flat probability distribution between 0 and 1. For any primary mass $M_\mathrm{prim}$, $\langle Y_\mathrm{sec}\rangle$ represents the average yields of secondary stars in the mass range $M_\mathrm{sec}=[0-M_\mathrm{prim}]$. For $M_\mathrm{sec}$ below 10\,$M_\odot$, which is the assumed transition mass between super-AGB and massive stars (\citealt{2015MNRAS.446.2599D}), no $^{26}$Al is ejected. For $M_\mathrm{sec}>10\,$M$_\odot$, we use the single-star yields for a star with mass $M_\mathrm{sec}$ without the enhancement factor. Assuming a binary fraction $h=[0.5-0.9]$, our effective yields are shown in the upper panel of Figure\,\ref{fig_SSP}.\\
\indent We introduced these effective binary yields at solar metallicity (Z=0.014) into the stellar population code \texttt{SYGMA} \citep{2018ApJS..237...42R}, assuming the initial mass function of \cite{2001MNRAS.322..231K} from 0.1 to 100\,M$_\odot$.  The total mass of our stellar population was set to 1\,M$_\odot$, so that our results can be scaled and applied to any population mass. We took the stellar lifetimes from the NuGrid massive star models \citep{2018MNRAS.480..538R} and expelled all the $ ^{26} $Al at the end of the life of the star. For any given stellar mass $M$, the binary contribution included in the effective yields (see Equation\,\ref{Benoitsequation}) represents the $^{26}$Al ejected per binary system having a primary star with a mass $M$. Therefore, since $Y_\mathrm{eff}$ do not only account for the yields of single stars, but also for the yields of binary systems including the contribution of secondary stars, we introduced the correction factor $1+h\langle q\rangle$ at the denominator of Equation\,\ref{Benoitsequation}. This regulates $Y_\mathrm{eff}$ once multiplied with the initial mass function, and ensures that the total mass of our stellar population is normalized to 1\,M$_\odot$.\\
\indent Assuming a binary fraction $h=[0.5-0.9]$ with $\langle q\rangle=0.5$, the time evolution of the mass of $^{26}$Al ejected by the stellar population is shown in the bottom panel of Figure\,\ref{fig_SSP}. After the initial rise of $^{26}$Al, which is caused by the ejecta of the most massive stars, the amount of $^{26}$Al starts to decline due to radioactive decay. The inclusion of the binary effective yields significantly affect the amount of $^{26}$Al produced by the winds when stars with initial mass below $\sim$\,40\,M$_\odot$ start to contribute after $\sim$\,10\,Myr.\\
\indent However, the inclusion of effective binary yields affects the total ejected mass of $^{26}$Al only by about $5-10$\%. This is due to two reasons, (i) the dominant contribution to the wind for $ ^{26} $Al is from the most massive stars, which do not have any binary enhancement factor for their yields, and (ii) even when the binary effect is taken into account the wind ejects $\sim$\,3 times less $^{26}$Al than the yields of \cite{LandC2018}, which include both the wind and the supernovae components (violet dashed line in bottom panel of Figure\,\ref{fig_SSP}).
This suggests that even when including the enhancement to the wind yield of 
$^{26}$Al due to the fact that massive stars are likely born as 
binaries, the major contribution to the total abundance of $^{26}$Al 
produced by a stellar population could still come from core-collapse supernovae. We note that binary interactions can also modify the supernova yields. Therefore, the comparison with \cite{LandC2018} should be taken with caution.\\
\indent Our preliminary conclusion is that mass loss from interacting binaries does not have a strong impact on the Galactic $^{26}$Al abundance and 
$^{60}$Fe/$^{26}$Al ratio observed via $\gamma$-ray spectroscopy, and 
that solutions to possible mismatches between models and observations 
are to be looked for within the nucleosynthesis occurring just before or 
during the core-collapse supernova. This conclusion is preliminary 
because it needs to be tested against a more complete exploration of the 
parameter space (for example the initial mass ratio, the stellar metallicity and the effect of 
rotation) and of the binary scenarios (for example, the 
effect of reverse mass-transfer). Once a more complete set of yields is 
available, we will introduce it into the galactic chemical evolution 
code \texttt{OMEGA} \citep{2017ApJ...835..128C} to address more 
accurately the impact of binary stars on the total mass of $^{26}$Al in 
the Milky Way, and its ratio to $^{60}$Fe.\\
\indent In relation to the presence of $^{26}$Al in the early Solar System, the 
upper panel of Figure\,\ref{fig_SSP} shows that binary stars could have a 
significantly impact. One of the currently favoured scenarios for the 
origin of $^{26}$Al in the early Solar System attributes such origin to 
the winds of one or more massive stars, see e.g. \cite{Gaidos2009, Gounelle2012, Young2014}. One issue with this idea is that, in the case of single 
stars, only those with initial mass larger than roughly 30\,M$ _{\odot} $
produce enough $ ^{26} $Al in the wind to provide a plausible solution, and these stars are 
rare. Our calculations, on the other hand, show that also stars of lower 
mass, which are more common, can expel significant amount of $^{26}$Al 
via winds if they are in a binary system. The implications of this 
result need to be considered carefully in relation to $^{26}$Al in the 
early Solar System, in terms of both the potential stellar source populations and the timescales 
of the ejection as compared to star formation timescales.
\section{Conclusion}\label{sec:Conclusion}
We have computed the $ ^{26} $Al yields from massive non-rotating single and binary stars with the aim of investigating the potential impact of binary interactions on the $ ^{26} $Al yields of massive stars. We have compared these results to each other and to the results of other single non-rotating star studies (\citealt{LandC2006, LandC2018, WandH2007} and \citealt{Ekstrom2012}). We also compared the results of two of the binary systems to the results found by \cite{BraunandLanger}. Our conclusions are that:
\begin{itemize}
	\item The primary stars in binary systems give a higher $ ^{26} $Al yield by up to a factor 100 higher than single stars for masses up to 35-40\,M$ _{\odot} $,  while above 45\,M$ _{\odot} $ the yields become comparable to or lower than the yields found for the single stars.
	\item Our synthetic approach (semi-numerical scheme), where we artificially remove the envelope to simulate binary mass-transfer, represents an upper limit to the $ ^{26} $Al yield, since they strip away more mass than the fully evolved binaries (numerical) and this happens instantaneously instead of gradually over time. The numerical binary yields are also an upper limit since we used fully non-conservative mass-transfer.
	\item When considering the effect of binary yields on the total $^{26}$Al abundance produced by a stellar population, our preliminary conclusion is that the total $^{26}$Al abundance is still dominated by core-collapse supernovae.
\end{itemize}
Future work will include investigations of:
\begin{itemize}
	\item the influence of the reverse mass-transfer, the mass-transfer efficiency, and varying the initial mass ratio,
	\item the influence of rotation and metallicity, in both the single and the binary stars,
	\item more complete models of the Galactic abundance of $^{26}$Al, including a wider exploration of the parameter space(as listed in the two points above),
	\item the impact of our results on the $ ^{26} $Al production in OB-associations and comparison to $ \gamma $-ray observations of such regions.
	\item the impact of our results on the potential origin of $^{26}$Al in the early Solar System from the winds of massive stars,
\end{itemize}
Also, a better determination of the $^{25}$Mg+p reaction and the branching factor to the ground state of $^{26}$Al will allow us to provide more accurate results.\\
We also plan to expand the SNB scheme into a more realistic scheme, i.e., including the orbital adjustment when the SNB primary loses mass, comparable to what is used in population synthesis codes. With this implementation, we will explore more of the binary parameter space, since from Sections\,\ref{RevMT}-\ref{Qs} it is clear that further effects of binary evolution cannot be ignored and apply the SNB scheme to single star models calculated with different codes. The $ ^{26} $Al yields from the secondary stars will be explored as well, by combining the detailed simulations with a binary population synthesis code, which can take the effects of common envelopes and supernova explosions into account. 
\section*{Acknowledgements} \label{sec:thanks}
We are grateful to George Meynet, Roland Diehl, and Moritz Pleintinger for discussion and suggestions and thank Moritz Pleintinger for sharing his table to produce Figure\,\ref{AllTogether}, and Abel Schootemeijer for sharing his implementation of the wind scheme for MESA. We also thank J\'{o}zsef Vink\'{o} for support and discussion. We thank Norbert Langer for giving us the periods for his binary systems. We thank the MESA team for making their code publicly available. We also thank the referee for all their comments and help with improving this paper. This work is supported by the ERC via CoG-2016 RADIOSTAR (Grant Agreement 724560). BC acknowledges support from the National Science Foundation (USA) under grant No. PHY-1430152 (JINA Center for the Evolution of the Elements). EL acknowledges support from National Natural Science Foundation of China (11505117) and Natural Science Foundation of Guangdong Province (2015A030310012).
\software{MESA (\citealt{MESA1, MESA2, MESA3, MESA4}), \texttt{SYGMA} (\citealt{2018ApJS..237...42R}), \texttt{OMEGA}(\citealt{2017ApJ...835..128C})}
\bibliographystyle{apj}
\bibliography{references}
\appendix
\section{Results: 20\,M$ _{\odot} $}\label{20MSUN}
In this section, we cover in detail one of the systems that was presented by \cite{BraunandLanger}. The system is a 20\,M$ _{\odot} $ primary star with an 18\,M$ _{\odot} $ companion. The period is not specified in their paper, though it was confirmed to be a Case B system (\citealt{BraunandLanger}; Langer, private communication). Here, we considered multiple periods for Case B and we also included Case A systems.\\
\indent This section is structured as follows; In Section\,\ref{20MsunSingle}, we describe the evolution of the single star using the Hertzsprung-Russell diagram (HRD) and Kippenhahn diagram (KHD) of this star (Figure\,\ref{20ALL}a,b). In Section\,\ref{20MsunBins} we describe three systems with different periods and cases of mass transfer. Also for these systems we show HRDs and KHDs (Figure\,\ref{20ALL}c-h). The $ ^{26} $Al yields for the numerical single stars and binaries are tabulated in Appendix\,\ref{BigTable} as well as the SNB yields. Note that the decay of the $ ^{26} $Al in the interstellar medium has not been taken into account in the $ ^{26} $Al yields. In Appendix\,\ref{BigTable} the duration of the core hydrogen burning and core helium burning are given, as well as the sizes of the cores at the end of these burning cycles. Further, the total duration of the simulation is given and the total amount of mass lost during the simulation.
\subsection{Single star}\label{20MsunSingle}
\begin{figure*}\begin{center}
	\includegraphics[trim = 2mm 5mm 7mm 13mm, clip, width=0.45\textwidth]{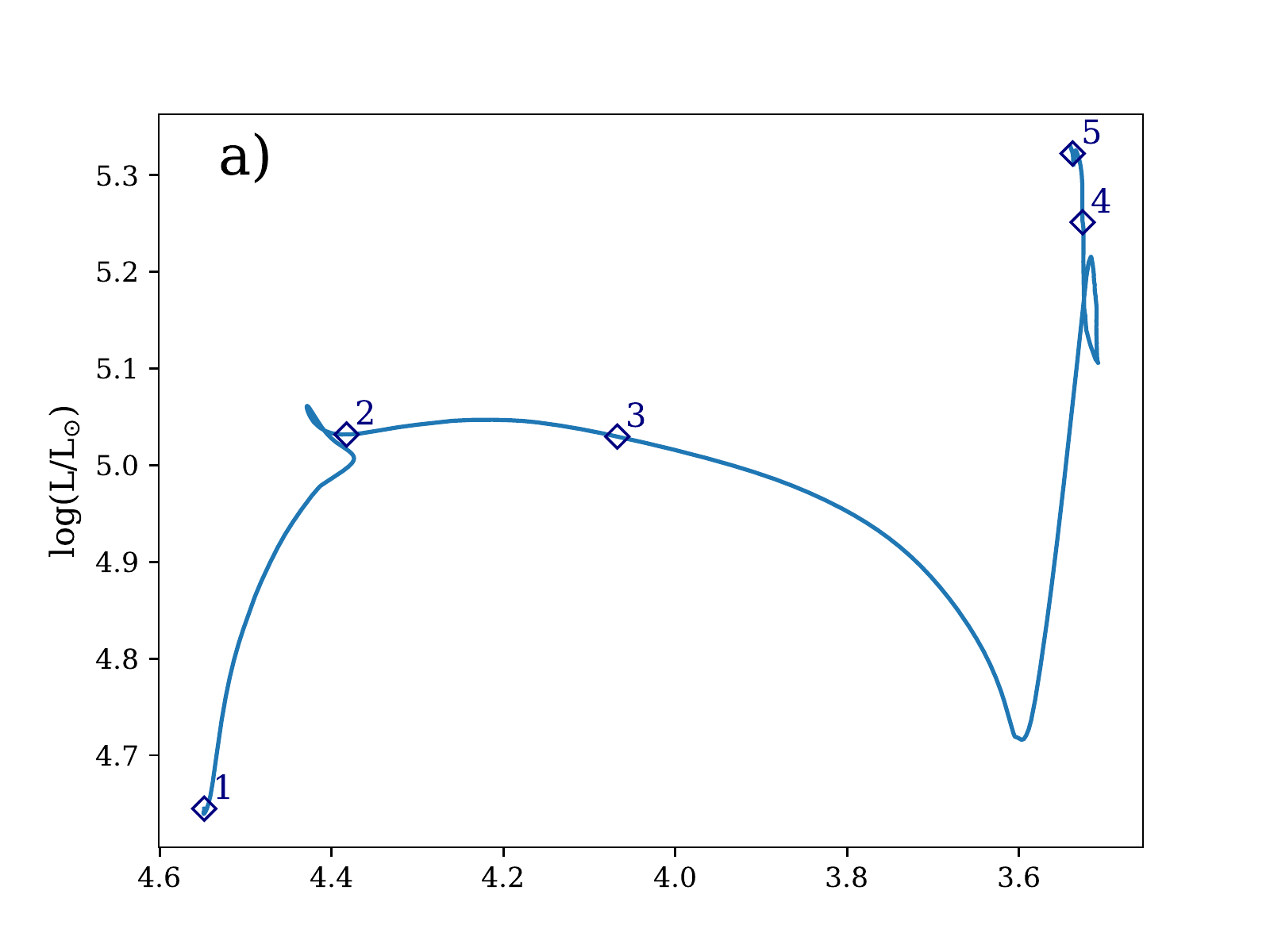}
	\includegraphics[trim = 2mm 5mm 7mm 13mm, clip, width=0.45\textwidth]{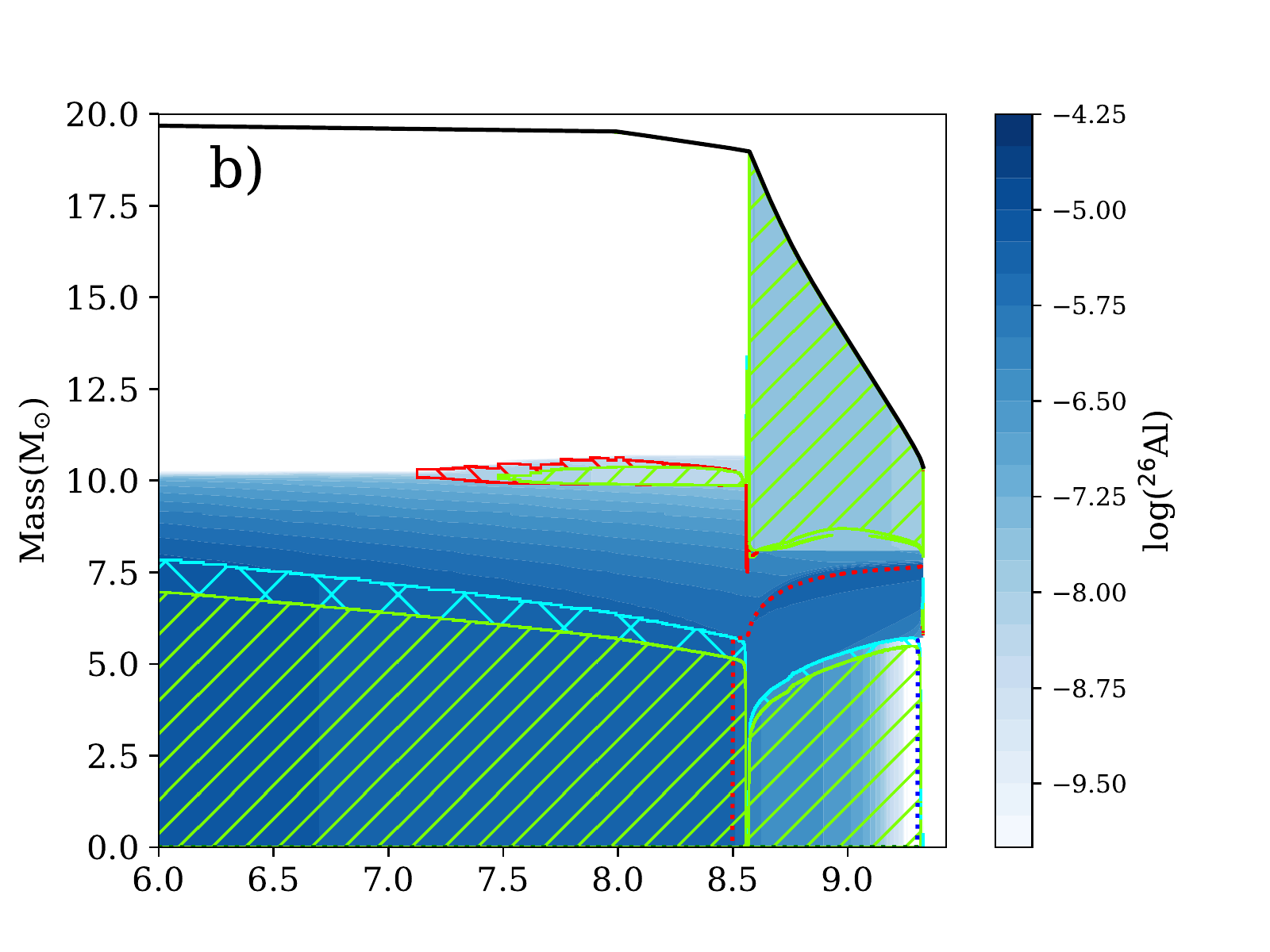}\\
    \includegraphics[trim = 2mm 5mm 7mm 13mm, clip,width=0.45\textwidth]{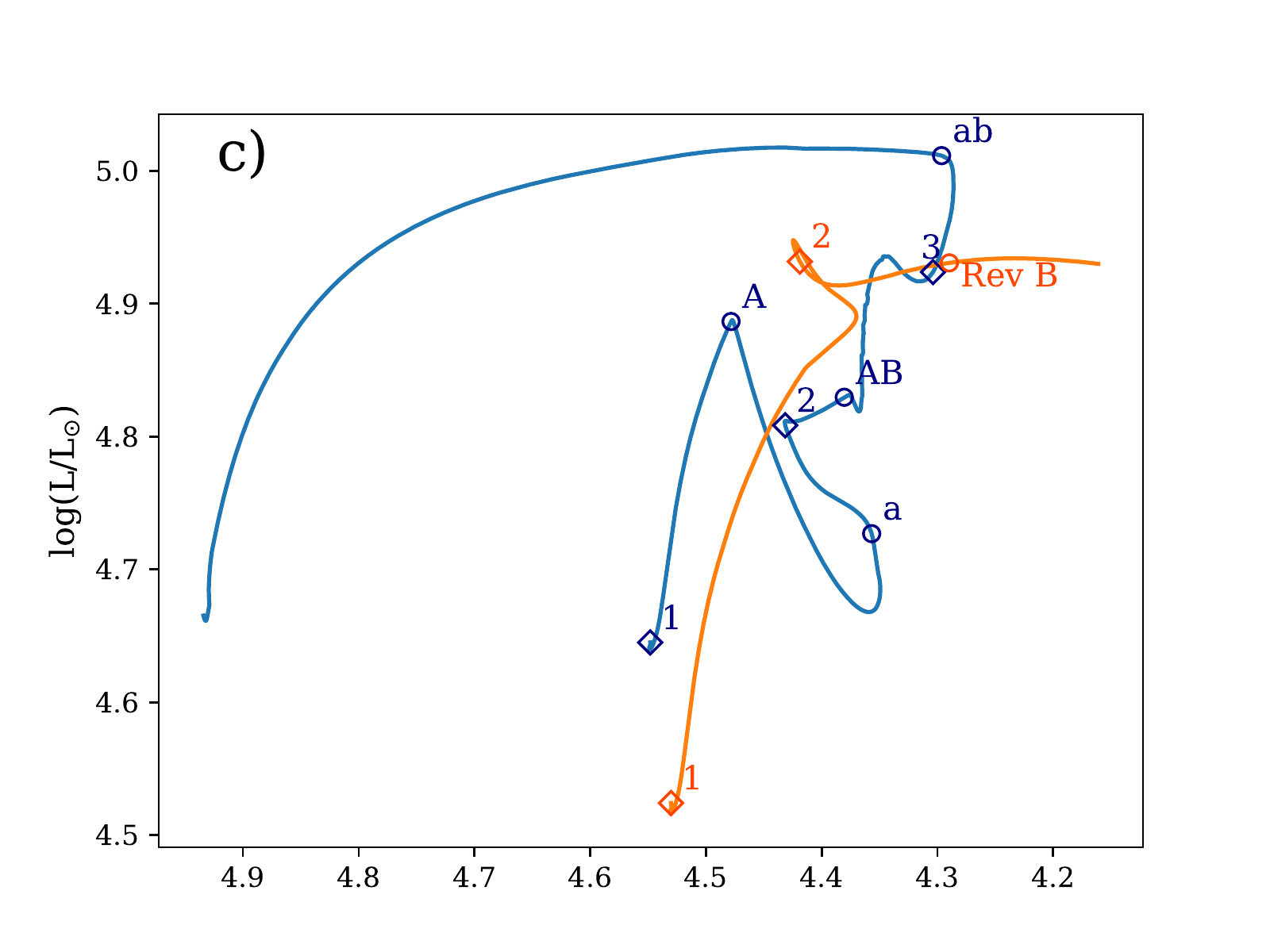}
    \includegraphics[trim = 2mm 5mm 7mm 13mm, clip,width=0.45\textwidth]{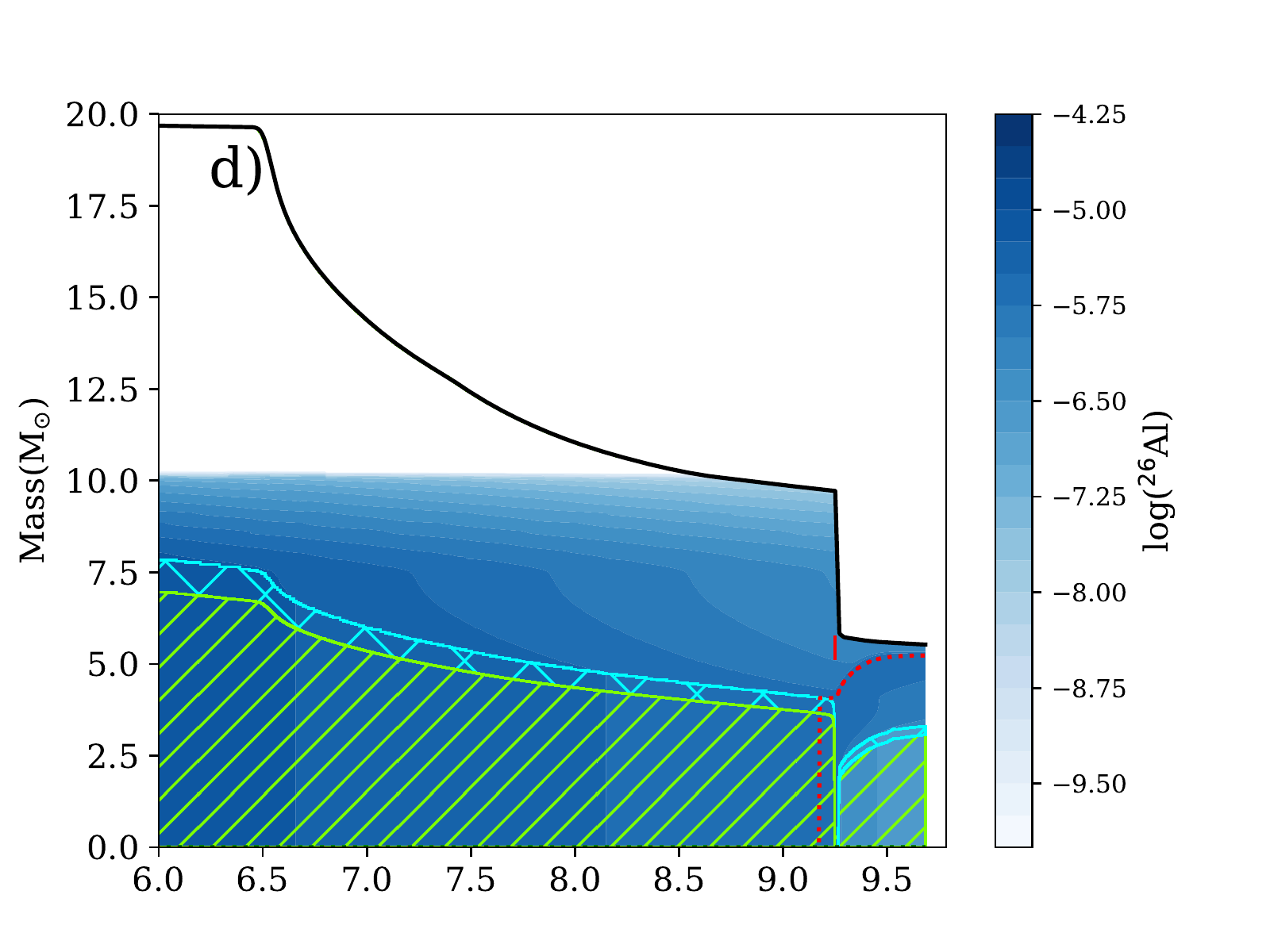}\\
	\includegraphics[trim = 2mm 5mm 7mm 13mm, clip,width=0.45\textwidth]{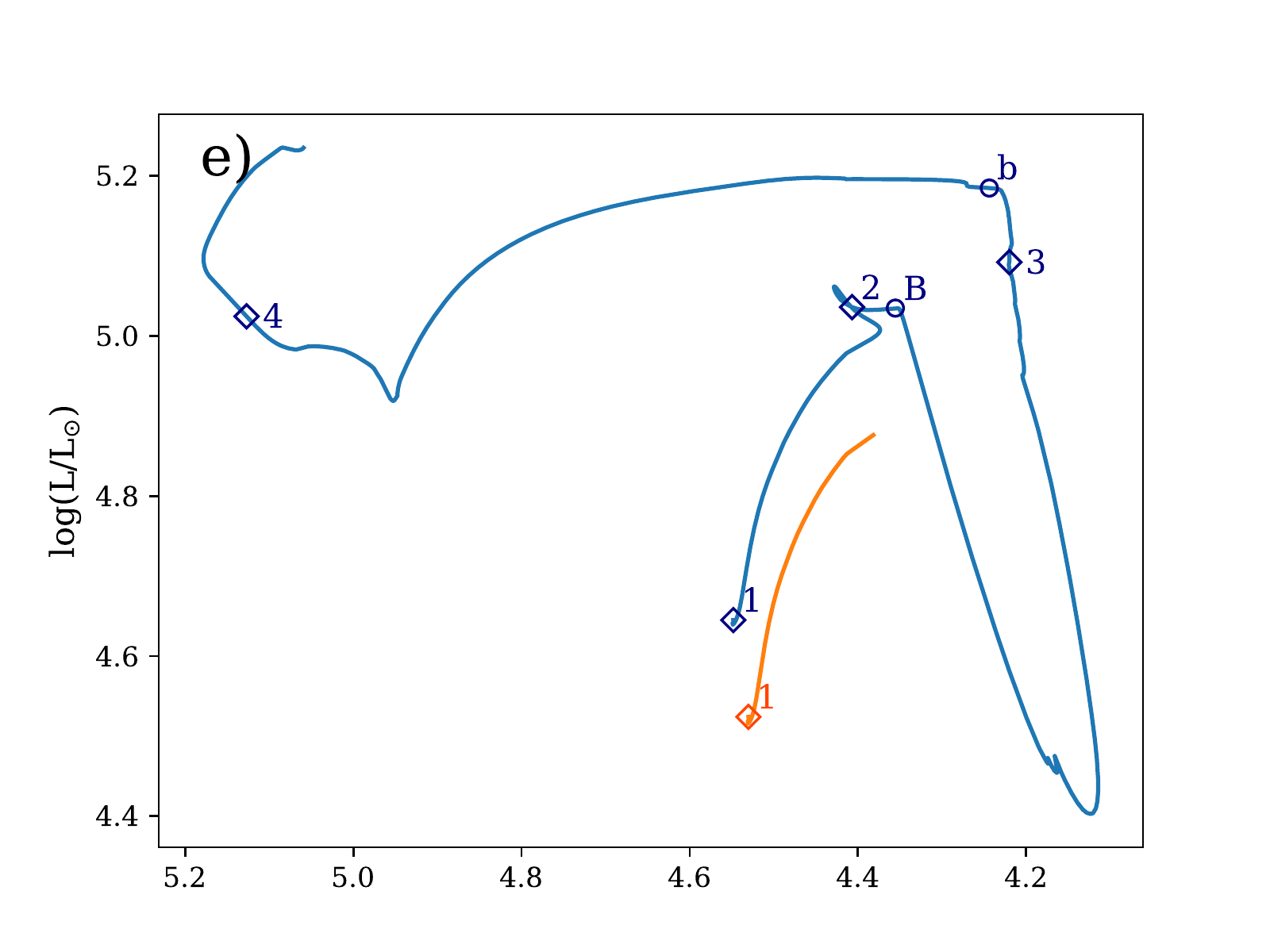}
	\includegraphics[trim = 2mm 5mm 7mm 13mm, clip,width=0.45\textwidth]{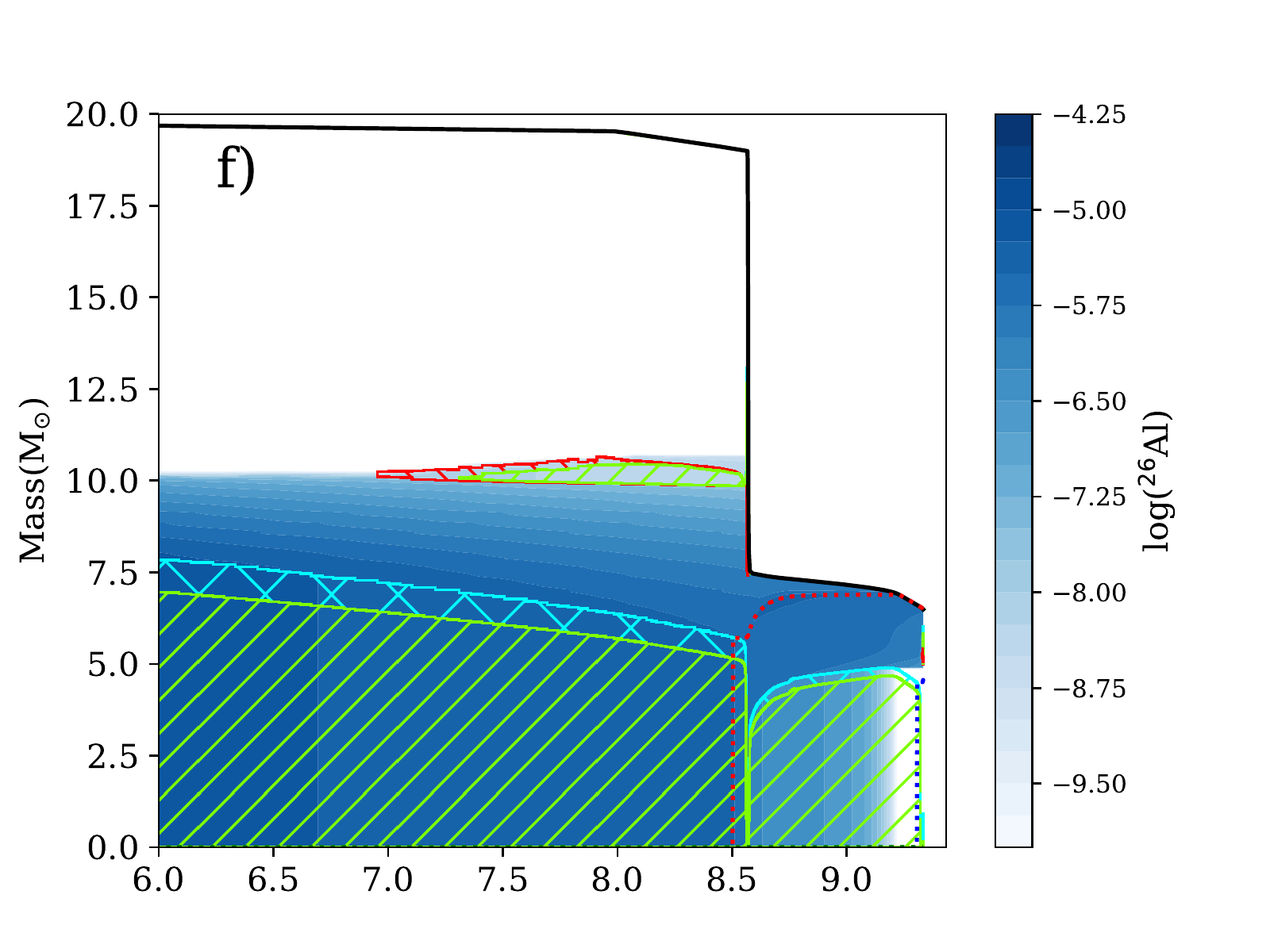}\\
	\includegraphics[trim = 2mm 2mm 7mm 13mm, clip,width=0.45\textwidth]{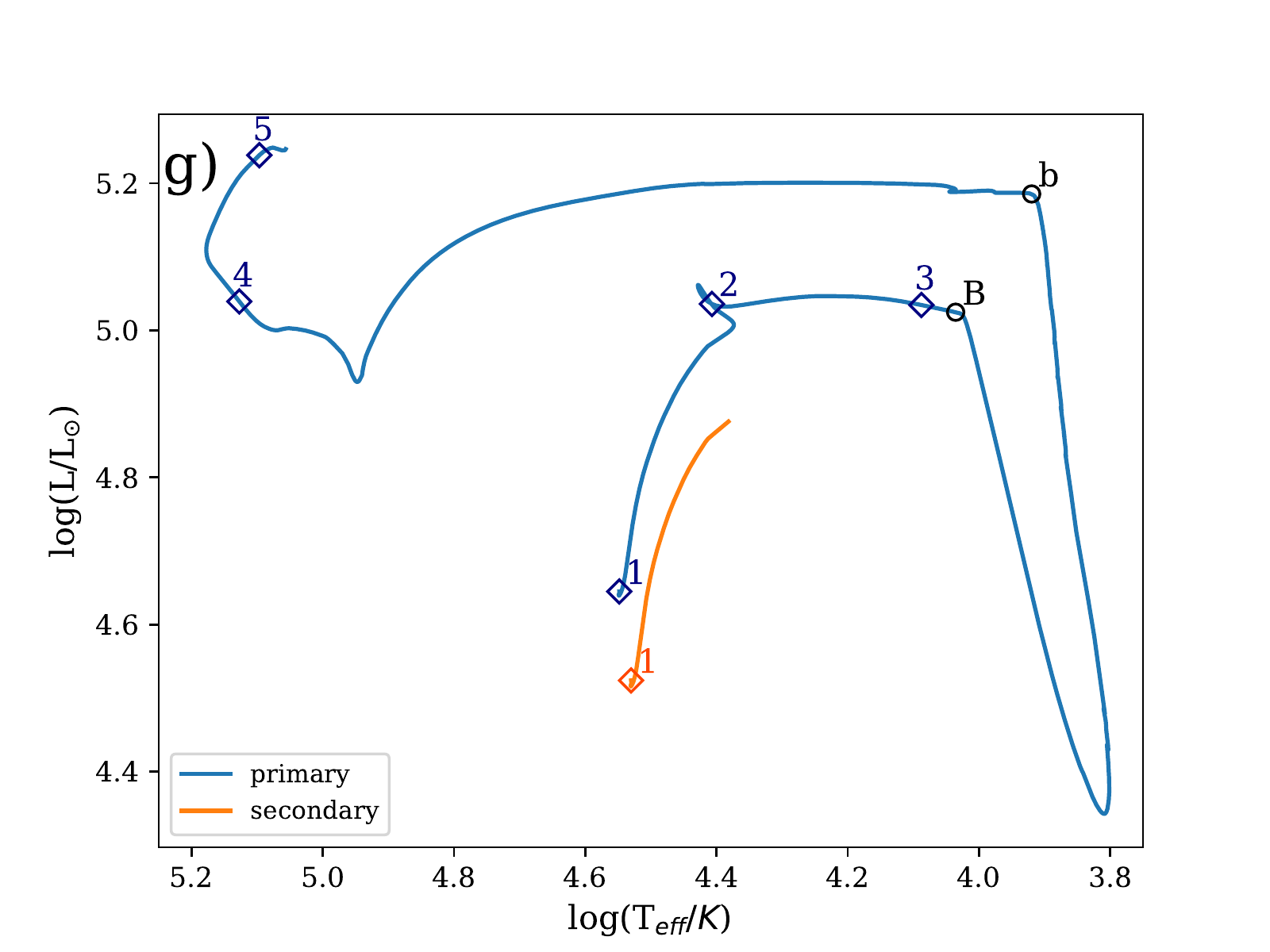}
	\includegraphics[trim = 2mm 2mm 7mm 13mm, clip,width=0.45\textwidth]{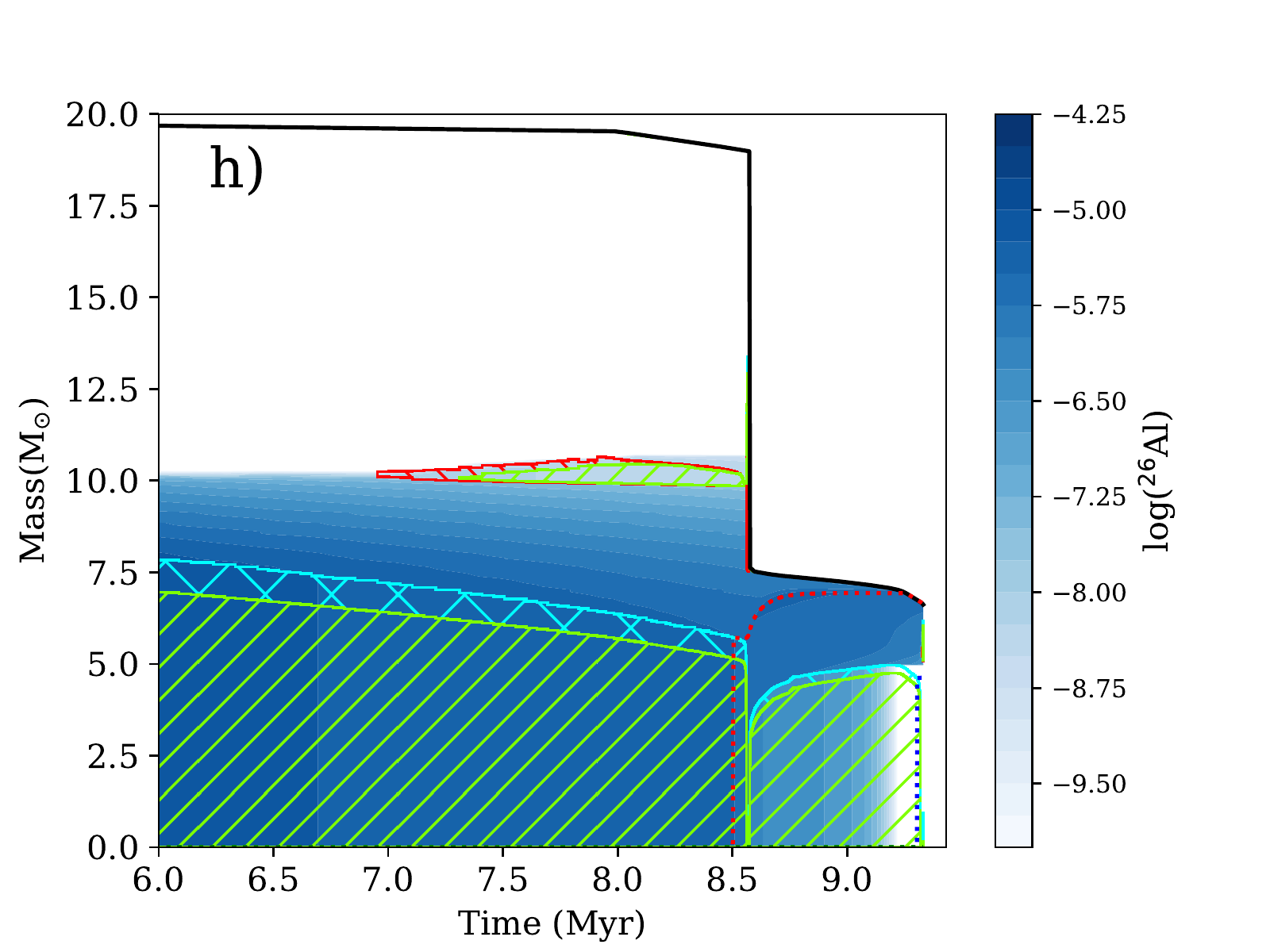}\vspace{-2mm}
	\end{center}
	\caption{HRDs (left panels) and KHDs (right panels) for the 20\,M$_{\odot}$ single star (a,b) and binary systems with a period of 2.5(c,d), 7.4(e,f), and 66.6(g,h) days. The main stages of stellar evolution are indicated with numbers and open diamonds on the track. Point 1 is the start of the main sequence. At Point 2 the hydrogen-burning phase has ended and the star leaves the main sequence. At Point 3 helium is ignited in the core. At Point 4 the helium burning phase ends. At Point 5 carbon burning has begun. We indicate the stages of binary evolution with the open circles and letters on the track. The mass transfer phases start at the capitals and end at the lower cases. For the KHDs all colours and shadings are the same as in Figure\,\ref{Semi-Numerical}.}
	\label{20ALL}
\end{figure*}

Figure\,\ref{20ALL}a shows the HRD for the 20\,M$ _{\odot} $ single star. In the figure, specific times in the stellar evolution are indicated by numbers. Figure\,\ref{20ALL}b shows the KHD of the 20\,M$ _{\odot} $ star with the $ ^{26} $Al content on the colour scale. The $ ^{26} $Al content reaches a maximum value in the center early on in the main sequence, within 1\,Myr. After this, the $ ^{26} $Al is decaying, but there is still production through proton capture on $ ^{25} $Mg. After 3\,Myr the decay is stronger than the production and the $ ^{26} $Al content starts to go down. The decay of $ ^{26} $Al is better visible in the area where the convective core has retreated. At the end of the main sequence the top layer has gone through several half-lives and the $ ^{26} $Al content has gone down. The layers below this have been part of the convective hydrogen burning core for longer, and therefore the $ ^{26} $Al content is higher in these layers.\\
\indent As soon as the star moves off the main-sequence, Point 2 in Figure\,\ref{20ALL}a, the mass-loss rate increases. This is visible in the KHD (Figure\,\ref{20ALL}b) by the steep decrease in total mass during the helium-burning phases. The first core convective area is the central hydrogen burning, the second is the helium burning. Between the main-sequence and the end of helium burning, Point 4, the star loses 9.43\,M$ _{\odot} $. The majority of the mass loss takes place after hydrogen burning, when the star is a red supergiant. From Figure\,\ref{20ALL}b it is clear the majority of the $ ^{26} $Al-rich region is not expelled by the mass loss from the winds. The darker blue area corresponding to the hydrogen-burning shell indicates that $ ^{26} $Al is produced there. However, this $ ^{26} $Al will not be expelled by the wind. The $ ^{26} $Al in these regions will be expelled by the subsequent supernova, which will happen in a few thousand years after the end of the simulation. The yields from the supernova will be larger than the yields from the wind. Figure\,\ref{20ALL}b also shows that during helium burning $ ^{26} $Al is destroyed in the core of the star. This destruction takes place via neutron capture reactions, (n,p) and (n,$ \alpha $), producing $ ^{26} $Mg and $ ^{23} $Na, respectively. These neutrons are produced by the $ ^{13} $C\,($ \alpha $,n)\,$ ^{16} $O reaction \citep{LandC2006}, and the $ ^{22} $Ne($ \alpha $,n)$ ^{25} $Mg reaction \citep{Pignatari2010}, depending on the temperature.
\subsection{Detailed description of selected numerical binaries}\label{20MsunBins}
In this section we look at three binary systems in detail, representing three different cases of mass transfer: a Case A, an early Case B, and a late Case B. All primaries have a mass of 20\,M$ _{\odot} $, all secondaries have a  mass of 18\,M$ _{\odot} $.
\subsubsection{Case A}\label{20MsunCaseA}
As an example of Case A mass transfer, the system with a period of 2.5 days is described here. Both stars start on the main sequence, where the heavier star has a higher luminosity and effective temperature. In the HRD (Figure\,\ref{20ALL}c) this point in time is indicated by 1 for the primary (blue track) and for the secondary (orange track). The mass transfer starts while the primary is still on the main sequence, at the point indicated with A in the figure, and it ends at the point indicated by a. The drop in the luminosity is caused by the mass loss. The main effect of mass transfer occurring early in the evolution is the shrinking the hydrogen burning core. As can be seen in the KHD of the primary star (Figure\,\ref{20ALL}d) the core becomes about half the original mass. This leads to a longer hydrogen burning phase for this star than for the single star by $ \approx $\,0.7\,Myr, which can be seen from comparing Figure\,\ref{20ALL}b and \ref{20ALL}d. The length of the main sequence can also be found in Appendix\,\ref{BigTable}.\\
\indent At the end of the mass-transfer phase the primary is still burning hydrogen in its core. The end of the main sequence is indicated by Point 2 on the HRD in Figure\,\ref{20ALL}c. A second phase of mass transfer starts at Point AB in the HRD. This mass transfer phase is called Case AB, because it takes place during the same phase in the stellar evolution as a Case B, after hydrogen burning (see next sections), but had a Case A mass-transfer phase preceding it. During this phase the star is out of thermal equilibrium, which results in a faster mass-loss rate than for the earlier mass-loss phase. This can be seen by comparing the gradual mass loss between 6.5-8.5\,Myr, and the sharp decrease of the mass at $ \sim $9.25\,Myr in Figure\ref{20ALL}d.
At Point 3 in Figure\,\ref{20ALL}c helium is ignited in the core. The mass transfer continues during the first part of helium burning and stops as soon as the star has regained equilibrium (Point ab in the HRD). At this point nearly the whole hydrogen envelope is lost (Figure\,\ref{20ALL}d at t\,=\,9.25Myr). During helium burning, after the mass transfer, the star moves to the left in the HRD, to higher effective temperatures because the mass of the hydrogen-rich envelope is decreasing as a result of hydrogen-shell burning and mass loss through winds. During this phase the last of the hydrogen envelope will be lost and the hydrogen shell will be extinguished, though the simulation ends before this has taken place. Because the mass transfer is assumed here to be fully non-conservative, the secondary does not accrete mass. Because the main-sequence lifetime of the primary star has been extended by the first mass-transfer phase, the secondary evolves off the main sequence before either of our stopping criteria (carbon ignition or 10$ ^{4} $ models) are met. As the secondary moves of the main sequence, it starts expanding and fills its own Roche lobe, leading to a case of reverse mass-transfer while the primary is burning helium. This mass-transfer phase is called reverse Case B. As soon as the secondary star overfills its Roche lobe, the simulation is stopped because our focus on the primary stars. In Section\,\ref{RevMT} we will briefly discuss the case of reverse mass-transfer.\\
\indent This system gives an $ ^{26} $Al yield of 1.63$ \times $10$ ^{-6} $\,M$ _{\odot} $ (Table\,\ref{OneBigTable}). This is lower than the yield given by the semi-numerical binary, 7.69$ \times $10$ ^{-6} $\,M$ _{\odot} $, by a factor of $ \sim $5. The reason for this is the difference in the mass loss between the SNB and the numerical binary. For the SNB the whole envelope is stripped in one go, while the numerical binary undergoes two mass-transfer phases. Even though the numerical binary loses more mass, the most $ ^{26} $Al-rich region is expelled during the Case AB, and at this point $ ^{26} $Al has already decayed substantially. Compared to the single star, the yield of the binary system is about one order of magnitude larger. By comparing Figure\,\ref{20ALL}b and d, it becomes clear that even though the single star mixes the $ ^{26} $Al through the whole envelope, the Case AB mass transfer reaches deeper layers of the star and the primary loses almost 5\,M$ _{\odot} $ more material than the single star. This leads to the higher yields for the binary system.
\subsubsection{Early Case B}\label{20MsunEarlyB}
When the mass transfer occurs soon after the end of hydrogen burning, it is called early Case B mass transfer, and an example of this is the system with a period of 7.4 days. This system goes through one mass-transfer event. This mass transfer is rapid, which explains the sudden decrease in mass at $ \sim $8.5\,Myr in Figure\,\ref{20ALL}f. The star is strongly out of thermal equilibrium, which leads to the strong decrease in luminosity between Point B and b in Figure\,\ref{20ALL}e. The ignition of helium, at Point 3 in the HRD shown in Figure\,\ref{20ALL}e, is just before the mass-transfer phase ends at Point b. When the star starts to regain equilibrium it detaches from its Roche lobe and the evolution continues towards higher effective temperatures. The luminosity decreases because the hydrogen-burning shell is stripped away due to mass loss by winds during core helium burning, as can be seen in Figure\,\ref{20ALL}f by the star becoming smaller than the hydrogen-depleted core, indicated by the red dotted line. For this system, the difference between the yield from the semi-numerical binary is smaller than for the Case A system because a fraction of the $ ^{26} $Al has already decayed by the time the mass transfer occurs. The SNB and the numerical binary give a yield of 4.49$ \times $10$ ^{-6} $\,M$ _{\odot} $ and 2.39$ \times $10$ ^{-6} $\,M$ _{\odot} $, respectively. The SNB yield is $ \sim $\,2 larger. The total mass loss for the SNB is $ \sim $14.25\,M$ _{\odot} $, which is a combination of wind loss before Roche lobe overflow and then the binary mass loss. The numerical binary loses mass due to winds as well, and then goes through Roche lobe overflow, losing a total amount of $ \sim $13.51\,M$ _{\odot} $. However, only $ \sim $11.5\,M$ _{\odot} $ is lost during the Roche lobe overflow, and the remaining amount is lost through winds after. These factors combined lead to a lower yield for the numerical binary compared to the SNB. Compared to the single star, the yield of the binary system is an order of magnitude larger, and the mass lost is 4\,M$ _{\odot} $ more for the primary star than for the single star.
\subsubsection{Late Case B}\label{20MsunLateB}
The system with a period of 66.2 days is an example of late Case B mass transfer. From the definition, Case B mass transfer takes place between core hydrogen- and core helium-burning. However, as can be seen from Figure\,\ref{Semi-Numerical}, the expansion of the star does not stop at helium ignition, but continues until the star has reached the red supergiant branch. Therefore, some systems start their mass transfer during or shortly after the ignition of helium, where the ignition of helium is defined as the point where the luminosity from the triple-$ \alpha $ reaction is larger than the luminosity from the pp-chain. These systems fall between the definitions of Case B and Case C. Here we will refer to these systems as late Case B systems.\\
\indent When comparing the KHD for this system, Figure\,\ref{20ALL}h, with the KHD in Figure\,\ref{20ALL}f, the difference is very small, all Case B mass transfer phases occur within a timespan of 0.01\,Myr. However, when comparing the HRDs, Figs.\,\ref{20ALL}e and \ref{20ALL}g, the difference becomes clear. In the latter figure, the primary star of the binary evolves further along the Hertzsprung gap than in the former. This leads to a small difference in the final yield for the stars, 2.17$ \times $10$ ^{-6} $\,M$ _{\odot} $ for the 66.2 day period and 2.39$ \times $10$ ^{-6} $\,M$ _{\odot} $ for the 7.4 day period (Table\,\ref{OneBigTable}). This difference is rather small, because both systems are very similar up to the moment of mass transfer and the times at which the mass transfer starts are close. The difference is that for the wider system the envelope is slightly less stripped. Compared to the single star, the yield of the binary system is an order of magnitude larger, and the mass lost is 4\,M$ _{\odot} $ more for the primary star than for the single star.
\section{Results: 50\,M$ _{\odot} $}\label{50MSUN}
In this section, we cover in detail the other system that was presented by \cite{BraunandLanger}. The system is a 50\,M$ _{\odot} $ primary star with an 45\,M$ _{\odot} $ companion. The period is not specified in the paper, though it was confirmed to be a Case B system as well (\citealt{BraunandLanger}; Langer, private communication). We consider multiple periods, just as for the 20+18\,M$ _{\odot} $ system in Section\,\ref{20MsunBins}.\\
\indent This section is structured as follows; In Section\,\ref{50MsunSingle}, we describe the evolution of the single star using the HRD and KHDs of this star, Figure\,\ref{50ALL}a,b. In Section\,\ref{50MsunBins} we describe three systems with different cases of mass transfer. For these systems we show HRDs and KHDs, Figure\,\ref{50ALL}c-f. The yields for the numerical stars, both single and binary, are tabulated in Appendix\,\ref{BigTable}.
\subsection{Single star}\label{50MsunSingle}
\begin{figure*}\begin{center}
	\includegraphics[trim = 2mm 2mm 7mm 13mm, clip, width=0.45\textwidth]{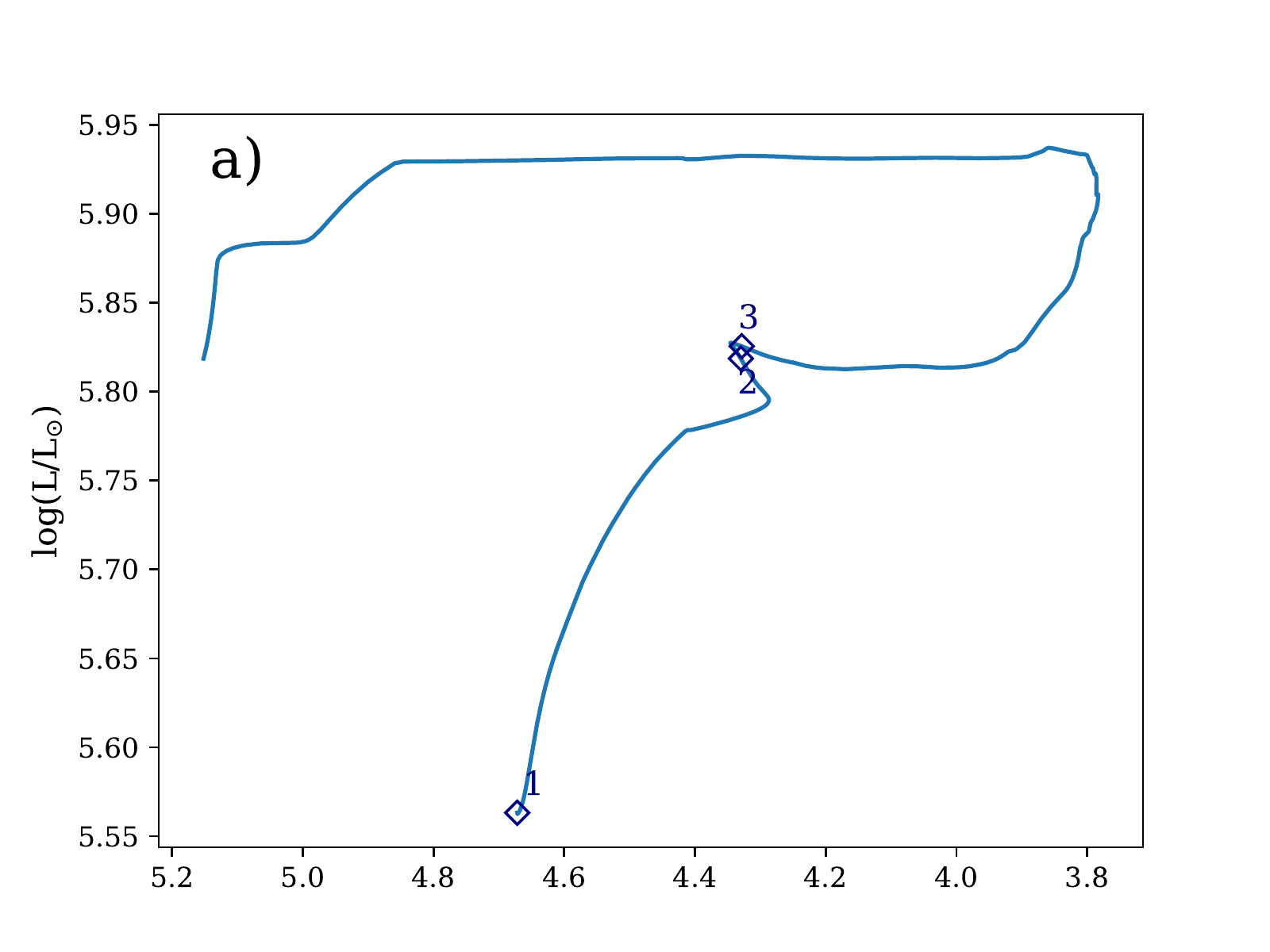}
	\includegraphics[trim = 2mm 2mm 7mm 13mm, clip, width=0.45\textwidth]{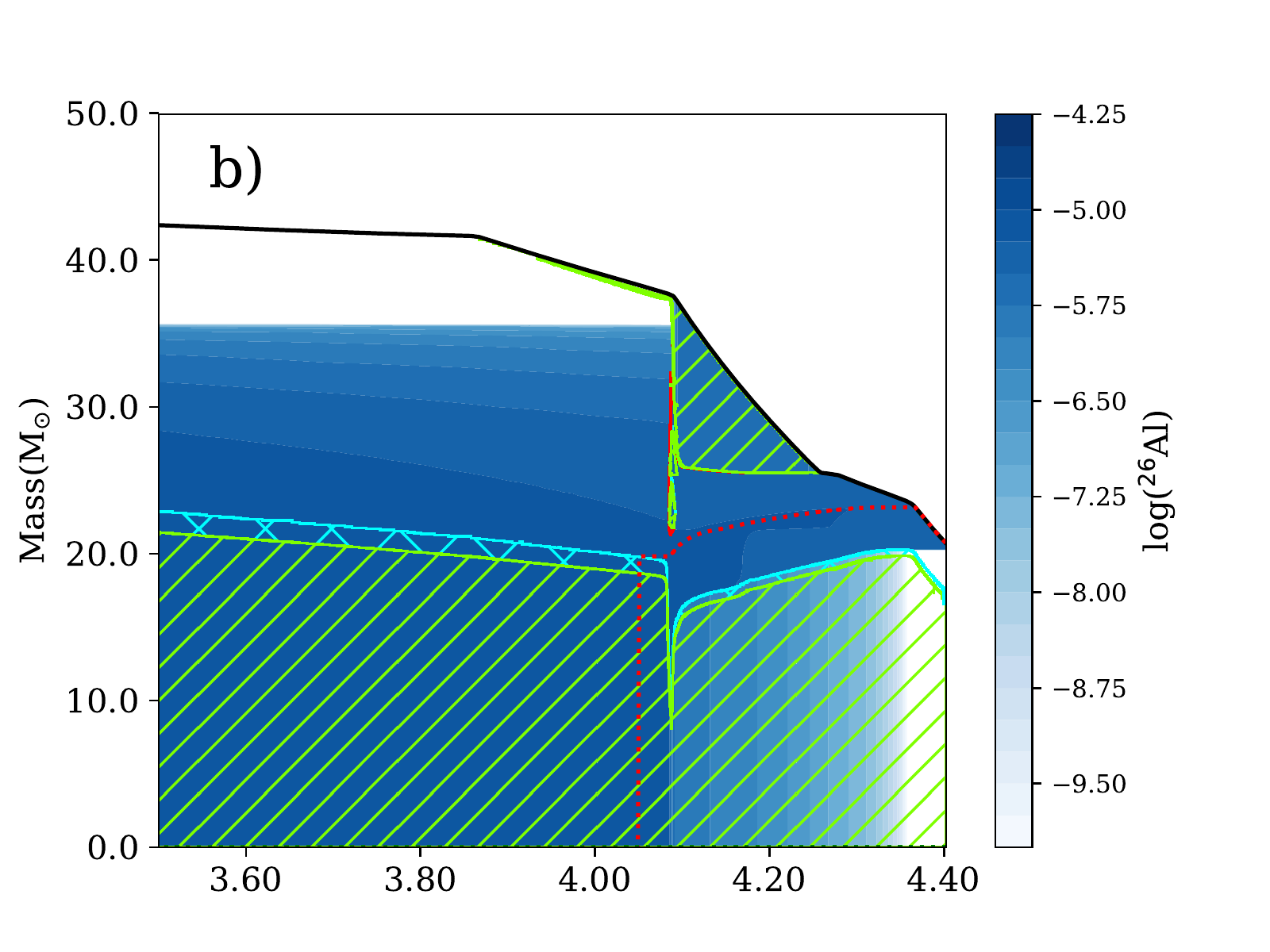}\\
    \includegraphics[trim = 2mm 2mm 7mm 13mm, clip,width=0.45\textwidth]{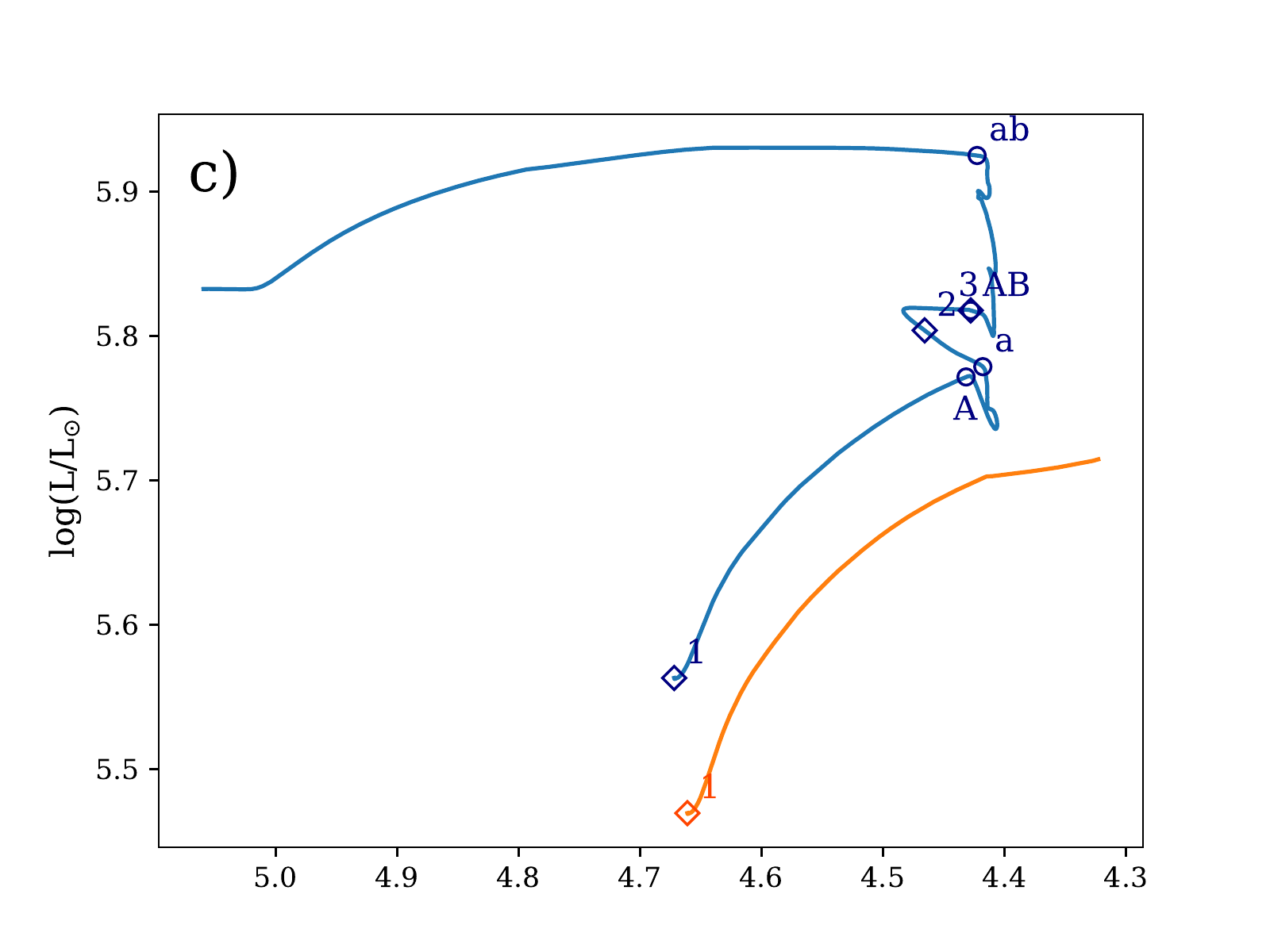}
    \includegraphics[trim = 2mm 2mm 7mm 13mm, clip,width=0.45\textwidth]{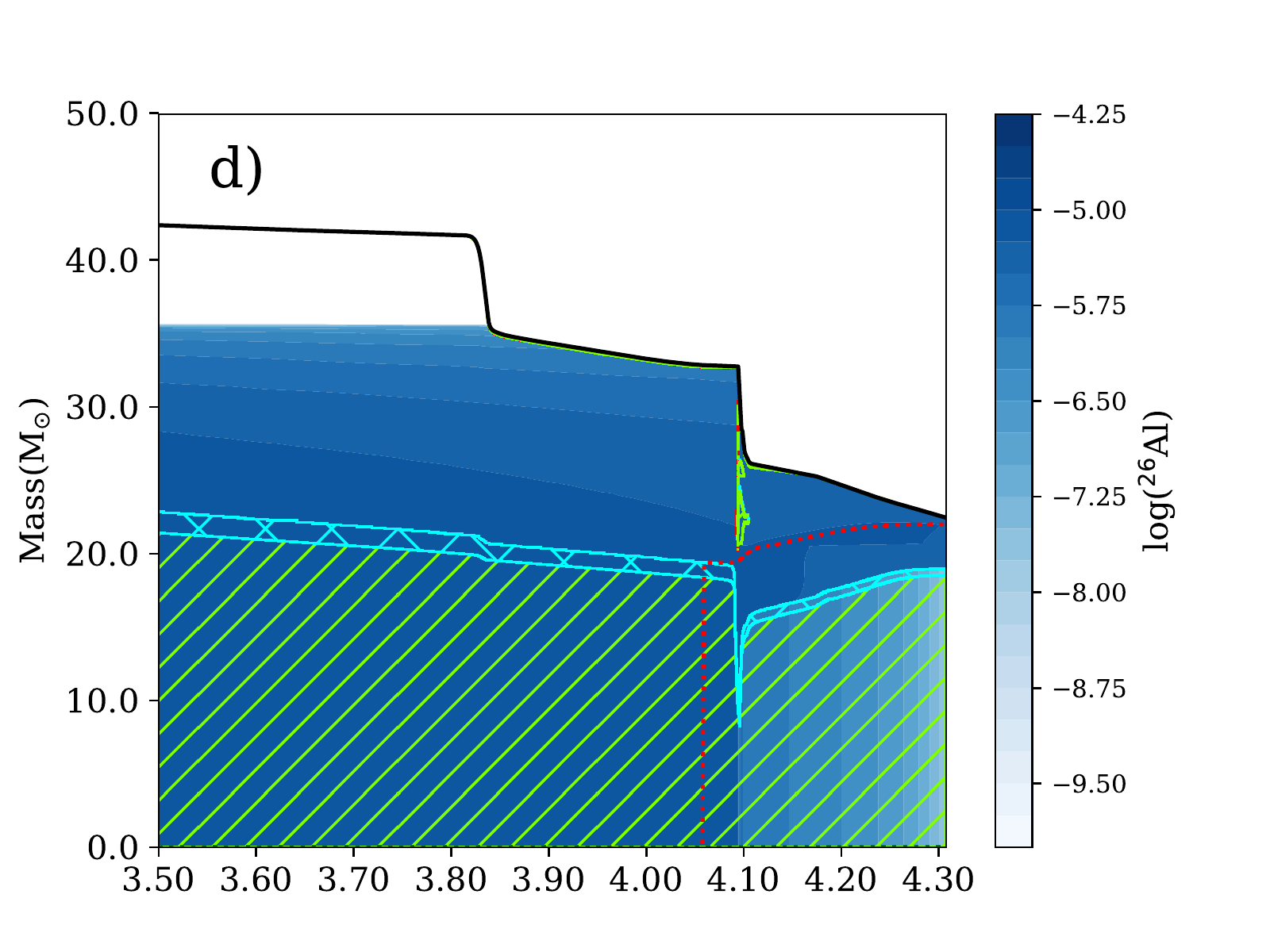}\\
	\includegraphics[trim = 2mm 2mm 7mm 13mm, clip,width=0.45\textwidth]{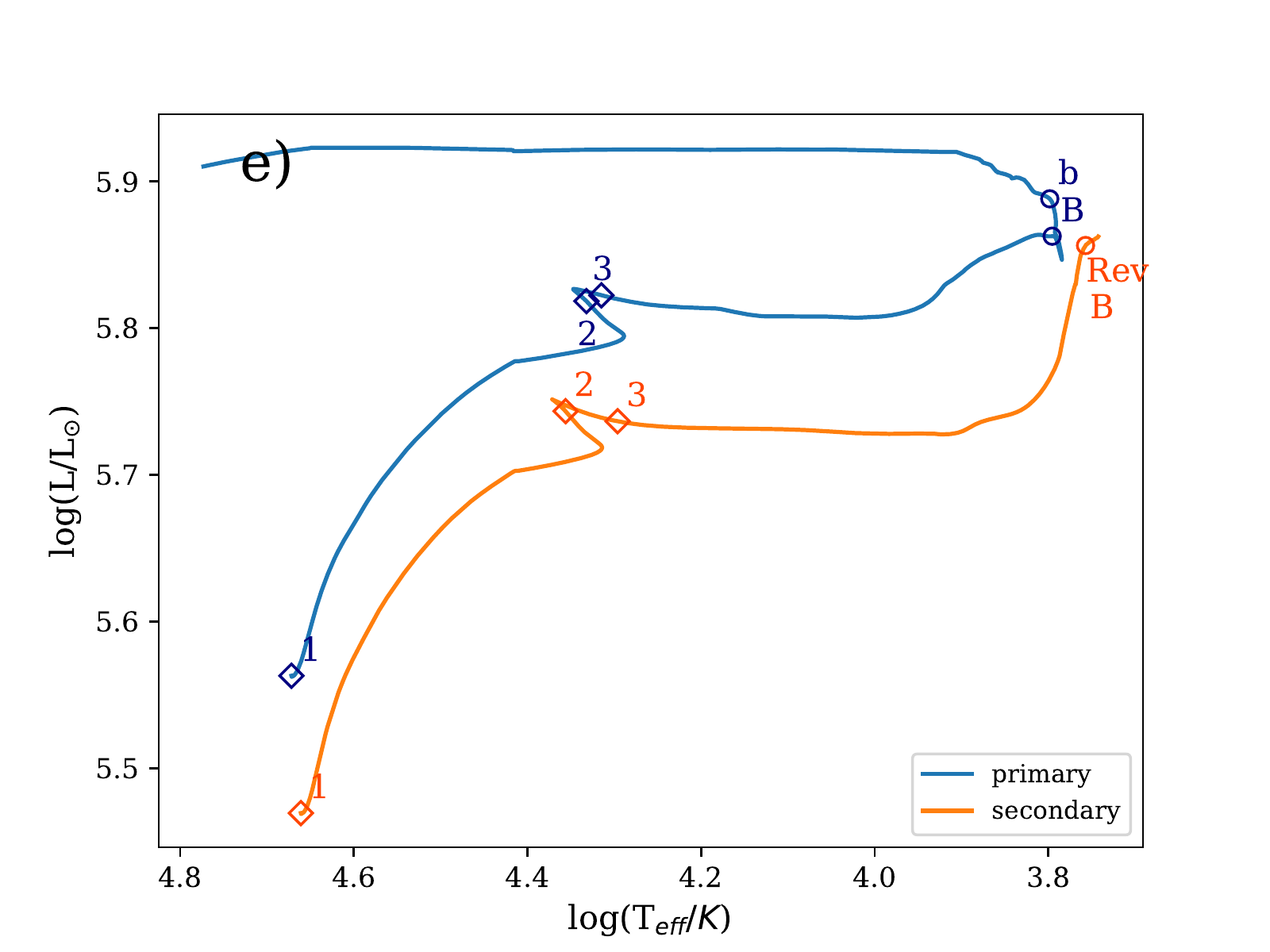}
	\includegraphics[trim = 2mm 2mm 7mm 13mm, clip,width=0.45\textwidth]{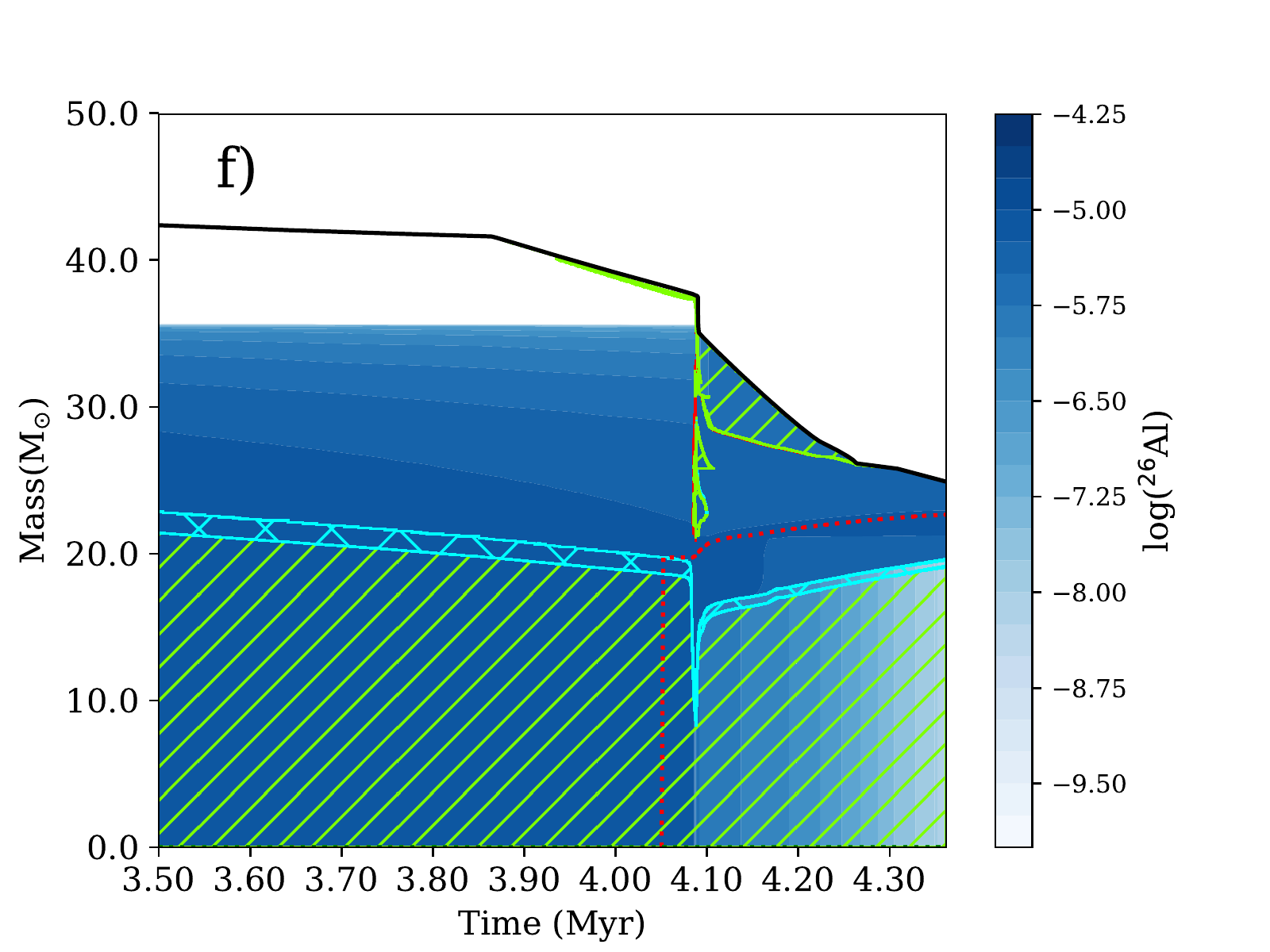}\\\vspace{-2mm}
	\end{center}
	\caption{HRDs (left panels) and KHDs (right panels) for the 50\,M$_{\odot}$ single star (a,b) and binary systems with a period of 8.1(c,d), and 72.3(e,f) days. The main stages of stellar evolution are indicated with numbers and open diamonds on the track. Point 1 is the start of the main sequence. At Point 2 the hydrogen-burning phase has ended and the star leaves the main sequence. At Point 3 helium is ignited in the core. We indicate the stages of binary evolution with the open circles and letters on the track. The mass transfer phases start at the capitals and end at the lower cases. For the KHDs all colours and shadings are the same as in Figure\,\ref{Semi-Numerical}.}
	\label{50ALL}
\end{figure*}

Figure\,\ref{50ALL}a shows the HRD for the single star of 50\,M$ _{\odot} $. As before, the important stages of stellar evolution are indicated by numbers. Figure\,\ref{50ALL}b the KHD for this star with the $ ^{26} $Al content on the colour scale. Even before the main sequence ends, mass-loss rate increases (at $ \sim $3.85\,Myr in Figure\,\ref{50ALL}b). This is due to a change in the opacity of the envelope at T$ _{eff} $\,$\approx$\,25000K, also known as the bi-stability jump (see Table\,2 of \citealt{Vink2000}). As soon as the star moves off the main sequence, Point 2 in Figure\,\ref{50ALL}a, the mass-loss rate increases even more, as can be seen in Figure\,\ref{50ALL}b at $ \sim $\,4.1\,Myr. The mass-loss rate is higher for this star than for the 20\,M$ _{\odot} $ star in Section\,\ref{20MsunSingle} because its luminosity is higher, and by the time the whole envelope is stripped away, the hydrogen-burning shell is extinguished and then this region is stripped away as well. What is left is the hydrogen-depleted core of the star, and the star has become a Wolf-Rayet star. The star loses about 29\,M$ _{\odot} $ through these stellar winds, more than half of its mass. The $ ^{26} $Al yield for this single star is 4.47$ \times $10$ ^{-5} $\,M$ _{\odot} $ (Appendix\,\ref{BigTable}).\\
\indent As can be seen in Figure\,\ref{50ALL}b, the star does not reach the end the helium-burning phase during the simulation due to convergence issues. However, from Figure\,\ref{50ALL}b it can also be seen that very little $ ^{26} $Al is left in the star at this point in time.
\subsection{Detailed description of selected numerical binaries}\label{50MsunBins}
In this section we look at two binary systems in detail, representing two different cases of mass transfer, a Case A and a late Case B. All primaries have a mass of 50\,M$ _{\odot} $, all secondaries have a mass of 45\,M$ _{\odot} $. \cite{BraunandLanger} found that for masses $ \gtrsim $ 40\,M$ _{\odot} $, $ \gtrsim $\,2 times higher than 20\,M$ _{\odot} $, the binary systems give lower yields than the single stars. In this section we discuss this finding as well.
\subsubsection{Case A}\label{50MsunCaseA}
As in Section\,\ref{20MsunCaseA}, the system described in this section undergoes Case A mass transfer. The system has a period of 8.1 days. Both stars start out on the main sequence in Figure\,\ref{50ALL}c at the point indicated by 1. The mass transfer starts at Point A, before the primary has finished hydrogen burning (Point 2) and ends at the point indicated by a. Compared to the primary in Figure\,\ref{20ALL}c, the drop in the luminosity is a lot smaller. The hydrogen-burning core becomes less massive than for the single star. A secondary mass transfer starts at Point AB, which is after helium ignition (Point 3) and ends at Point ab. As explained in Section\,\ref{20MsunLateB}, we refer to this mass transfer as a Case AB. The mass transfer stops when the primary star contracts as it regains equilibrium due to helium burning in its core. The star moves to the left in the HRD, to higher effective temperatures. During this phase, additional  mass is lost on top of the mass lost due to the binary interaction. This is visible in Figure\,\ref{50ALL}d, where the the first of the near vertical decreases corresponds to fast phase of the Case A mass transfer, the slower decrease after is a combination of the slower phase of the Case A mass transfer and the wind loss, the second near vertical decrease corresponds with Case AB mass transfer. All mass lost after this is lost through winds. The total amount of mass lost by the primary star in the binary system is lower than the mass lost by the single star, $ \sim $27.5\,M$ _{\odot} $, leading to a 15\% smaller $ ^{26} $Al yield of 3.82$ \times $10$ ^{-5} $\,M$ _{\odot}$ less. However, this star does not get as far into helium burning as the single star before the calculation stops, which can be seen from comparing Figure\,\ref{50ALL}b and d. Additional mass will be lost in the wind during the rest of the helium-burning phase, bringing the yields closer to each other. This is different from the 20\,M$ _{\odot} $, where the difference was an order of magnitude in the other direction. This is because the mass-loss history between the single star and the Case A primary is more similar, and the mass loss through winds and the mass loss through Roche lobe overflow have a similar effect. This result is in agreement with \cite{BraunandLanger}.
\subsubsection{Case B}
As in Section\,\ref{20MsunLateB}, the system described in this section undergoes late Case B mass transfer. The system has a period of 72.3 days. In the HRD (Figure\,\ref{50ALL}e) the mass-transfer phase (points B to b) is not as recognizable as in the previous HRDs. There is only a small dip in the luminosity. The secondary overflows the Roche lobe at the end of the simulation, leading to reverse Case B mass transfer. Compared the Case A system described above, the binary star loses even less mass, $ \sim $\,25\,M$ _{\odot} $. The yield is also lower, 2.59$ \times $10$ ^{-5} $M$ _{\odot} $, $ \sim $\,43$ \% $ less than the single star yield. However, a large amount of $ ^{26} $Al is left in the envelope, which could increase the yield to a similar level as the single star if we continue the helium-burning phase. For the Case B system the mass-loss history is almost identical to the mass-loss history of the single star. This is because the mass transfer happens during a phase where the star is already rapidly losing mass due to the red supergiant wind, which can be seen from comparing Figure\,\ref{50ALL}b and d. This makes the effect of the mass transfer rather small. This can also be seen when comparing Figure\,\ref{50ALL}a and e. The tracks in the HRD are very similar. The main effect of the mass transfer is that the red supergiant phase is shortened, turning the primary into a Wolf-Rayet star at an earlier stage. For the Wolf-Rayet star, the mass loss rate is lower than for the red supergiant. This explains why the primary star of the binary loses less mass than the single star.
\section{Tablulated results for all models}\label{BigTable}
In Table\,\ref{OneBigTable} all $ ^{26} $Al yields for both the single as binary star models as well as more information on the evolutionary phases of the stars can be found.
\begin{table*}[t]
\caption{M$ _{ini} $ is the initial (primary) mass in M$ _{\odot} $. For all binary systems the mass ratio is the same, q=0.9. P$ _{init} $ is the initial period in days. For the single stars there is no period. The case of the mass transfer is given in the next column. t$ _{H} $, t$ _{He} $, and t$ _{tot} $ are the durations of hydrogen burning, helium burning, and the simulation in Myr, respectively. M$ _{H} $ and M$ _{He} $ are the masses of the hydrogen-depleted core and the helium-depleted core at the end of the corresponding burning phases in M$ _{\odot} $. Y$ _{c} $ gives the final helium abundance if helium burning was not finished. $\Delta$M is the total mass lost in M$ _{\odot} $. $ ^{26} $Al and $ ^{26} $Al (SNB) give the $ ^{26} $Al yields for the numerical stars and the SNBs, respectively.}
\begin{center}\begin{tabular}{cccccccccccc}
\hline 
M$ _{ini} $&  P$ _{init} $& Case&t$ _{H} $& M$ _{H} $&t$ _{He} $ & M$ _{He} $&Y$ _{c} $&t$ _{tot} $ & $  \Delta$M&$ ^{26}Al $ &$ ^{26}Al $(SNB) \\
(M$ _{\odot} $) & (days)&& (Myr) & (M$ _{\odot} $)&(Myr) &(M$ _{\odot} $)&&(Myr) & (M$ _{\odot} $)&(M$ _{\odot} $) &(M$ _{\odot} $) \\
\hline
10 & - &-& 23.19& 1.82& 2.29& 1.52& -&25.61& 1.01& 2.62e-10&-\\
& 2.8$ ^{1} $&A& 24.71& 1.10& 3.17& - &0.18& 28.05& 8.10& 5.22e-08&8.13e-07\\
& 4.9&B& 23.19&1.82 & 2.60& 1.17&-& 25.97&7.60 &5.56e-08 &5.93e-07\\
& 13.1&B&23.19&1.82 &2.56 & 1.19&-& 25.93& 7.56&2.60e-08&6.23e-07\\
\hline
15 & - &-&12.20 & 3.62 & 1.09& 3.40&-& 13.33& 3.99&9.06e-09 &-\\
& 3.8&A&12.54 & 2.62& 1.36&2.095 &-& 13.97& 11.31& 9.74e-07 &2.81e-06\\
& 6.7&B& 12.20& 3.62&1.14 &2.79 &-&13.39 & 10.43&2.98e-07 &1.95e-06\\
& 16.8&B& 12.20& 3.62& 1.13& 2.83&-& 13.38&10.37 &2.59e-07&1.87e-06\\
\hline 
20 & -&-&8.56 & 5.73 & 0.75 &5.65 &-& 9.33 & 9.62   &2.01e-07& -\\ 
  & 2.5$ ^{1,2} $ &A& 9.25&4.07 & 0.42 & - & 0.52&9.68 &14.47&1.63e-06&7.69e-06\\
  & 5.1 &A&8.66 &4.72 & 0.79&  3.73&-& 9.48 &14.34&3.51e-06&5.86e-06\\
  & 6.2 &B$ ^{6} $& 8.56&  5.71& 0.75 & 4.46 &-& 9.33 &13.53&2.43e-06&5.65e-06\\ 
  & 7.4 &B&8.56 &  5.73& 0.75 &  4.48&-& 9.33 &13.51&2.39e-06&4.49e-06\\ 
  & 18.4 &B& 8.56&5.73 & 0.75 & 4.71& -&9.33 &13.27&1.99e-06&4.38e-06 \\
  & 66.2 &B& 8.56& 5.73& 0.75 &4.62&-& 9.33 &13.38&2.17e-06&4.20e-06\\
\hline
25 & - &-&6.81 & 7.94 & 0.60 & 7.95 &-& 7.43 & 14.13 &1.32e-06&-\\
  & 2.7$ ^{2} $ &A&7.24 & 6.20& 0.32& - &0.49& 7.56 &16.95&4.24e-06 &7.35e-06\\
  & 6.7$ ^{2} $ &A&6.85 & 7.06 &  0.45 & - & 0.18&7.31 &16.72&6.68e-06 &1.07e-05\\ 
  & 8.9$ ^{4} $ & B&-&  -& - & - & - &-&-&failed&8.59e-06\\ 
  & 17.8$ ^{4} $ &B& -& -& - & - & - &-&-&failed&8.39e-06\\
  & 71.3&B& 6.81&7.94 & 0.60 & 6.58 &-& 7.42 &16.22&5.62e-06&8.12e-06\\
\hline
30 & - &-&5.81& 10.26 & 0.54 &10.27  &-& 6.36 & 15.91 &3.27e-06& -\\
& 2.8$ ^{1,3} $ &A& -&- & - & - & - &-&13.38&2.79e-07 &2.36e-05\\
  & 8.4$ ^{2} $ &A&5.82 & 9.54 & 0.35& - &0.23& 6.18 &19.16&1.14e-05&1.73e-05\\ 
  & 10.1$ ^{1} $ &B$ ^{6} $& 5.81& 10.26 & 0.55 &8.88 &-& 6.37 &18.38&1.04e-05&1.71e-05\\ 
  & 12.2$ ^{1} $ &B&5.81 & 10.26& 0.55 & 8.94 &-& 6.36 &17.88&1.00e-05&1.45e-05\\
  & 30.3 &B&5.81 & 10.26 &0.53& 9.15 &-&6.36 &18.18&9.25e-06&1.41e-05\\ 
  & 75.4$ ^{1} $ &B& 5.81&  10.26 & 0.55 & 9.24&-& 6.37&17.83&7.91e-06&1.40e-05\\ 
\hline 
35 & - &-&5.16& 12.61 &  0.48& 12.57 &-& 5.65 &18.74 & 7.52e-06&-\\
& 2.9$ ^{1,3} $ &A& -&-& - & - & - &-&14.52&7.04e-07 &3.58e-05\\
  & 8.8$ ^{2} $ &A& 5.16& 11.98 & 0.28& - &0.30& 5.45 &21.59&1.77e-05&2.58e-05\\ 
  & 10.6$ ^{2} $ &B$ ^{6} $& 5.16& 12.57 & 0.30 & - & 0.25&5.46 &21.41&1.70e-05&2.54e-05 \\ 
  & 12.7$ ^{1,2} $ &B& 5.15& 12.64& 0.43 & - & 0.08& 5.59&19.04&1.17e-05&2.17e-05\\
  & 31.5$ ^{1,2} $ &B& 5.15& 12.64& 0.44 &  -& 0.04&5.59 &20.11&1.27e-05&2.14e-05\\ 
  & 78.6$ ^{1,2} $ &B& 5.15& 12.64& 0.44 & - &  0.04&5.59 &20.01&1.22e-05&2.10e-05\\
\hline 
40$ ^{2} $  & - &-&4.70&  15.00& 0.35 & -&0.12& 5.06 &20.57&1.06e-05&-\\
& 3.1$ ^{4} $ &A& -& -& - & - & - &-&-&failed&4.88e-05\\
  & 7.6$ ^{2} $ &A& 4.71&  14.38& 0.06& -&0.83  &4.78 &21.61&1.55e-05&3.82e-05\\ 
  & 15.8$ ^{1,2} $ &B$ ^{6} $& 4.70& 15.00 & 0.36& -&0.13 & 5.06 &21.15&1.92e-05&3.55e-05\\ 
  & 20.4$ ^{2} $ &B& 4.70& 15.00& 0.24 & -& 0.35& 4.94&21.50&1.73e-05&3.11e-05\\
  & 32.8$ ^{2} $ &B& 4.70& 15.00 & 0.09 & -& 0.74& 4.79 &21.21&1.44e-05&3.10e-05\\ 
  & 81.7$ ^{2} $ &B& 4.70& 15.00& 0.12 & - &0.68&4.82 &21.01&1.35e-05&3.06e-05\\
\hline
\end{tabular}\end{center}
\label{OneBigTable}
\end{table*}

\begin{table*}[t]
\caption{Table\,\ref{OneBigTable} continued.}
\begin{center}\begin{tabular}{cccccccccccc}
\hline 
M$ _{ini} $&  P$ _{init} $& Case&t$ _{H} $& M$ _{H} $&t$ _{He} $ & M$ _{He} $&Y$ _{c} $&t$ _{tot} $ & $  \Delta$M&$ ^{26}Al $ &$ ^{26}Al $(SNB) \\
(M$ _{\odot} $) & (days)&& (Myr) & (M$ _{\odot} $)&(Myr) &(M$ _{\odot} $)&&(Myr) & (M$ _{\odot} $)&(M$ _{\odot} $) &(M$ _{\odot} $) \\
\hline
45$ ^{2} $ & - &-&4.35& 17.36 & 0.23 & - &0.34& 4.58 &22.34 &1.53e-05&-\\
& 3.2$ ^{4} $ &A& -& -& - & - & - &-&-&failed&6.54e-05\\
  & 7.8$ ^{3} $ &A& -&-  & - & - & - &-&16.28&1.54e-06&5.14e-05\\ 
  & 19.5$ ^{2} $ &B$ ^{6} $& 4.35& 17.36 & 0.30 & -& 0.19& 4.65 &22.04&2.01e-05&4.73e-05\\ 
  & 23.4$ ^{2} $ &B& 4.35& 17.36& 0.29 & - &0.20& 4.65&22.22&2.08e-05&4.20e-05\\
  & 42.0$ ^{2} $ &B& 4.35& 17.36 &  0.09& - & 0.74&4.44 &22.68&1.79e-05&4.18e-05\\ 
  & 69.9$ ^{2} $ &B&4.35 & 17.36& 0.12 & -& 0.63& 4.48 &22.59&1.73e-05&4.17e-05\\
\hline
50$ ^{2} $ & -&- &4.08& 19.82& 0.32& -&0.12& 4.40 &29.24 & 4.47e-05 &-\\
  & 8.1$ ^{2} $ &A& 4.09& 19.39 & 0.21 & -& 0.36& 4.31 &27.49&3.82e-05&6.76e-05\\ 
  & 14.0$ ^{2} $ &A&4.09 & 19.58 &  0.17& - &0.45& 4.26 &27.28&3.67e-05&6.29e-05\\ 
  & 21.7$ ^{1,2} $ &B&4.08 &19.75 & 0.27 & -&0.22 & 4.35 &25.51&3.02e-05&6.14e-05\\
  & 29.1$ ^{1,2} $ &B&4.08& 19.75 &  0.27& -& 0.22& 4.35 &25.28&2.92e-05&5.48e-05\\ 
  & 72.3$ ^{2} $ &B&4.08 & 19.75& 0.27& - &0.21& 4.36 &25.03&2.59e-05&5.46e-05\\
  \hline 
60$ ^{2} $ & - &-&3.70& 24.53 & 0.29 & - &0.13& 3.99 &34.01 & 6.83e-05&-\\
& 3.5$ ^{1,3} $&A & -& -& - & - & - &-&18.09&4.33e-06&1.24e-04\\
  & 7.2$ ^{1,3} $&A & -& - & - & - & - &-&19.65&6.21e-06& 1.05e-04\\ 
  & 14.9$ ^{1,2} $&B$ ^{6} $ &3.70 & 24.53 & 0.21 & - &0.30 &3.91 &28.81&4.28e-05&9.53e-05\\ 
  & 17.8$ ^{2} $ &B& 3.70&24.53 &0.20 & - &0.32&3.90&34.75&7.65e-05&7.91e-05 \\
  & 37.0$ ^{2} $&B& 3.70& 24.53 & 0.21 & -& 0.31&3.91&34.42&7.43e-05&7.87e-05\\ 
  & 92.2$ ^{1,2} $ &B& 3.70& 24.53& 0.21 & - &0.30&3.91&28.29&3.90e-05&7.80e-05\\
\hline
70$ ^{2,5} $ &- &-& 3.43 & 29.66 &0.27 & -& 0.13& 3.70& 39.30& 9.89e-05&-\\
\hline
80$ ^{2,5} $ &- &-& 3.23 & 34.56 & 0.26& - &0.14& 3.49& 44.62& 1.40e-4&-\\
\hline
\end{tabular}\end{center}
$ ^{1} $This run was terminated early due to reverse mass transfer.\\
$ ^{2} $This run was terminated before the end of helium burning.\\
$ ^{3} $This run was terminated before the end of hydrogen burning.\\
$ ^{4} $This run was terminated due to convergence issues.\\
$ ^{5} $This run was calculated for Section\,\ref{RATES}.\\
$ ^{6} $These systems are Case A according to the SNBs, but in the detailed simulations they are Case B
\label{OneBigTableCont.}
\end{table*}
\end{document}